\documentclass[twocolumn,showpacs,preprintnumbers,amsmath,amssymb,pre,longbibliography]{revtex4-1}

\usepackage{blindtext}
 
\usepackage{amsfonts}
\usepackage{color}    
\usepackage{graphicx}
\usepackage{braket} 
\usepackage{dcolumn}
\usepackage{bm}
\usepackage{subfigure}
\usepackage{amssymb}
\usepackage{multirow}
\usepackage{natbib}
\usepackage{amsmath}
\usepackage{mathtools}

\newcommand{\db}[2][]{\text{d}^{#1}#2}
\DeclarePairedDelimiter\Avr{\langle}{\rangle}%
\newcommand{\sgn}{\text{\normalfont Sgn}}

\begin{document}

\title{Dynamic properties and the roton mode attenuation in the liquid $^3\text{He}$: an \textit{ab initio} study within the self-consistent method of moments}


\author{A.V. Filinov}
\email{filinov@theo-physik.uni-kiel.de}
\affiliation{Institut für Theoretische Physik und Astrophysik, Christian-Albrechts-Universität zu Kiel, Kiel, Germany}

\author{J. Ara}
\affiliation{Instituto de Tecnolog\'{\i}a Qu\'{\i}mica,
Universitat Polit\`{e}cnica de Val\`{e}ncia-Consejo Superior de Investigaciones Cient\'{\i}ficas, Valencia, Spain}

\author{I.M. Tkachenko}
\email{imtk@mat.upv.es}
\affiliation{Departament de Matem\`{a}tica Aplicada, Universitat Poli\`{e}cnica de Val\`{e}ncia, Valencia, Spain}
\affiliation{Al-Farabi Kazakh National University, Almaty, Kazakhstan}

\date{24.01.2023}

\begin{abstract}

The dynamic structure factor and the eigenmodes of density fluctuations in the uniform liquid $^3$He are studied using a novel non-perturbative approach. This new version of the self-consistent method of moments invokes up to nine sum rules and other exact relations involving the spectral density, the two-parameter Shannon information entropy maximization procedure, and the \textit{ab initio} path integral Monte Carlo (PIMC) simulations which provide crucial reliable input information on the system static properties.
Detailed analysis of the dispersion relations of collective excitations, the modes' decrements and the static structure factor (SSF) of $^3$He at the saturated vapor pressure is performed. The results are compared to available experimental data~\cite{Albergamo2007,fak1994}. The theory 
reveals a clear signature of the roton-like feature in the particle-hole segment of the excitation spectrum with a significant reduction of the {\it roton} decrement in the wavenumber range $1.3\AA^{-1} \leq q\leq 2.2 \AA^{-1}$. 
The observed roton mode remains a well defined collective excitation even in the particle–hole band, where, however, it is strongly damped. Hence, the existence of the {\it roton-like} mode in the bulk liquid $^3$He is confirmed like in other strongly interacting quantum fluids~\cite{Noz2004}. The phonon branch of the spectrum is also studied with a reasonable agreement with the same experimental data being achieved. 

The presented combined approach permits to produce \textit{ab initio} data on the system dynamic characteristics in a wide range of physical parameters and for other physical systems.
\end{abstract}

\maketitle

\section{Introduction}

In the last decades, much effort has been devoted to the investigation of the  dynamics of strongly correlated Bose and Fermi liquids, with fluid bosonic $^4$He and fermionic $^3$He systems being the most representative examples deeply studied both theoretically and experimentally, and a great deal of information about these systems is nowadays accessible, see ~\cite{book_gases,book_plasma,filinov.2012pra,PhysRevA.87.033624,filinov.2016pra,dornheim.2021cpp} and references therein. 
The most relevant information on the dynamics of density fluctuations in quantum liquids can be extracted from the dynamic structure factor $S(q,\omega)$, which is experimentally measurable by means of the inelastic neutron scattering, see ~\cite{Godfrin2012}. 
The deepest and most accurate microscopic description of the dynamical response of liquid helium {\it in the limit of zero temperature} has been obtained within the sophisticated though perturbative correlated basis functions (CBF) theory ~\cite{CBF-book, PhysRevA.26.3536}. In particular, the progressive and continuous development of this theory has stimulated recently the production of new experimental data sets for $S(q,\omega)$ with the improved  experimental resolution in inelastic neutron scattering, see ~\cite{Godfrin2012} and references therein. Some tiny and delicate features of the excitation spectrum  (i.e. roton, double plasmon, etc) were found to be in a nice agreement with recent experimental data~\cite{Krotscheck2011, Schoerkhuber-book}. 

On the other hand, the most accurate tools to deal with {\it finite-temperature} properties are the quantum Monte Carlo (QMC) methods. In particular,  for bosonic systems, the zero-temperature and finite-temperature path integral Monte Carlo simulations demonstrated their very accurate predictive power for the equation of state and structural properties~\cite{RevModPhys.67.279}.
QMC methods are not restricted to a low-coupling limit and include, from the first principles, the exchange-correlation contributions to the thermodynamic observables. Moreover, the ensemble averages (population of excited states) can be accounted for very efficiently via the sampling of the statistical density matrix implemented by the path integral Monte Carlo (PIMC) method~\cite{RevModPhys.67.279,dornheim_physrep_18,filinov.2012pra,filinov-etal.2021ctpp}.

One main drawback of the statistical approach within the QMC method is its inability  to provide a direct access to the real-time evolution, similar to the simulations of classical systems using the molecular dynamics. In the latter case, the dynamic structure factor (DSF) can be resolved by a Fourier transform of the intermediate scattering function $F(q,t)$ propagating in real time $t$. Instead, in the framework of the Feynman's path integral formulation of quantum mechanics, a system of quantum particles is represented as an ensemble of trajectories propagating in the imaginary time 
$i \hbar \tau$ such that $0\leq \tau \leq 1/k_B T$, and the analogous intermediate scattering function (ISF) is sampled in the imaginary time domain so that the dynamic response can be accessed via the inverse Laplace transform of the ISF,  
\begin{eqnarray}
F(\mathbf{q},\tau)=\Avr{\rho_{\mathbf{q}}(0)\rho_{\mathbf{-q}}(\tau)}=\int\limits_{-\infty}^{\infty} S(\mathbf{q},\omega)\, e^{-\hbar \tau \omega}\, \db \omega,\ 
\label{Fqw}
\end{eqnarray}
with, $\rho_{\mathbf q}(\tau)=\sum_{i=1}^N e^{-i \mathbf{q}\,\mathbf{r}_i(\tau)}/N$, being the $N$-particle density operator 
at a certain value of the real parameter 
$\tau$.

It is well known, however, that this
inverse transformation is an ill-posed problem since a finite statistical noise always present in QMC data makes it practically impossible to extract a unique solution for $S(q,\omega)$ even in a uniform system. 

This long-standing problem related to the inversion of the QMC data has lead, recently, to the development of a class of elaborate regularization techniques relying on a set of physically justified constrains, which significantly reduce a number of possible solutions. Several methods have been suggested in recent years. To mention a few: the stochastic optimization (STO) method~\cite{mishchenko.prb2000,filinov.2012pra}, the method of consistent constraints~\cite{SCC13}, the GIFT method (genetic inversion via falsification of theories)~\cite{vitali.2010prb}, and the techniques based on the dynamic local field correction (DLFC) to the random phase approximation~\cite{dornheim.2018prl}. 
One of the mostly used approaches is the maximum entropy (ME) method which is based on the maximization of the entropy function with some \textit{a priori} expected behaviour~\cite{silver.1990prb}.

On the other hand, for classical systems, there also exists a broad class of reconstruction approaches 
which rely solely on the static properties of a system: 
the non-perturbative self-consistent method of moments (SCMM)~\cite{arkhipov-etal.2017prl, arkhipov-etal.2020pre} and the memory-function formalism~\cite{GV-book}.

For the theoretical approaches the most challenging is the ability to accurately resolve either rather sharp quasiparticle peaks (energy resonances) superimposed on a broad continuum formed by corresponding multi-particle excitations, like the phonon-maxon-roton eigenmode in $^4$He systems at very low temperatures, or, in the opposite case, predict sufficiently smooth density distributions without non-physical oscillations in the fermionic systems similar to $^3$He and the uniform electron gas (UEG), where the particle-hole decay processes dominate the excitation spectrum. In particular, both the GIFT and the STO methods have demonstrated their advantage over the ME techniques for $^4$He in the superfluid phase, while the DLFC-based reconstruction proved to be superior for the UEG in the warm dense matter regime.  

Below we present a new approach based on the combination of the SCMM which includes the physically relevant information (low and high energy excitations) via a set of sum rules (the response function frequency power moments), and the dynamical information contained in the intermediate scattering function (ISF), $F(q,\tau)$. We demonstrate, in this case, a significant improvement and stability of the present SCMM+ISF inversion method in comparison to the 
application of the stochastic multidimensional optimization~\cite{mishchenko.prb2000,filinov.2012pra}.
The temperature and density dependence of $S(q,\omega)$ for $^3$He systems can be studied within the present approach "from first principles", with a future goal to extent the method to more broad practical applications and validation via a direct comparison with available experimental data~\cite{Glyde-book, Dobbs-book}, see also~\cite{Ara_2022}.

The rest of the paper is organized as follows. The setup of the physical problem is provided in the next Section. The quantum version of the SCMM is discussed in Sec.~III. In Sec.~IV, we report the results achieved for the dynamic response, excitation spectrum and the dispersion relations.  Our main conclusions are summarised in Sec. V. Some computational and mathematical details are provided in the Appendices.

\section{Physical model}

Since the pioneering studies of "different types of waves that can propagate in a Fermi liquid and their absorption, both at absolute zero and at non-zero temperatures" initiated by L.D. Landau back in 1956-57~\cite{Landau1, Landau2}, see also~\cite{Abrikosov:1958} and~\cite{PN-book}, understanding of the dynamics of correlated many-body
quantum systems is still a challenge \cite{Godfrin2012}. The theory of Fermi liquids of Landau was intrinsically a perturbative theory, but no small parameters can be found in a typical "contemporary" Fermi liquid. Indeed, in $^3$He with the density of atoms corresponding 
to the saturated vapor pressure $\rho=1.6355\times 10^{22}$ $cm^{-3}$, the Wigner-Seitz radius is $2.4$ $\AA$, 
the Fermi wavenumber $k_{F}=7.9\times 10^7$ $cm^{-1}$\ while 
the de Broglie wavelength is $8.4$ $\AA$. This results in the Brueckner (density) parameter $r_{s}=4.62$ and the Fermi temperature $T_{F}=4.9521$ $K$. Hence, for $T=2-5$ $K$, the degeneracy temperature parameter, $\theta =T/T_{F}$, is about $1$ or smaller. These conditions correspond to the warm dense matter regime nowadays closely studied  using the quantum Monte-Carlo methods~\cite{dornheim-etal.2022sr,VFilinovHe3}. Besides, we understand that $^3$He at the above conditions is a strongly interacting Fermi fluid and that it is important to study its properties in order to understand general characteristics of Fermi liquids when the exchange-correlation effects strongly influence the properties of the quasi-particle excitation spectrum, going far beyond the standard Landau description.

This observation motivated us to employ here the {\it non-perturbative} self-consistent method of moments, whose classical version has proved its viability in a number of previous publications~\cite{arkhipov-etal.2017prl, arkhipov-etal.2020pre, arkhipov-etal.2010pre, ara-etal.2021pop}. 
This theoretical approach not only requires much less computation time, but is also free from the typical computational limitation with respect to the range of variation of the thermodynamic parameters, as it is shown below. 

In our calculations an accurate potential of inter-atomic interaction in $^3$He constructed by fitting to the second virial coefficient~\cite{JChemPhys79} and valid in the temperature range from $1.5$~K to $1475$~K is employed, see Appendix B. 

Computational details of the path integral Monte Carlo simulations we carry out are provided in Appendix A. There we discuss the severity of the fermion sign problem in the fermionic simulations of $^3$He and benchmark our results against the recent PIMC data by Dornheim {\it et al.}~\cite{dornheim-etal.2022sr}. 

In addition, we conduct quantum simulations with Boltzmann statistics (of distinguishable particles) which allows us to establish the importance of Fermi statistics with variation 
 in the temperature $1.5\text{K} \leq T\leq 6$K  (or in the degeneracy factor $0.43 \leq \theta \leq 1.3$) for the evaluation of most relevant thermodynamic properties, such as the static structure factor $S(q)$, the static density response function $\chi(q,0)$, the kinetic and the potential energy, and the momentum distribution function $n(q)$. Like in the path integral simulations carried out by D. Ceperley~\cite{RevModPhys.67.279} for condensed helium $^4$He, we find no significant impact of the exchange effects at low temperatures $T< 2$K, except of $n(q)$. The details are provided in Appendix A. Therefore, to avoid the fermion sign problem and to study the dynamical properties in $^3$He at $T=1.5$K (comparable to experimental temperatures) within our novel approach based on the method of moments (see below) we use as the input the static characteristics evaluated with Boltzmann statistics. 
For $T\geq 2$K full fermionic PIMC simulations are feasible, and no further approximations are involved.    

As an alternative route to overcome the fermion sign problem, V.~Filinov {\it et al.}~\cite{VFilinovHe3} have recently proposed a new approach where the exchange interaction effects are included via a positive semidefinite Gram determinant which significantly simplifies the calculations. The verification of this approach can be an interesting topic for a future analysis.

\section{Self-consistent method of moments}\label{SSMMSec}

From the mathematical point of view the method of moments reduces the solution of a certain (dynamical) physical problem to the solution of the truncated Hamburger problem of moments~\cite{shohat-book, akhiezer-book, krein-book} for the spectral density designed to make the latter as simple as possible. The truncated Hamburger problem
consists in the reconstruction of a positive spectral density by a limited number of its power frequency moments which are effectively the sum rules known independently. Excluding some specific, though mathematically important, cases~\cite{PhysRevE.71.066501} such a problem has an infinite number of solutions parameterized by the so called Nevanlinna (parameter) function (NPF). Nevertheless, additional physical considerations permit to choose an adequate NPF and construct a unique relatively simple solution which proves to be in (even quantitative) agreement with available experimental or simulation data~\cite{tkachenko-book}.

Contrary to the Stieltjes, Hausdorff~\cite{krein-book}, and the physically motivated "local"~\cite{pamm.2014} problems, the distribution density (the spectral density) is
extended in the Hamburger problem to the whole real axis of frequencies so that if we select a symmetric spectral density the working formulas simplify significantly (see, nevertheless~\cite{chapter2}).

We wish to point out that the mathematical foundations of the method of moments are firmly established in a number of theorems, 
see~\cite{nevanlinna-book, krein-book, UTFC} and~\cite{sb1, sb2}; applications of this approach in plasma physics were described in~\cite{tkachenko-book, PST18}, and (in the 2d geometry) in~\cite{PhysRevA.46.7882}, for 3d model one-component~\cite{PhysRevE.48.2067} and two-component~\cite{PhysRevE.76.026403, PhysRevE.81.026402} Coulomb systems, see also~\cite{PhysRevE.90.053102}.
In the case of classical plasmas the method has been 
justified by a quantitatively successful comparison to the simulation data~\cite{arkhipov-etal.2017prl, arkhipov-etal.2020pre}. The present work is practically the first attempt to extend the self-consistent method of moments to 
the dynamical properties of non-Coulomb and quantum systems, 
see also our recent development for the quantum uniform electron gas~\cite{UEGarXiv}.

The self-consistency of the moment approach means, from our point of view, that its only input are the static data, i.e., the static structure factor and the static dielectric permeability. The efficiency of the moment approach depends not only on the number of sum rules employed in the calculations, but, even more, on the model of the non-phenomenological Nevanlinna parameter function, though the moment conditions are satisfied independently of the (mathematically correct) model of the latter. 


Precisely, in the present paper we consider the response function $\chi \left( q,\omega \right) $ related to the dynamic
structure factor $S\left( q,\omega \right) $ by the fluctuation-dissipation
theorem,%
\begin{equation}
\operatorname{Im}\chi \left( q,\omega \right) =-\frac{\pi \rho }{\hbar }\left( 1-\exp
\left( -\beta \hbar \omega \right) \right) S\left( q,\omega \right) ,
\label{fdt}
\end{equation}%
(the inverse temperature $\beta ^{-1}=k_{B}T$, $k_{B}$ and $\hbar $\ being the Boltzmann and Planck constants), and introduce the spectral density (the distribution function) in the form 
\begin{equation}
M\left( q,\omega \right) =-\frac{\operatorname{Im}\chi \left( q,\omega \right) }{\pi
\omega }=\frac{\rho\left[1-\exp \left( -\beta \hbar \omega \right)\right] }{\hbar \omega }%
S\left( q,\omega \right) ,  \label{M}
\end{equation}%
which is the positively defined (even) function of the frequency $\omega $.

\subsection{Frequency power moments}\label{FreqMom}

By definition, its power moments or sum rules are
\begin{equation}
\mu _{\ell }\left( q\right) =\int_{-\infty }^{\infty }\omega ^{\ell }M\left(q,\omega \right) d\omega \ , \label{moms}
\end{equation}
and we take into consideration a limited odd number of them, $\; \ell =0,1,\ldots ,2\nu$ with $\nu =0,1,2,\ldots ,$
and with vanishing (due to the symmetry of the spectral density $M\left(q,\omega \right)$ ) odd-order moments,
\begin{equation*}
\mu _{1}\left( q\right) =\mu _{3}\left( q\right) =\ldots=\mu _{2\nu+1}\left( q\right) =0.
\end{equation*}
By virtue of the detailed balance condition, 
\begin{equation*}
S(q,-\omega)=\exp \left( -\beta \hbar \omega \right) S(q,\omega), 
\end{equation*}

\begin{equation*}
\mu _{\ell }\left( q\right) =\frac{\rho}{\hbar}
\left[ 1+\left( -1\right) ^{\ell }\right]\int_{-\infty }^{\infty }\omega ^{\ell -1}S\left( q,\omega \right) d\omega
 \ . 
\end{equation*}%
In particular, due to the $f$-sum rule \cite{PN-book}, 
\begin{equation*}
\mu _{2}\left( q\right)=\frac{2\rho}{\hbar }\int_{-\infty
}^{\infty }\omega S\left( q,\omega \right) d\omega  =\frac{\rho q^{2}}{m} \ . 
\end{equation*}
The rest of the moments can be calculated independently in terms of the system static characteristics.
An explicit expression for the fourth moment $\mu _{4}\left( q\right)$ is provided in Appendix B. 

\subsection{Nevanlinna's formula}

Given $\left( 2\nu +1\right) $ moments, Nevanlinna's theorem~\cite{nevanlinna-book, krein-book, akhiezer-book} establishes a bijection between the spectral density and the so-called
Nevanlinna (parameter) function $Q_{\nu }\left( z;q\right) $, a response (Nevanlinna class~\cite{krein-book}) function which, in addition, must satisfy the following asymptotic condition: 
\begin{equation*}
\lim_{z\rightarrow\infty }\frac{Q_{\nu }\left( z;q\right) }{z}=0\ ,\quad \operatorname{Im}z>0:
\end{equation*}%
\begin{equation}
\int_{-\infty }^{\infty }\frac{M\left( q,\omega \right) d\omega }{z-\omega }=%
\frac{E_{\nu +1}\left( z;q\right) +Q_{\nu }\left( z;q\right) E_{\nu }\left(
z;q\right) }{D_{\nu +1}\left( z;q\right) +Q_{\nu }\left( z;q\right) D_{\nu
}\left( z;q\right) }\ .  \label{Nt}
\end{equation}%
Notice that Nevanlinna's theorem can be proved on the basis of
the technique of generalized resolvents of M.G. Krein, see \cite{UTFC}.
Further details of the method of moments can be found in \cite{PhysRevE.71.066501, PST18, sb1, sb2}.

The coefficients of the linear-fractional transformation \eqref{Nt} between the spectral density $M\left( q,\omega \right) $ and the Nevanlinna parameter function
$Q_{\nu }\left( z;q\right)$ are orthogonal polynomials $D_{\nu }\left( \omega ;q\right)$ (with
the weight $M\left( q,\omega \right) $), 

\begin{equation*}
\int_{-\infty }^{\infty }D_{\nu }\left( \omega ;q\right) D_{\nu ^{\prime
}}\left( \omega ;q\right) M\left( q,\omega \right) d\omega =\left\Vert
D_{\nu }\left( \omega ;q\right) \right\Vert ^{2}\delta _{\nu ^{\prime }\nu
}\ ,
\end{equation*}
and the polynomials $E_{\nu }\left( z;q\right) $ which are their conjugate ones:
\begin{equation*}
E_{\nu }\left( z;q\right) =\int_{-\infty }^{\infty }\frac{D_{\nu }\left(
\omega ;q\right) -D_{\nu }\left( z;q\right) }
{\omega -z}M\left( q,\omega
\right) d\omega \ ,\quad \operatorname{Im}z>0.
\end{equation*}%

It is obvious that both sets of polynomials depend only on the 
moments (sum rules) and this is why the spectral density constructed using the Navanlinna formula (\ref{Nt}) with a mathematically correct Nevanlinna function $Q_{\nu }\left( z;q\right) $ satisfies the sum rules automatically. The latter non-phenomenological parameter is formally equivalent to the 
dynamical local-field correction (with respect to the random-phase approximation (RPA)) 
function, but we do not have to oblige it to satisfy the higher-order sum rules, as it is usually done in order to go beyond the RPA.
The (monic) polynomials $D_{\nu }\left( \omega ;q\right)$ can be easily constructed from the basis of the space of polynomials $\left\{ 1,\omega
,\omega ^{2},\omega ^{3},...\right\} $ by the Gram-Schmidt procedure. 
The explicit form of both sets of polynomials involved in the present work (for $\nu =2$ and $\nu =4$ ) can be found in Appendix C.

Being the response (Nevanlinna) function, the density response function satisfies the Kramers-Kronig relations so that in the upper half-plane of the complex frequency, $\operatorname{Im}z>0,$
\begin{equation}
\chi \left( q,z\right) =\frac{1}{\pi }\int_{-\infty }^{\infty }\frac{\operatorname{Im}\chi \left( q,\omega \right) d\omega }{\omega -z}= \label{chi1}
\end{equation}%
\begin{equation}
-\mu _{0}\left( q\right)+z\frac{E_{\nu +1}\left( z;q\right) +Q_{\nu }\left( z;q\right) E_{\nu}\left( z;q\right) }{D_{\nu +1}\left( z;q\right) +Q_{\nu }\left( z;q\right)
D_{\nu }\left( z;q\right) }\ ,  \label{chi2}
\end{equation}%

and, in particular, 

\begin{equation*}
\mu _{0}\left( q\right) =-\lim_{z\rightarrow 0}\chi \left( q,z\right) \equiv
-\chi \left( q,0\right) \ .
\end{equation*}%

It is important for the comparison to the simulation or experimental data that if we can invoke $\left( 2\nu +1\right) $
moments or sum rules, it stems from the Nevanlinna theorem \eqref{Nt} that 

\begin{equation}
M\left( q,\omega \right) =-\operatorname{Im} N_{\nu }\left( q,\omega \right)/\pi ,   \label{Mnu}
\end{equation}%
where
\begin{equation}
N_{\nu }\left( q,\omega \right) = \frac{E_{\nu +1}\left(
\omega ;q\right) +Q_{\nu }\left( \omega ;q\right) E_{\nu }\left( \omega
;q\right) }{D_{\nu +1}\left( \omega ;q\right) +Q_{\nu }\left( \omega
;q\right) D_{\nu }\left( \omega ;q\right)},   \label{Nnu}
\end{equation}%
or that in this approximation 
\begin{equation}
S\left( q,\omega \right) =\frac{\hbar \omega \operatorname{Im} N_{\nu }\left( q,\omega \right)}{\pi \rho \left[\exp \left( -\beta \hbar
\omega \right) -1\right]} . \label{SNnu}
\end{equation}%

Thus, the calculation of the DSF is reduced to the knowledge of the Nevanlinna parameter function
$Q_{\nu }\left( \omega;q\right)$.

In what follows it is convenient to introduce, in addition to the moments, the set of characteristic frequencies 
\begin{equation}
\omega _{j}^{2}\left( q\right) =\frac{\mu _{2j}\left( q\right) }{\mu
_{2j-2}\left( q\right) }=\frac{\int_{-\infty }^{\infty }\omega
^{2j-1}S\left( q,\omega \right) d\omega }{\int_{-\infty }^{\infty }\omega
^{2j-3}S\left( q,\omega \right) d\omega }\ ,\quad j=1,2,...,\nu.
\label{omega_set}
\end{equation}

Due to the Cauchy-Schwarz-Bunyakovsky inequalities, the conditions 
\begin{eqnarray}
0 \leq \omega_{1}\left( q\right) \leq \omega_{2}\left( q\right) \leq ... \leq \omega_{j}\left( q\right) 
\label{CSB}
\end{eqnarray}
should be satisfied to warrant the fulfillment of the required mathematical properties
of the Nevanlinna and the spectral density response function $\chi \left( q,\omega \right). $ 
Mathematically admissible equalities in \eqref{CSB} can occur only under very specific physical conditions not valid here.
Observe that in the present work we employ in (\ref{chi1} - \ref{SNnu}) the five-moment approximation ($\nu=2$) with the Nevanlinna parameter function determined in the nine-moment approximation ($\nu=4$), which implies that the asymptotic expansion ($\ref{high_w}$) is exactly applicable.

\subsection{The dynamical Nevanlinna function}

In our studies of classical one-component 
plasmas~\cite{PhysRevE.81.026402, arkhipov-etal.2017prl, arkhipov-etal.2020pre}, 
due to the effective absence in the spectrum of the Rayleigh
zero-frequency mode, it was enough to work
in the 5-moment static approximation, i.e. invoking only 5 moments
($\nu =2$) and modelling the Nevanlinna function by its
zero-frequency limiting value, 
\begin{equation}
Q_{2}\left( z;q\right) =Q_{2}\left( 0;q\right) =ih_{2}\left( q\right) 
\label{sa}
\end{equation}%
with the positive static parameter 
\begin{equation*}
h_{2}\left( q\right) =\frac{\omega _{2}^{2}\left( q\right) }{\sqrt{2}\omega
_{1}\left( q\right) } \label{h2} .
\end{equation*}%
These simple approximations do not seem to be sufficient to describe complicated dispersion and decay processes in liquid $^{3}$He. 
Thus, in the present work, we go beyond the static approximation (\ref{sa}) and use four additional sum rules. This permits to construct a dynamical 5-moment 
Nevanlinna function $Q_{2}\left( z;q\right) $, see Appendix C for more details.
Now, in the nine-moment approximation the dynamic structure factor, see Eq.~(\ref{SNnu}), can be explicitly evaluated using the resulting
\begin{equation}
Q_{2}\left( \omega ;q\right) =-\frac{\Omega _{1}^{2}\left( \omega
+ih_{4}\right) }{\omega \left( \omega +ih_{4}\right) -\Omega _{2}^{2}}. \label{Q2}
\end{equation}%
We introduce here the following static parameters determined only by the set of the characteristic frequencies $\{\omega_1\left( q\right),\omega_{2}\left( q\right),\omega_{3}\left( q\right),\omega_{4}\left( q\right)\}$, %
\begin{equation*}
h_{4}=\frac{\omega _{3}^{2}\left( \omega _{4}^{2}-\omega _{3}^{2}\right)
\left( \omega _{2}^{2}-\omega _{1}^{2}\right) }{\omega _{1}\sqrt{2\left(
\omega _{3}^{2}-\omega _{1}^{2}\right) \left( \omega _{3}^{2}-\omega
_{2}^{2}\right) ^{3}}}\ ,
\end{equation*}%
\begin{equation*}
\Omega _{1}^{2}=\frac{\omega _{2}^{2}\left( \omega _{3}^{2}-\omega
_{2}^{2}\right) }{\omega _{2}^{2}-\omega _{1}^{2}}\ ,
\end{equation*}%
\begin{equation*}
\Omega _{2}^{2}=\frac{\omega _{3}^{2}\left( \omega _{4}^{2}-\omega
_{3}^{2}\right) }{\left( \omega _{3}^{2}-\omega _{2}^{2}\right) }-\frac{%
\omega _{1}^{2}\left( \omega _{3}^{2}-\omega _{2}^{2}\right) }{\left( \omega
_{2}^{2}-\omega _{1}^{2}\right) }>0.
\end{equation*}%

Finally, the asymptotic expansion of the spectral function (\ref{chi1}) once more shows that the specific
(mathematically correct) form of the Nevanlinna parameter function, 
in particular $Q_{2}\left(\omega ;q\right)$, (\ref{Q2}), does not influence the corresponding high-frequency asymptotic behaviour of $\chi \left( q,\omega\right)$:

\begin{eqnarray}
\lim\limits_{\omega \rightarrow\infty}\chi\left( q,\omega\right)\simeq &&\frac{\omega _{1}^{2}}{\omega^{2}}
+\frac{\omega _{1}^{2}\omega _{2}^{2}}{\omega^{4}}+
\frac{\omega _{1}^{2}\omega _{2}^{2}\omega _{3}^{2}}{\omega^{6}}
+\frac{\omega _{1}^{2}\omega _{2}^{2}\omega _{3}^{2}\omega _{4}^{2}}{\omega^{8}}+ \nonumber \\
&&\frac{iH}{\omega^{9}}+O\left( \frac{1}{\omega^{10}}\right),\label{high_w}
\end{eqnarray}

where
\begin{equation}
H=\frac{h_{4}\omega _{1}^{2}\omega _{2}^{2} \left( \omega _{1}^{2}\omega _{2}^{4}+
\omega _{2}^{2}\omega _{3}^{4}-2\omega _{1}^{2}\omega _{2}^{2}\omega _{3}^{2}+
\omega _{1}^{2}\omega _{3}^{2}\omega _{4}^{2}-\omega _{2}^{2}\omega _{3}^{2}\omega _{4}^{2}\right)}
{\omega^{9}\left( \omega _{2}^{2}-\omega _{1}^{2}\right)} ,
\end{equation}%
which implies that at high frequencies the DSF decreases at least as $\omega^{-9}$ so that its seventh power 
frequency moment or the moment $\mu_8\left( q\right)$
converge. 

In what follows, we employ the nine-moment density response function (\ref{chi2}) and the dynamic structure factor (\ref{SNnu}) (the nine-moment approximation, 9MA) to study the eigenmodes and other dynamical characteristics of the liquid $^{3}$He.

\subsection{Shannon information entropy}\label{Ent_Sec}
As it was pointed out in the Introduction, the problem of reconstruction of the spectral density from a set of static characteristics can lead to a quite broad class of mathematical solutions. This class can be significantly reduced by invoking the higher-order moments which specify the high-frequency behaviour, see Eq.~(\ref{high_w}). 

We understand that the frequencies $\omega_{3(4)}\left( q\right)$, formally introduced above, are determined by the three- and four-particle static correlation functions. The $\textit{ab initio}$ QMC data for them can be achieved though precise expressions for the higher-order moments in terms of these correlation functions are not yet available. The precision of the latter seems to be a problem and, as we show, to achieve quantitative agreement with the simulation data, we need to possess highly precise values of the sixth and the eighth moments. This is why we determine the values of the frequencies $\omega_{3(4)}\left(q\right)$ by means of the Shannon information entropy maximization (EM) procedure~\cite{shannon.1948,khinchin.53umn,zubarev-book,jaynes.1957pr}, see also \cite{tkachenko-book}. As an independent check of the applicability of this approach, we verify the reconstructed spectral density, $S(q,\omega)$, by the evaluation of the ISF using the r.h.s. of Eq.~(\ref{Fqw}). This model ISF is compared to the {\it physical} one obtained in our PIMC simulations~\cite{filinov-etal.2021ctpp} using the sampled particle trajectories $\{\mathbf{r}_1(\tau),\mathbf{r}_2(\tau),\ldots,\mathbf{r}_N(\tau),\; 0\leq \tau\leq  \beta\}$ and the corresponding autocorrelation function of the density operator $\rho_{\mathbf{q}}(\tau)$. The particle configurations (or the microstates in the statistical canonical ensemble) are sampled from the $N$-particle density matrix and, in principle, contain full physical information determined by the exchange-correlation effects and the density fluctuations induced in the system with the model Hamiltonian, $\hat H =\sum \hbar^2/2m \nabla^2+ \sum V(r_{ij})$, with $V(r_{ij})$ being the $^3$He interatomic interaction potential~\cite{JChemPhys79} specified below in Eqs.(\ref{V1})-(\ref{Vint}). 

This comparison allows us to judge on the validity of the entropy principle to describe the quasiparticle spectrum in its different $q$-wavenumber domains, i.e. in the phonon, the roton and the single-particle range. This analysis will be presented in detail in Sec.~\ref{results}.

Thus, within the EM procedure, we introduce the two-parameter Shannon entropy functional defined  by the spectral density function
\begin{eqnarray}
E\left(q;\tilde\omega\right) 
=-\int\limits_{-\infty}^{\infty}M\left(q,\omega;\tilde\omega\right) \ln \left[ M\left(q,\omega;\tilde\omega\right) \right]
\db\omega ,\
\label{sh0}
\end{eqnarray}%
where $\tilde\omega=\{\omega_1,\omega_{2},\omega_{3},\omega_{4}\}$,
and resolve the corresponding maximization problem with respect to
$\omega_{3(4)}(q)$, with $\omega_{1(2)}(q)$ being fixed by the known sum rules.
To solve the extremum conditions for two unknown frequencies, 
\begin{eqnarray}
\int\limits_{-\infty }^{\infty } &&\left\{ \frac{\partial M\left(q,\omega;\tilde\omega\right)}{\partial \omega_{3(4)}}\ln \left[
e M\left(q,\omega;\tilde\omega\right) \right] \right\}
\db\omega=0 \ ,
\label{sh1}
\end{eqnarray}%
we employ the Newton-Raphson method. As the starting points in the gradient descent method the higher characteristic frequencies $\omega_{3(4)}$ are chosen randomly in the range $\hbar \omega_2 <\hbar \omega_{3(4)} \leq 1500$K and are required to satisfy the (exact) inequality conditions~(\ref{CSB}). 
The Hessian of the entropy~(\ref{sh0}) was studied to warrant the satisfaction of the maximization condition.

\section{Results}\label{results}

\begin{figure}[]
\hspace{-0.2cm}
\includegraphics[width=0.48\textwidth]{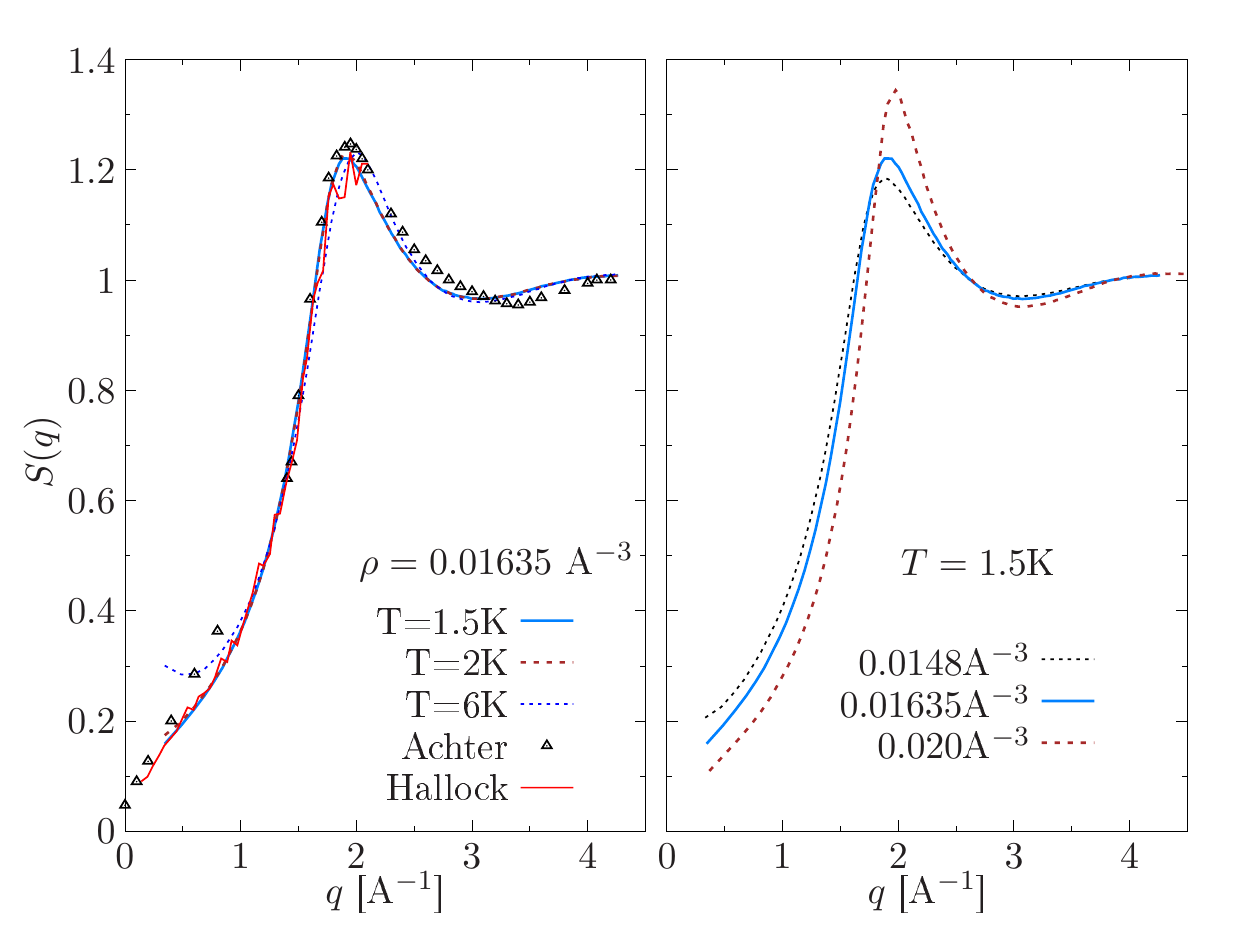}
\vspace{-0.5cm}
\caption{{\it Left:} The static structure factor $S(q)$ for the density $\rho=0.01635 \AA^{-3}$, i.e., at the saturated vapor pressure (SVP) and three temperatures $T=1.5, 2$ and $6$K. The PIMC data are compared to the X-ray scattering data by Achter~\cite{Achter1969} ($T=0.56$~K) and Hallock~\cite{Hallock1972} ($T=0.40$~K). 
{\it Right:} Analysis of the density dependence: the SVP $S(q)$ at $T=1.5$~K 
and two different densities, $\rho=0.0148 \AA^{-3}$ and $\rho=0.020 \AA^{-3}$. 
See more details in Fig.~\ref{fig:SkShann}.}  
\label{fig:Sk}
\end{figure}

\subsection{Static properties}

In Fig.~\ref{fig:Sk} we compare our PIMC results with the experimental data obtained for the static structure factor (SSF) using the X-ray scattering method~\cite{Achter1969,Hallock1972}. Though the experiments were performed at a lower temperature 
($T\sim 0.5$ $K$) than the present PIMC simulations ($T=1.5$ and $2$ $K$), the experimental data and 
the theoretical predictions are found to be in a very good agreement: the present PIMC data are very close 
to the measurements of Ref.~\cite{Hallock1972} for $q\lesssim 2 \AA^{-1}$. 
On the other hand, we believe that the observed discrepancy with the data of 
Ref.~\cite{Achter1969} 
for larger momenta under the SVP conditions, i.e, for $\rho=0.01635 \AA^{-3}$, 
can be attributed to the density effect. 
An example of a system with a slightly lower (higher) density, $\rho=0.0148 \AA^{-3}$ ($\rho=0.02 \AA^{-3}$), at the temperature $T=1.5$ $K$ is demonstrated on the right-hand panel (see the dashed black and solid red curves). 
One can clearly observe that a more compressed $^3$He system 
possesses a higher amplitude in the first SSF peak and a slight shift of the subsequent minimum to higher $q$ values compared to the SVP density case, $\rho=0.01635 \AA^{-3}$ (solid blue line). 

As to the temperature effects analysed on the left-hand panel, we can conclude that they are indeed very weak for $T\lesssim 2$ $K$ and $q \gtrsim 1 \AA^{-1}$. 
The PIMC data at $T=1.5$K (Boltzmann statistics, $N=100$) and $T=2$K (Fermi statistics, $N=38$) nearly coincide with the experimental data~\cite{Hallock1972} even in the long-wavelength limiting case, 
where the SSF-value is specified by the isothermal compressibility. The $N$-dependence in $S(q)$ (the finite size effects) was found to be negligible (see Appendix A3).
Some noticeable temperature effects start to be observed 
at $q \lesssim 1 \AA^{-1}$ for temperatures $3K \leq T \leq 6K$.

\subsection{Shannon entropy maximization: solutions for higher frequency moments}\label{optSH}

\begin{figure}[]
\hspace{-0.2cm}
\includegraphics[width=0.48\textwidth]{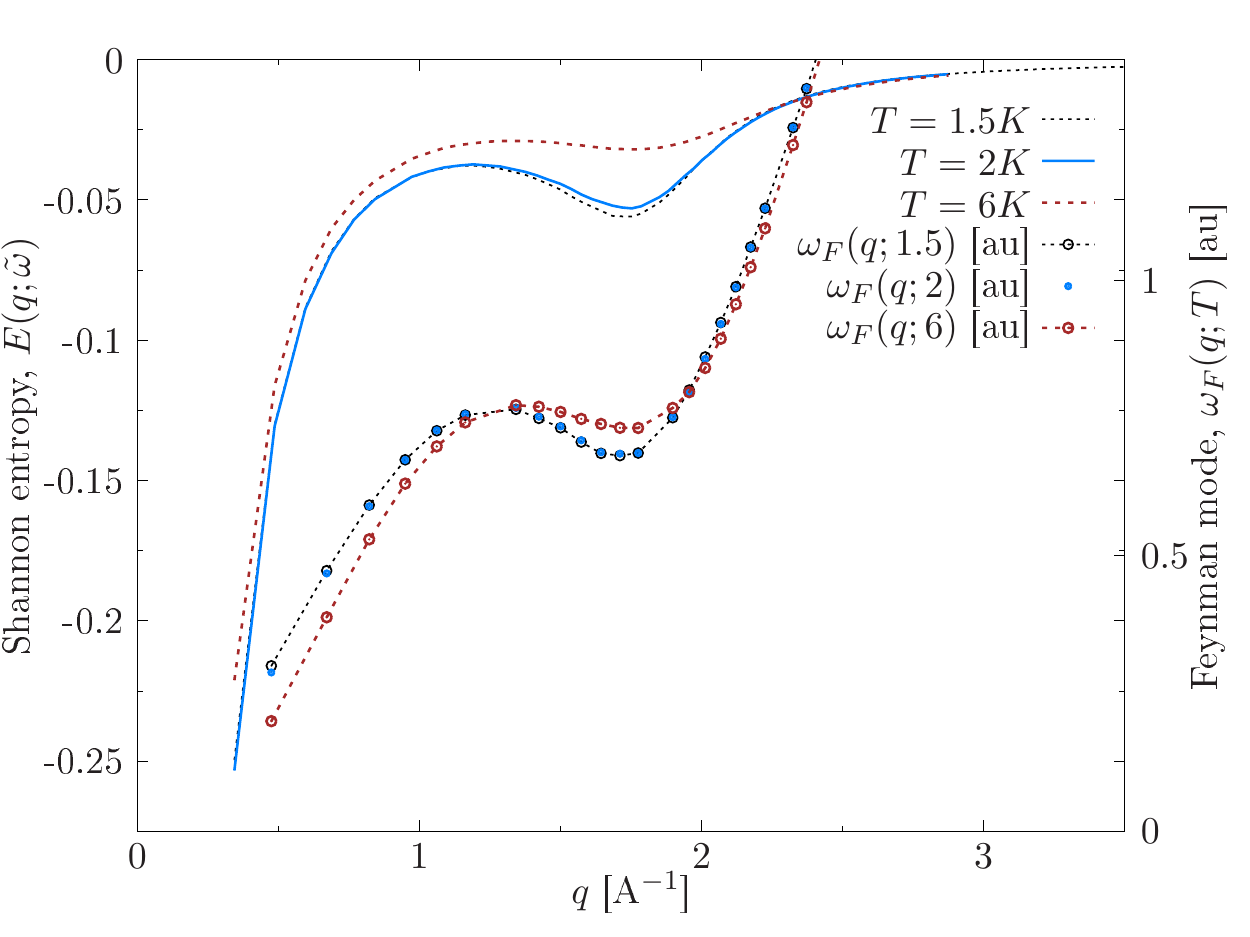}
\vspace{-0.5cm}
\caption{The Shannon information entropy $E(q;\tilde \omega)$, see Eq.~(\ref{sh0}), at the SVP density $\rho=0.01635 \AA^{-3}$ and temperatures $T=1.5, 2$ and $6$~K. The observed non-monotonic decay with a local maximum around $q\sim 1.8 \AA^{-1}$ is 
directly related to the expected for $^3$He phonon-maxon-roton dispersion relation estimated here from the Feynman relation, $\omega_F(q;T)=(\hbar q^2/2m)/S(q)$ (see the curves with the filled symbols).}
\label{fig:Entropy}
\end{figure}

The main goal of the application of the Shannon information entropy maximization procedure discussed in Sec.~\ref{Ent_Sec}, is to impose additional physically justified restrictions on a wide class of mathematical solutions which satisfy the known set of power moments $\{\mu_0,\mu_2,\mu_4\}$ exactly. Since within the present approach 
we wish to reconstruct the quasiparticle spectral density, possible physical solutions should belong to a class of smooth functions which can take into account possible decay and interaction processes of quasiparticles. As a result, sharp energy resonances in some range of variation of the wavenumber $q$ can be "washed out", but the center of mass of the resulting spectral density will be located near the position of the original dispersion relation. The half-width of the dynamic structure factor will characterize, in this case, the strength of the decay/interaction effects. This, in turn, should lead to a non-monotonic behavior of the Shannon entropy, as the latter is constructed from the distribution function proportional to $S(q,\omega)$, see Eq.~(\ref{M}). 

As an example, in Fig.~\ref{fig:Entropy} we present the wavenumber dependence of the Shannon information entropy $E(q;\tilde \omega (q))$ 
obtained by the solution of the maximization problem. The low-temperature ($T=1.5\text{K}, 2$K) and high-temperature ($T=6$K) cases are compared. 
First, at the  wavenumbers $q\lesssim 1 \AA^{-1}$ we observe a fast increase, 
which coincides with the phonon~\cite{Landau2, Pitaevskii:1967} part of the excitation spectrum roughly 
estimated here from the Feynman frequency $\omega_F(q)=(\hbar q^2/2m)/S(q)$.  
The expected increase in the decrement of the phonon mode with growing $q$ 
should smooth the peak of the phonon-mode down and lead to the increase in the entropy function. 
The observed maximum around the maxon segment ($q_M\sim 1.3 \AA^{-1}$) is followed by the local minimum around $q_R\sim 1.8 \AA^{-1}$ 
corresponding to the roton mode branch predicted by the Feynman frequency $\omega_F(q)$. As we see, comparing different temperature cases 
($T=1.5$K and $T=6$K), the depth of the roton minimum correlates with the amplitude of the entropy function around these wavenumber values. 

\begin{figure}[]
\hspace{-0.4cm}
\includegraphics[width=0.51\textwidth]{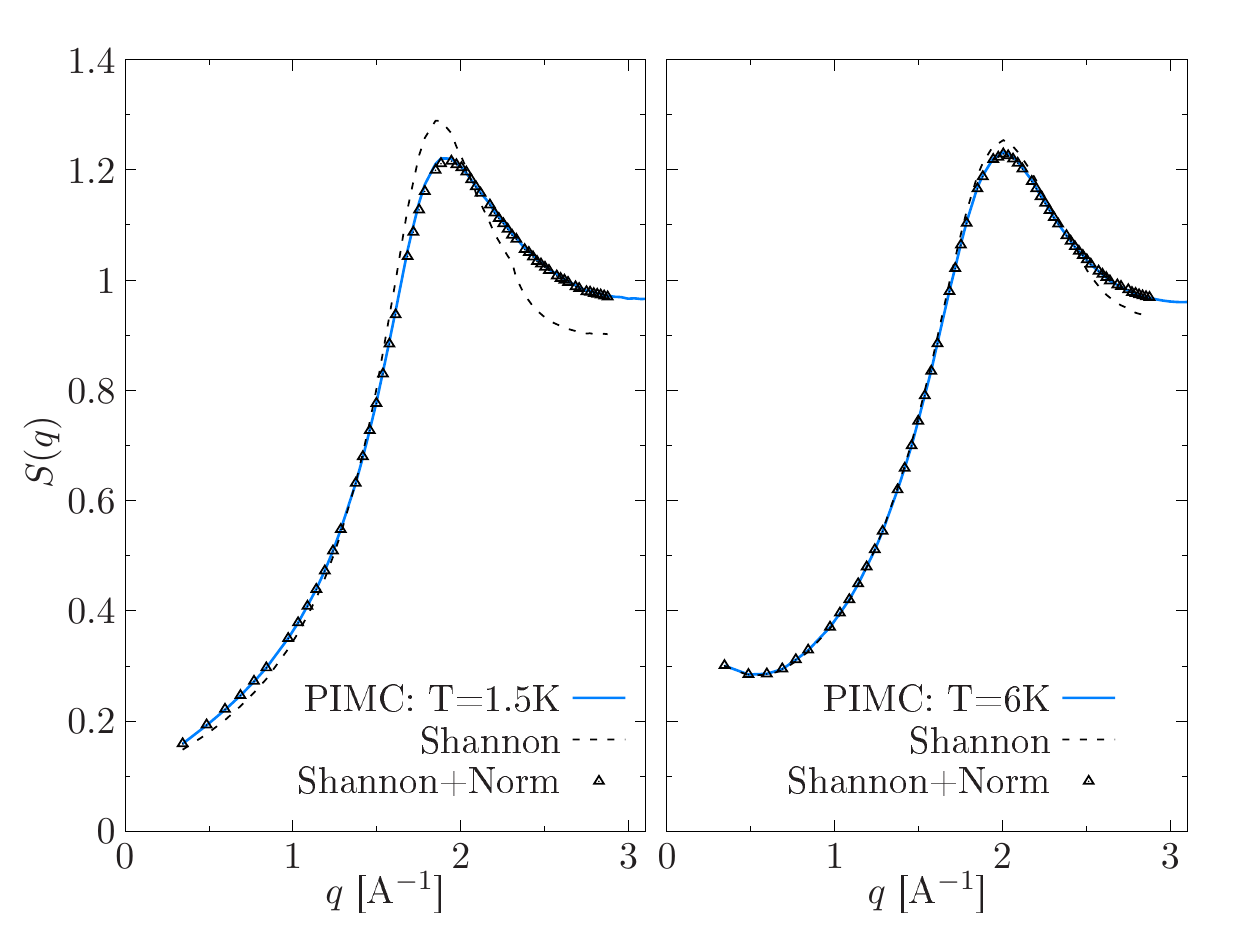}
\vspace{-0.2cm}
\caption{The static structure factor at the SVP density and temperatures $T=1.5$~K and $6$~K. The PIMC data (solid blue line) is compared to the 9MA result obtained with two Shannon entropy procedures: 1) the characteristic frequencies  $\omega_{3(4)}$ are used for the maximization of the information entropy $E(q;\tilde \omega)$ as {\it free independent parameters} (dashed black lines); 2) the same problem in solved with an additional requirement to satisfy the zero frequency moment for $S(q,\omega)$: the normalization factor, $\mu^{S}_0(q)=\int \db \omega \, S(q,\omega)=S(q)$, is fixed by the PIMC SSF (black triangles).}
\label{fig:SkShann}
\end{figure}

Looking for an additional confirmation of the physical consistency of the entropy principle, we have analysed the normalization of the reconstructed model dynamic structure factor, $S^{\text{9MA}}(q)$, versus the corresponding PIMC data. This comparison for two temperature cases is presented in Fig.~\ref{fig:SkShann}.
First, the low-temperature case ($T=1.5$ $K$) on the left panel reveals some noticeable distinction between 
the PIMC data and the results of the original Shannon procedure. Using the symmetric form of the spectral density, see Eq.~(\ref{M}), the normalization of the reconstructed DSF solution to $S(q)$ is not included in the analytical representation, Eqs.~(\ref{Mnu})-(\ref{Nnu}).
As a result, while the structure of the SSF PIMC curve is reproduced qualitatively quite well 
(with the maximum around $q\sim 1.8 \AA^{-1}$), the spectral density is slightly 
underestimated in the phonon- and the free-particle segments of the spectrum, see the Feynman mode 
in Fig.~\ref{fig:Entropy}, 
correspondingly for $q\lesssim 1 \AA^{-1}$ and $q\gtrsim 2 \AA^{-1}$. 
In contrast, around the roton minimum the maximization of the entropy functional leads to a much slower 
decay of $S(q,\omega)$ in comparison to the optimized solution (see below)  in the high-frequency limit, see Eq.~(\ref{high_w}).
Thus, the normalization factor is overestimated around the maximum.  

\begin{figure}[]
\hspace{-1cm}
\includegraphics[width=0.54\textwidth]{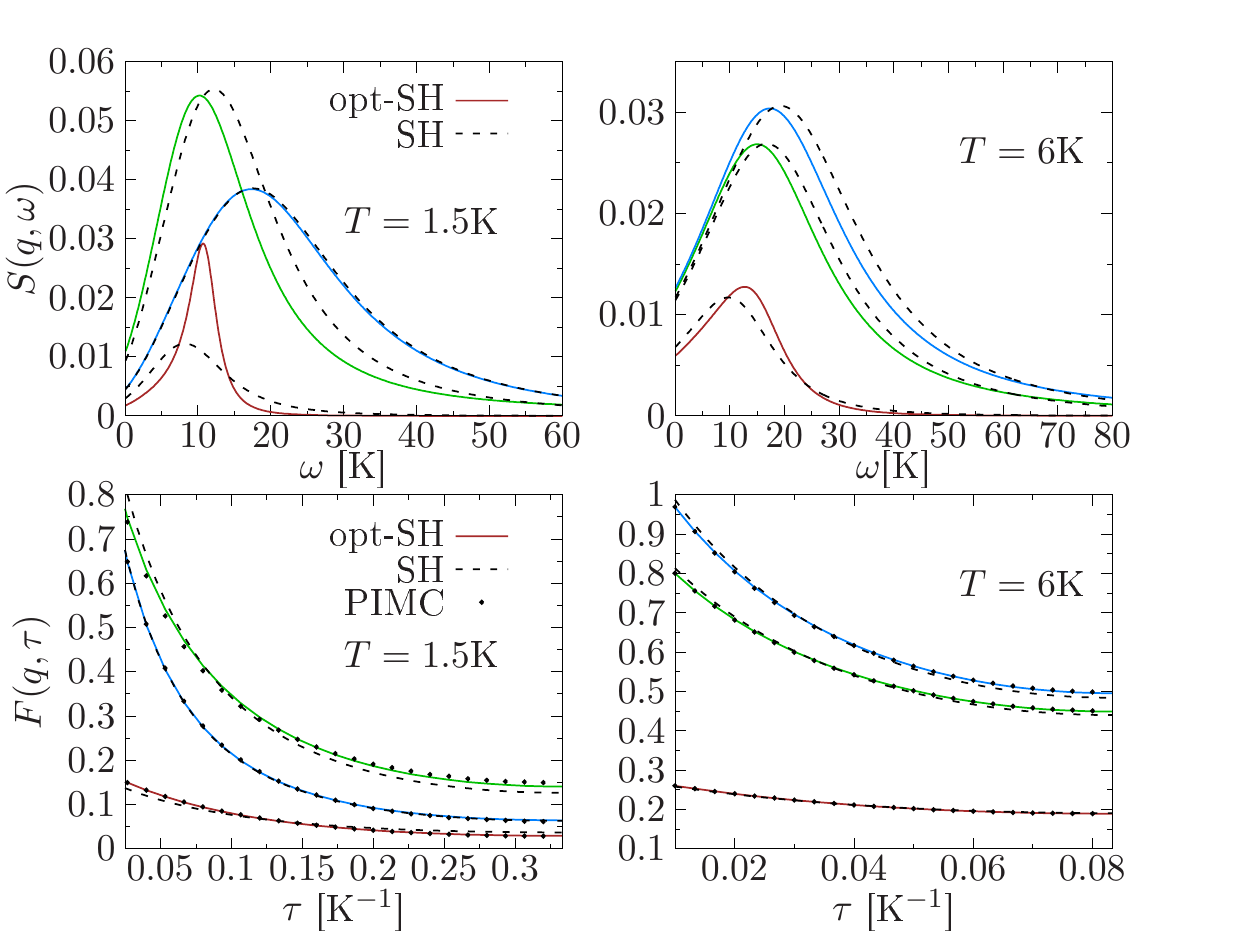}
\vspace{-0.4cm}
\caption{{\it Upper panels:} The dynamic structure factor $S(q,\omega)$ reconstructed using the Shannon ("SH") and the optimized Shannon ("opt-SH") procedures at temperatures $T=1.5$ $K$ and $6$ $K$. 
Three examples are shown which correspond to the wavenumbers $q [\AA^{-1}]=0.48, 1.78, 2.0$ ($T=1.5$ $K$) and $q [\AA^{-1}]=0.595, 1.68, 2.0$ ($T=6$K).  The spectral density increases with the wavenumber $q$.  
{\it Lower panels:} The model ISF ("SH" and "opt-SH") for $0.025 \leq \tau [K^{-1}] \leq 0.33$ compared to the PIMC data (solid dots) for the same wavenumbers.}
\label{fig:ISF_SH}
\end{figure}

Similar large deviations are observed once the Shannon solution is substituted to the r.h.s. of Eq.~(\ref{Fqw}), and the model imaginary time intermediate scattering function (ISF) is compared to the corresponding QMC data, $F^{\text{PIMC}}(\mathbf{q},\tau)$. This comparison is presented in Fig.~\ref{fig:ISF_SH} for the characteristic wavenumbers $q$ in the phonon, roton and free-particle regimes. Fortunately, the revealed discrepancies can be significantly reduced in the revised (or optimized) Shannon maximization procedure. In particular, we apply an additional restriction given by the DSF zero-order frequency moment: we search for an optimal solution $S^{*}(q,\omega;\tilde \omega)$ in the vicinity of an extremum of the entropy function $E(q;\tilde \omega)$ with a minimal deviation from the PIMC SSF:
\begin{eqnarray}
\min_{\{\omega_3(q),\omega_4{q}\}}\left| \int S^{*}(q,\omega;\tilde \omega) \db \omega - S^{\text{PIMC}}(q)\right|. \label{addnorm}    
\end{eqnarray}

The obtained solutions ("opt-SH") are presented in Fig.~\ref{fig:ISF_SH} and demonstrate the desired agreement with the ISF PIMC data. The above procedure has proved to be very efficient both for low- and high-temperature analysis. 
In particular, in the high-temperature regime ($T=6$ $K$), the original Shannon entropy optimization already 
reproduces the PIMC SSF very accurately (see the right-hand panel in Fig.~\ref{fig:SkShann}) and the additional minimization condition~(\ref{addnorm}) provides only minor corrections, 
mainly around the SSF maximum and for the wavenumbers $q \gtrsim 2 \AA^{-1}$.
The main reason for this observation is a quite smooth frequency dependence of the spectral density (see Figs.~\ref{fig:DSF_T6},~\ref{fig:DSF_T15} and Sec.~\ref{dyn_resp} 
for more details). 
The optimal conditions for the applicability of the entropy approach take place at high temperatures when the thermal effects and the quasiparticle decay processes broaden the decrements of the collective modes significantly, in particular, in the phonon and the roton part of the spectrum.  

In conclusion, our present analysis confidently confirms a strong predictive power of the entropy maximization principle. It permits to predict both the temperature and the density dependence of the SSF and the imaginary time density-density autocorrelation function (ISF) quite accurately.      

\subsection{Dynamical structure factor: theory vs experiment}

\begin{figure}[]
\hspace{-0.2cm}
\includegraphics[width=0.5\textwidth]{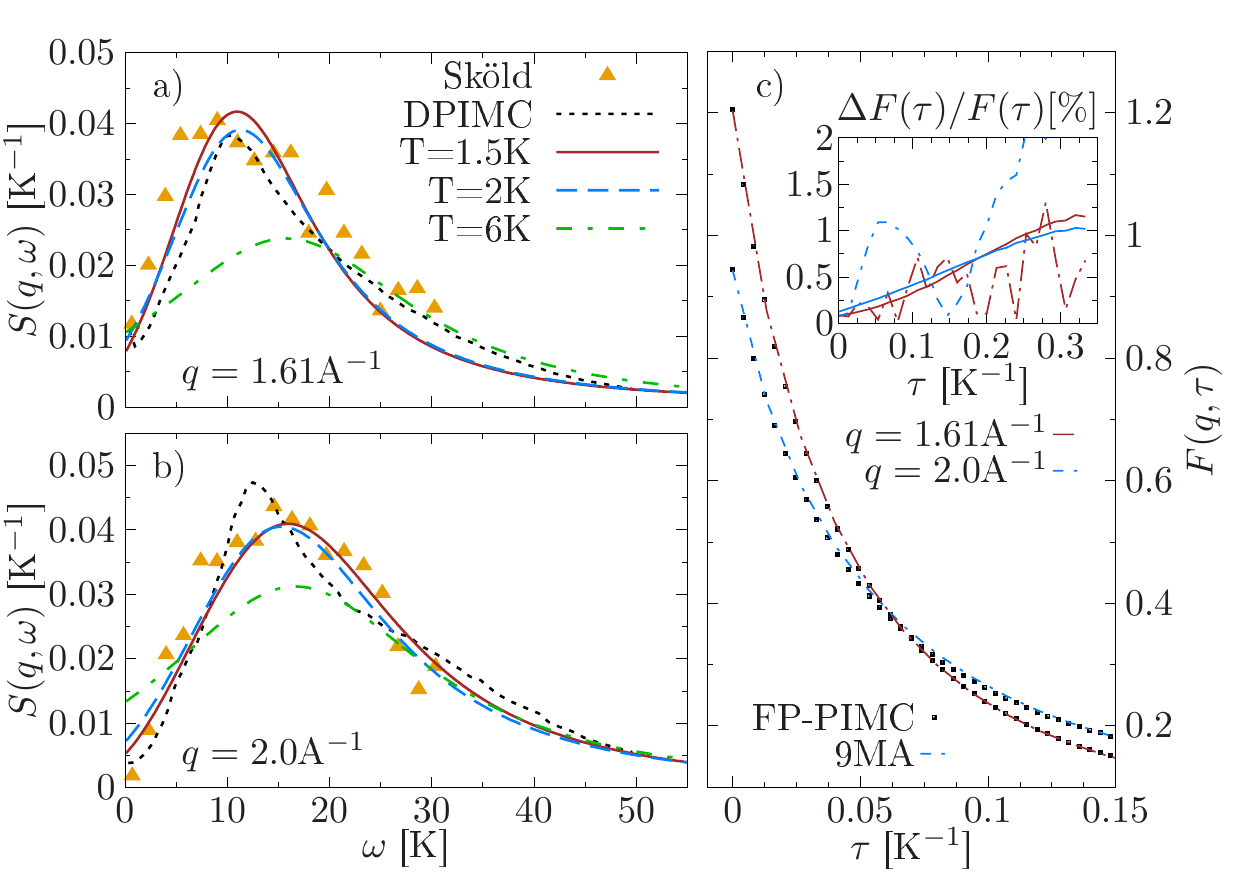}
\vspace{-0.6cm}
\caption{{\it Left}: Comparison of the reconstructed dynamic structure factor $S(q,\omega)$ for 
$q=1.61 \AA^{-1}$ (a) and $2.0 \AA^{-1}$ (b) with Boltzmann statistics at $T=1.5$K ($N=100$), and Fermi statistics at $2$K ($N=38$) and $6$K ($N=100$) to the experimental data by Sköld {\it et al}~\cite{Skold1980} ($T=1.2$K) and the theoretical DSF (DPIMC) by Dornheim {\it et al}~\cite{dornheim-etal.2022sr} reconstructed with the genetic algorithm scheme~\cite{vitali.2010prb}. (c) The intermediate scattering function $F(q,\tau)$ at $T=1.5$K (solid squares) 
obtained by the PIMC method (see Appendix A) and the 9MA-model via Eq.~(\ref{Fqw}) (dashed lines) for the corresponding wavevectors. {\it Insert:} the dashed lines stand for the relative deviations (in percentage points), $\Delta F(\tau)/F(\tau)$, 
between the 9MA and the FP-PIMC data. The solid lines signal the level of the statistical error in the PIMC data, 
$\delta F(\tau)/F(\tau)$.  
}
\label{fig:DSFExp}
\end{figure}

Certainly, experiments in neutron scattering capable to produce reliable data on the density-density dynamic structure factor of the liquid $^3$He are quite scarce. Moreover, typical experimental temperatures are well below $2$K, where the direct fermionic PIMC simulations are not accessible (see Appendix A2). For this reason, in the present analysis we follow the route proposed by Dornheim {\it et al}~\cite{dornheim-etal.2022sr} and reconstruct the $^3$He DSF at $T=1.5$K using as the input the static properties evaluated with Boltzmann statistics. The performed analysis presented in Appendix A4 reveals no noticeable effect of Fermi statistics on the static structure factor $S(q)$ and the static density response function $\chi(q,0)$ for $T\geq 2$K which are the key ingredients of the self-consistent method of moments, see Sec.~\ref{SSMMSec}. This justifies the involved approximation at $T=1.5$K. As a further confirmation, we observe no influence of the quantum statistical effects on the potential energy $\epsilon_p$ in a broad temperature range $1.5\text{K}\leq T\leq 6$K, see Fig.~\ref{fig:avrS} (right-hand panel), and in the kinetic energy per atom $\epsilon_k$ for $1.5\text{K}\leq T\leq 2.5$K. Hence, at these low temperatures the direct correlation contribution prevails in $\epsilon_p$, while the effects of Fermi statistics as observed in the momentum distribution $n(q)$ and in the deviation from the Boltzmann case,  $\Delta n(q)=n_B(q)-n_F(q)$, see Fig.~\ref{fig:nkDiffT}, mutually compensate in the momentum-averaged quantities like $\epsilon_k$.  

To demonstrate the performance of the new self-consistent method of moments
with the new dynamical Nevanlinna function, in Fig.~\ref{fig:DSFExp} we report our DSF results compared to the experimental 
data by Sköld {\it et al}~\cite{Skold1980} and the direct PIMC (DPIMC) reconstruction~\cite{dornheim-etal.2022sr} carried out at $T=1.2$K. 
Our theoretical predictions reproduce the position and the half-width of the experimental DSF very well. Moreover, at $q=2 \AA^{-1}$ we observe even better agreement to the experiment then with the DPIMC~\cite{dornheim-etal.2022sr}. The main reason, in our opinion, it that the presented analytical \textit{Ansatz} for $S(q,\omega)$, Eq.~(\ref{SNnu}), 
and the corresponding spectral density $M(q,\omega)$, Eq.~(\ref{M}), satisfy an extended set of frequency power moments (up to seven, plus two additional ones reconstructed via the optimized Shannon entropy, Eq.~(\ref{addnorm})). 
This significantly reduces the class of "physically relevant" solutions for the DSF. In addition, our solution includes the information about the fourth frequency moment $\mu_4(q)$, explicitly derived in Appendix B and estimated via the PIMC simulations, see Fig.~\ref{fig:C4mom}. As an independent check, in Fig.~\ref{fig:DSFExp}c we use the reconstructed $S(q,\omega)$ to reproduce the imaginary-time dependence of the intermediate scattering function $F^{\text{9MA}}(q,\tau)$, Eq.~(\ref{Fqw}), and compare it to the PIMC data, $F(q,\tau)$. The relative deviation between the two, $\Delta F(q,\tau)/F(q,\tau)$ versus $\tau$, is shown in the insert panel, and in the ideal case should not exceed the statistical uncertainty, $\delta F(q,\tau)/F(q,\tau)$, in the PIMC data~\cite{vitali.2010prb,filinov.2012pra,filinov.2016pra,dornheim-etal.2022sr}. The use of the optimized Shannon entropy maximization procedure (see Sec.~\ref{optSH} and Fig.~\ref{fig:ISF_SH}) allows us to satisfy this additional criteria with a high level of precision for $q=2.0\AA^{-1}$ and only slightly violate it (by $1\%$) for $q=1.61\AA^{-1}$. 

The temperature dependence of the reconstructed DSF is demonstrated by two additional spectra at $T=2$K and $6$K, obtained with Fermi statistics, in contrast to the lowest temperature case ($T=1.5$K). Notice the similarity of two DSF curves at $T=1.5$K (solid brown) and $2$K (dashed blue). First, this important result points out a weak temperature dependence of the theoretical $S(q,\omega)$ for $T\lesssim 2$K, and, second, implies a minor role played by the quantum statistics in the spectrum of collective excitations in $^3$He fluids at similar temperatures.  

\subsection{Dispersion equation and eigenmodes}

\begin{figure}[]
\hspace{-0.4cm}
\includegraphics[width=0.51\textwidth]{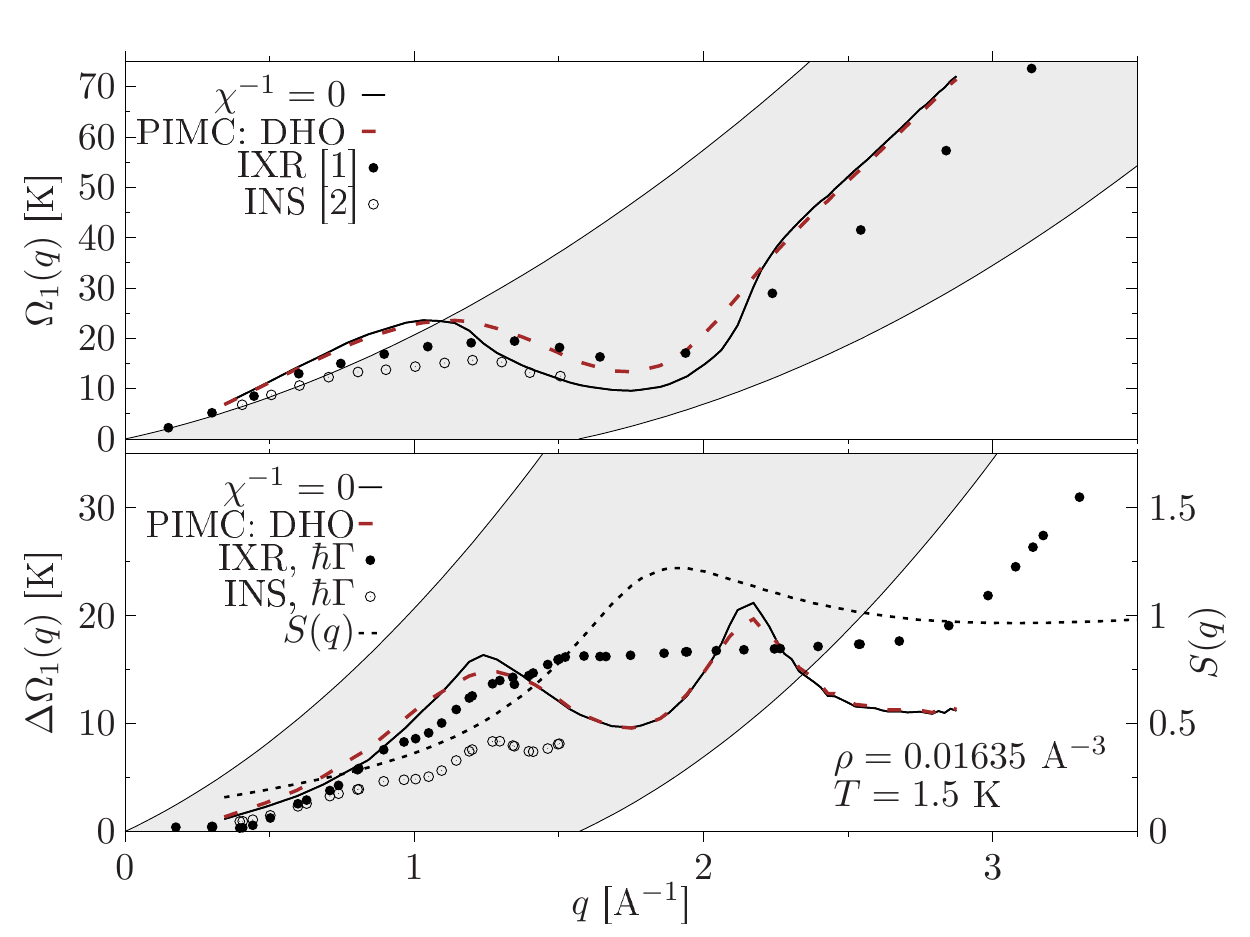}
\vspace{-0.3cm}
\caption{
{\it Upper panel:} The lower-energy mode dispersion $\Omega_1(q)$ obtained as an exact solution of the dispersion equation, 
i.e. one of the poles of the density response function $\chi \left( q,z \right)$, where $z=\Omega_{1}(q)-i \Delta \Omega_1(q)$. Simulation parameters: $T=1.5$K (Boltzmann statistics) and $\rho=0.01635$ [$\AA^{-3}$]. The theoretical dispersion corresponds to a combined solution obtained with two system sizes, $N=100$ and $N=38$. The shaded area represents the particle-hole band 
with the $^3$He bare mass. The symbols are the  experimental data for He$^{3}$ at the saturated vapor pressure (SVP) 
conditions: the IXR (Albergamo {\it et. al.}~\cite{Albergamo2007}) ($T=1.1$K) and the INS (Fåk {\it et. al.}~\cite{fak1994}) ($T=0.120$K). The IXR experiment uses the DHO ansatz~(\ref{DHOEq}) to fit the DSF experimental data.
{\it Lower panel:} Solid and dashed lines are the theoretical results for the decrement of the mode, $\Delta \Omega_1(q)$, 
and the full line width, $2\hbar\Gamma $, in the DHO model. The symbols represent the experimental data~\cite{Albergamo2007},~\cite{fak1994}. 
The dotted line corresponds 
to right-hand vertical axis and displays $S(q)$ from Fig.~\ref{fig:Sk} 
(the PIMC data).}
\label{fig:modesT15}
\end{figure}

The dispersion equation for the collective modes can be written as an equation for the poles of the complex density response function~(\ref{chi2}), 
\begin{eqnarray}
    \chi^{-1}(q,z)=0,
    \label{chi-1}
\end{eqnarray}
where $z$ is the complex frequency. 
Within the method of moments the above equation can be constructed in a closed analytical form and within the present model it leads to the fifth-order algebraic equation,
\begin{eqnarray}
z\left( z^{2}-\omega_{2}^{2}\left( q\right) \right) +Q_{2}\left( q,z\right)
\left( z^{2}-\omega_{1}^{2}\left( q\right) \right) =0 \ , \label{deQ2}
\end{eqnarray}
with (due to the specific properties of the involved orthogonal polynomials~\cite{akhiezer-book, krein-book}) five distinct complex solutions, two of them being always symmetric with respect to the real part of the frequency:
\begin{eqnarray}
&&z_{0}\left( q\right) =-i \Delta \Omega_0\left( q\right),\, \Omega_0=\text{Re}(z_0)=0,\label{diss0}\\
&&z_{\pm 1 (\pm 2)}\left( q\right) =\pm \Omega_{1(2)}(q)
-i\Delta \Omega_{1(2)}\left( q\right).\label{diss12} 
\end{eqnarray}%
Thus, there appear three eigenmodes: the diffusion mode $z_0(q)$ and two shifted modes $z_{\pm 1(\pm 2)}(q)$. The intrinsically negative imaginary parts of the solutions are defined by the decrements of the modes, $\Delta \Omega_0\left( q\right) $ and $\Delta \Omega_{1(2)}\left( q\right) $ (see discussion below).
 
In what follows we concentrate on the low-frequency solution, $\Omega_1(q)$, as in the $^3$He system it carries the main spectral weight in the dynamic structure factor.
The high-frequency solution $\Omega_2(q)$ can also be resolved for all wavenumbers $q$, but due to the large decrement, $\Delta \Omega_2(q) \sim \Omega_2(q)$, it should not be treated as a propagating collective mode, but rather as a contribution of a multi-excitation quasiparticle continuum; it constitutes nevertheless an important part of the DSF {\it physical} solution.

\begin{figure}[]
\hspace{-0.3cm}
\includegraphics[width=0.52\textwidth]{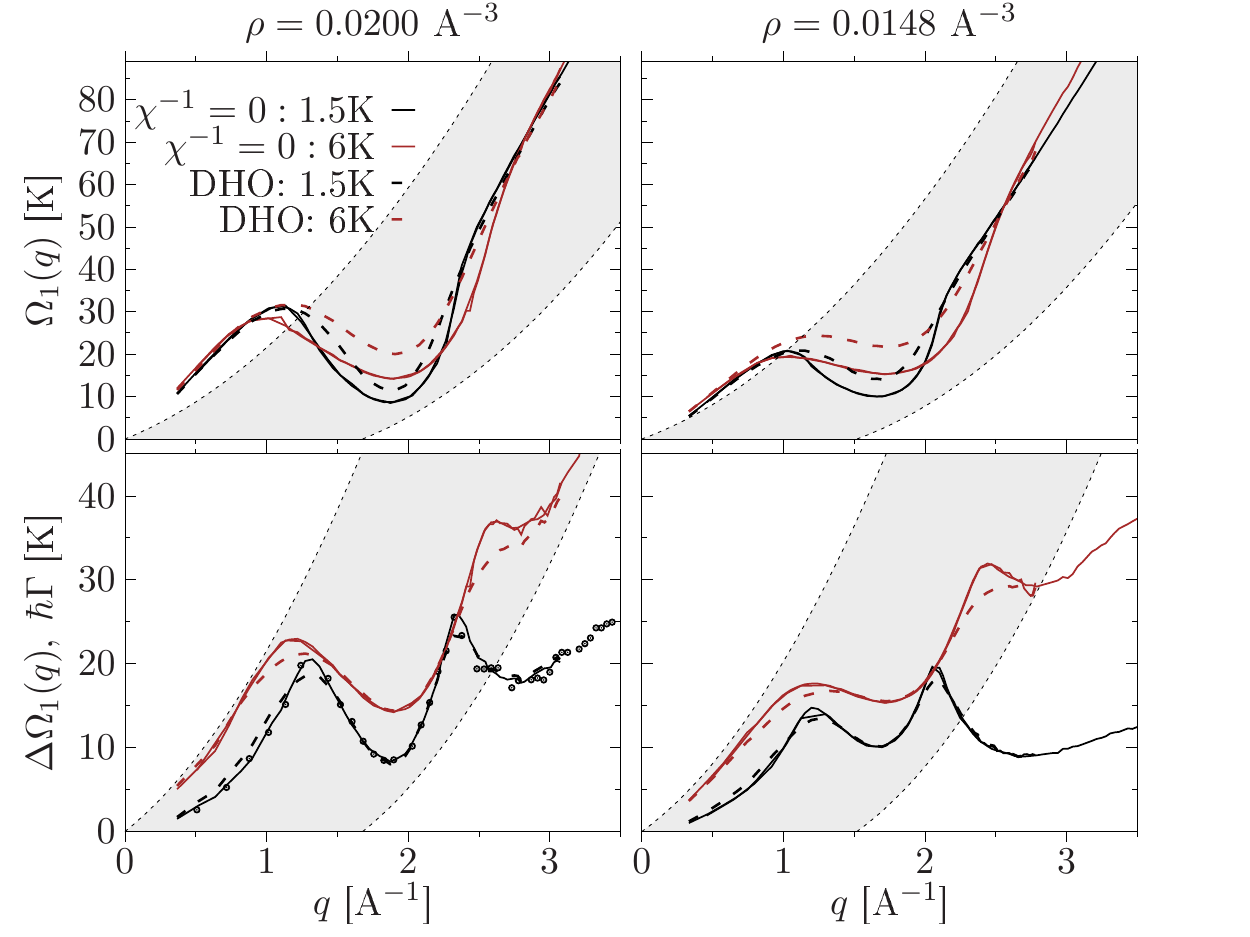}
\vspace{-0.3cm}
\caption{As in Fig.~\ref{fig:modesT15}: The low-energy solution of the 
dispersion equation $\chi^{-1} \left( q,z \right)=0$ 
(solid lines) in comparison to the results found within the DHO model (dashed lines). 
Two temperature cases ($T=1.5$K and $T=6$K) are compared at the 
densities $\rho=0.0200$ [$\AA^{-3}$] ({\it left-hand column}) 
and $\rho=0.0148$ [$\AA^{-3}$] ({\it right-hand column}).}
\label{fig:modesT6}
\end{figure}

\begin{figure}[]
\hspace{-0.8cm}
\includegraphics[width=0.52\textwidth]{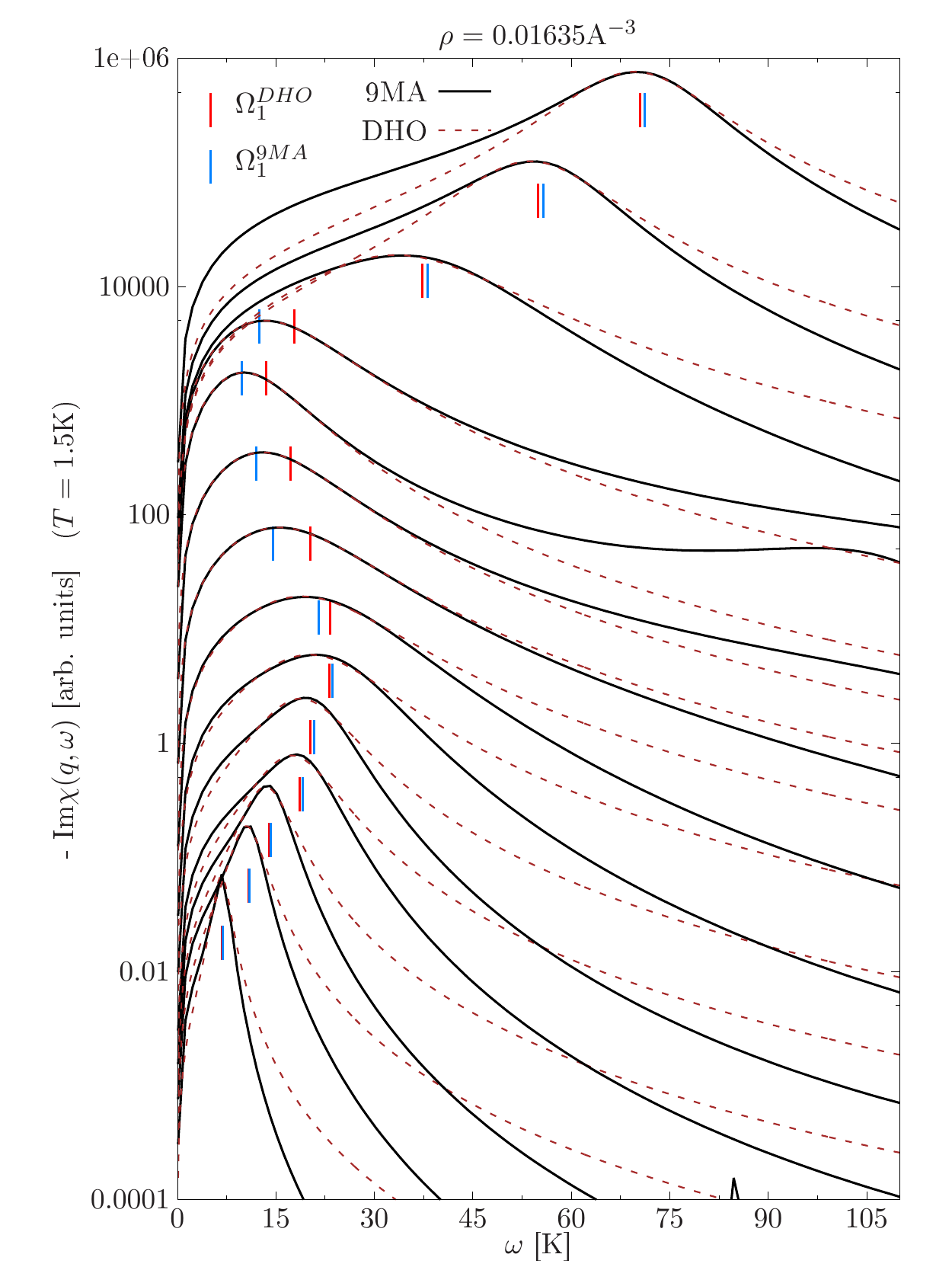}
\vspace{-0.5cm}
\caption{Imaginary part of the density response function, $\operatorname{Im} \chi(q,\omega)$, from the nine-moment model compared to the best fit within the DHO model~\cite{Albergamo2007}. Simulation parameters: $T=1.5$K and $\rho=0.01635$ [$\AA^{-3}$]. The selected wavenumbers $q$ (from bottom to top): $q[\AA^{-1}]=0.343,0.486,0.595,0.687.0.84,1.03,1.19,1.37,1.50,1.68,$
$1.94,2.25,2.57,2.85$.
For a better representation $\operatorname{Im} \chi(q,\omega)$ for each $q$ is multiplied by the factor of $4(8)$ with respect to the preceding one. The vertical arrows denote the corresponding frequencies $\Omega_1(q)$ resolved either as the solution of the dispersion equation, 
$\chi^{-1}=0$, or as a fitting parameter in the DHO model~(\ref{DHOEq}). 
The corresponding full dispersion relations are demonstrated in Fig.~\ref{fig:modesT15}.}
\label{fig:SqdynT15DHOfit}
\end{figure}

The dispersion of the lower-energy mode, $\Omega_{1}(q)$, is studied in detail in Figs.~\ref{fig:modesT15} and~\ref{fig:modesT6}. The DSF maxima at different wavenumbers are positioned very close to the dispersion curve obtained within the same approximation 
(see below Fig.~\ref{fig:SqdynT15DHOfit}). We can clearly observe the typical phonon-maxon-roton dispersion relation. However, in contrast to other strongly coupled classical and quantum liquids, like the bosonic $^4$He, here we observe strong damping effects once the dispersion curve enters into the particle-hole band (see the shaded area). These effects could not be revealed within the well-known quasi-localized charge approximation~\cite{Kalman_2010, Kalman_2012} which does not take the energy dissipation into account. 

Following the analysis of Ref.~\cite{Albergamo2007} we have fitted the reconstructed dynamical structure factor with a simple model based on the damped harmonic oscillator (DHO)
\begin{eqnarray}
S(q,\omega)=\frac{n(\omega)+1}{\pi} \frac{4 Z(q) \, \omega \Gamma(q)}{[\omega^2 -\Omega^2(q)]^2+4 \omega^2 \Gamma^2(q)}\label{DHOEq}   
\end{eqnarray}
with $n(\omega)$ being the Bose factor, $n(\omega)=(e^{-\beta \hbar \omega}-1)^{-1}$, and $Z(q)$ standing for the intensity factor.
This simple functional form allows to extract both the position of the maximum, $\Omega^{\text{DHO}}(q;T)$, and its decrement $2 \Gamma(q;T)$.
The detailed comparison of the DHO dispersion vs. the theoretical one, 
$\Omega_1(q)$, is presented in Figs.~\ref{fig:modesT15}, \ref{fig:modesT6} 
for three density cases, $\rho=0.0148; 0.01635;0.020$ [$\AA^{-3}$]. 
Our analysis clearly reveals a weak temperature dependence of 
the modes' properties $\{\Omega_1,\Delta \Omega_1\}$ at least for $T\lesssim 3$ $K$. Some significant temperature effects are observed at $T=6$ $K$ (see the solid brown curves in Fig.~\ref{fig:modesT6}).

Next, our low-temperature results in Fig.~\ref{fig:modesT15} are compared to the experimental data~\cite{Albergamo2007,fak1994} at the saturated vapor pressure
(SVP) conditions which correspond to the density $\rho=0.01635$ [$\AA^{-3}$].
We observe a reasonable agreement taking into account that the mode parameters 
in Refs.~\cite{Albergamo2007,fak1994} were determined with a finite experimental resolution. Moreover, the observed difference between the inelastic x-ray scattering (IXR) and the inelastic neutron 
scattering (INS) spectra for $q > 0.8 \AA^{-1}$, is due to the superposition of the coherent and incoherent signals in the INS DSF data which leads to the overlap of the sound mode peak and the spin wave one. In contrast, in the acoustic range ($q\lesssim 0.8 \AA^{-1}$) where both signals are well separated, the experimental data are in a good mutual agreement, and agree with our theoretical solution $\Omega_1(q)$ of the dispersion equation as well.  

The decrement of the quasiparticle mode, $\Delta \Omega_{1}(q)$, is presented in the lower panel. First, we note a quadratic scaling of the damping factor with the wavenumber $q$~\cite{Pitaevskii:1967} both in the theory and  the IXR~\cite{Albergamo2007} and  the INS~\cite{fak1994} experiments. Our theoretical result (solid line) represents the imaginary part of the complex frequency, $z_{+1}(q)=\Omega_1(q)-i \Delta \Omega_1(q)$, obtained as a solution of the dispersion equation~(\ref{deQ2}).
Up to the wavenumber $q\sim 1.3 \AA^{-1}$ our results for the damping coefficient $\Gamma(q)$ are in a reasonable agreement with the IXR data~\cite{Albergamo2007}. Next, we observe a non-monotonic behaviour of the damping factor with a significant reduction in the {\it roton branch of the spectrum}, $1.3^{-1} \AA \leq q \leq 2.2 \AA^{-1}$. The effect is reproduced in the low- and high-density cases, see Fig.~\ref{fig:modesT6}. In contrast, no similar non-monotonic $q$-dependence is observed in the experiment.

\begin{figure}[]
\hspace{-0.3cm}
\includegraphics[width=0.51\textwidth]{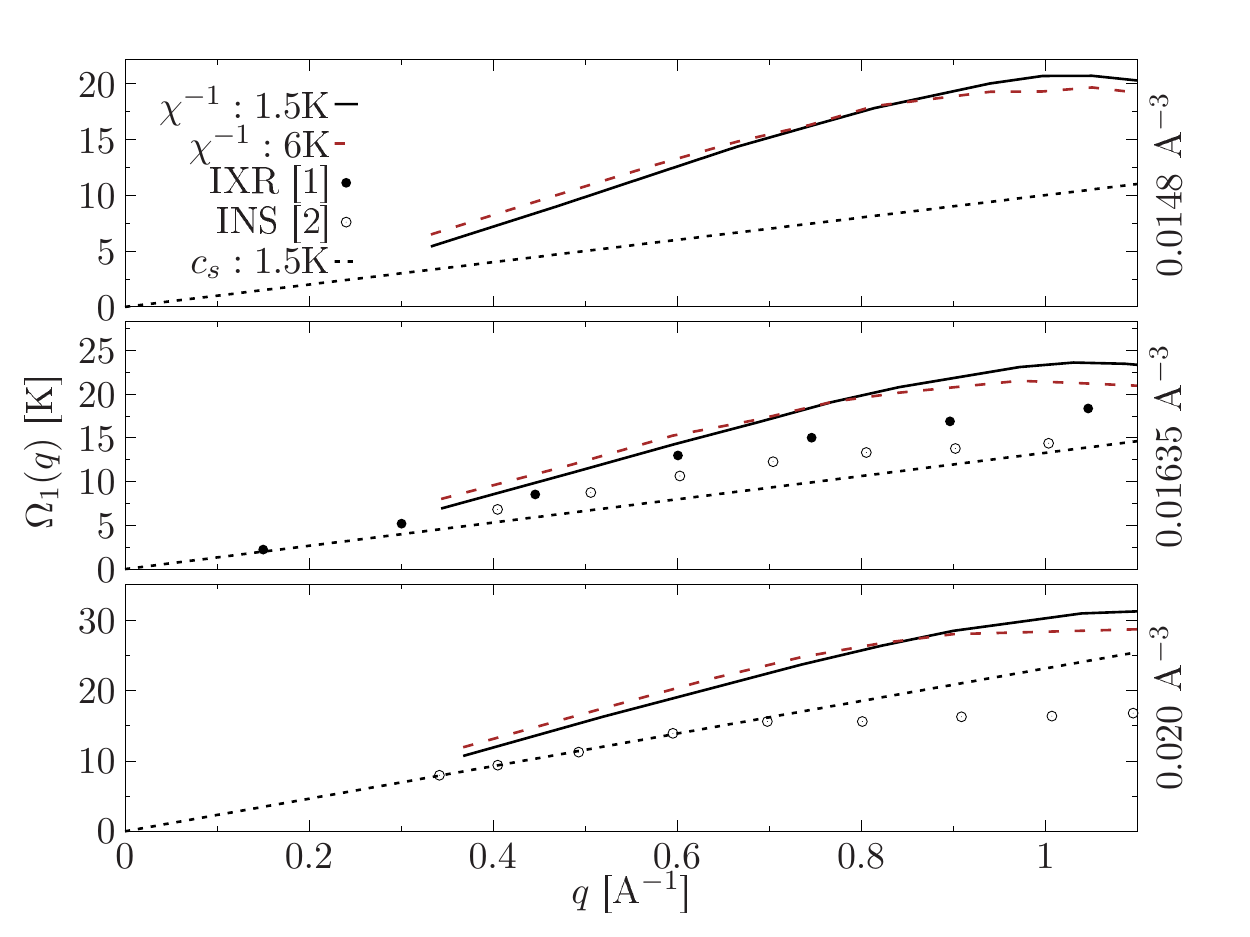}
\vspace{-0.5cm}
\caption{The long-wavelength behaviour of the theoretical dispersion relation $\Omega_1(q)$
at $T=1.5$K (solid black) and $T=6$K (dashed brown) vs. the experimental data (see the caption of Fig.~\ref{fig:modesT15}). The lowest wavenumber ($q=2\pi/L$) available from the simulations is limited by the system size $N\leq 100$. The dotted black curve in each panel represents the sound dispersion at $T=1.5$K, $\omega(q) = c_s(\rho,T) q$, with $c_s=(\kappa \rho \,m)^{-1/2}$ being the isothermal sound speed estimated from the compressibility relation,  $\kappa \rho=S(0)/T$.}
\label{fig:zerosound}
\end{figure}

In order to find a possible explanation of this disagreement, we have compared both theoretical dispersion relations, cf. $\Omega_1^{\text{9MA}}$ and $\Omega_1^{\text{DHO}}$ in Fig.~\ref{fig:SqdynT15DHOfit}. We found that exactly in the roton segment of the spectrum both models deviate significantly and the DHO predictions become unreliable: the resolved dispersion curve $\Omega_1^{\text{DHO}}$ overestimates the position of the DSF maximum, whereas the $\chi^{-1}=0$ solution, $\Omega_1^{\text{9MA}}$, stays always much closer to the spectral density maximum. This interpretation is further confirmed in the upper panel of Fig.~\ref{fig:modesT15} where the theoretical DHO model (dashed brown curve) agrees very well with the experimental one based on the IXR data~\cite{Albergamo2007}.

Further, in Fig.~\ref{fig:zerosound} we present the resolved long-wavelength behaviour of the theoretical dispersion $\Omega_1(q)$ in a more detailed form, again in comparison to the experiment~\cite{Albergamo2007,fak1994}. 
As the experimental analysis reports 
a weak temperature dependence of the phonon branch~\cite{PhysRevB.61.1421, fak1994}, in the simulations we used the temperature $T=1.5$K. This allowed us to reduce both the number of the high-temperature propagators (see Fig.~\ref{fig:EnergyDiffP}) and the computational costs. Indeed, we found no noticeable differences in the dispersion curves estimated at $T=1.5$K and $T=2$K. This fact permits to extrapolate our results to lower temperatures and compare our results to the experimental data resoled at $T \lesssim 1$K directly.  
The simulated system size ($N=100$) is not sufficient to reach the limiting form of the sound mode slope, though the experiment~\cite{Albergamo2007} seems to converge to the expected asymptotic behaviour. The theoretical sound speed in Fig.~\ref{fig:zerosound} increases with the density and reveals noticeable temperature effects. For the three specified densities (from top to bottom) and $T=1.5$K, the compressibility sum rule, $\kappa \rho=S(0)/ T$, allows us to resolve, respectively, $c_s=131(2),174(2),303(2)$ [m/s]. To obtain the zero momentum value of the SSF we performed additional PIMC simulations in the grand canonical ensemble similar to that of Ref.~\cite{worm2006}. At a higher temperature $T=6$K we get, $c_s=178(2),205(2),293(2)$ [m/s] (in the same order).

\subsection{Dynamical response in He$^{3}$: temperature and density dependence}\label{dyn_resp}

\begin{figure}[t]
\hspace{-0.90cm}
\includegraphics[width=0.51\textwidth]{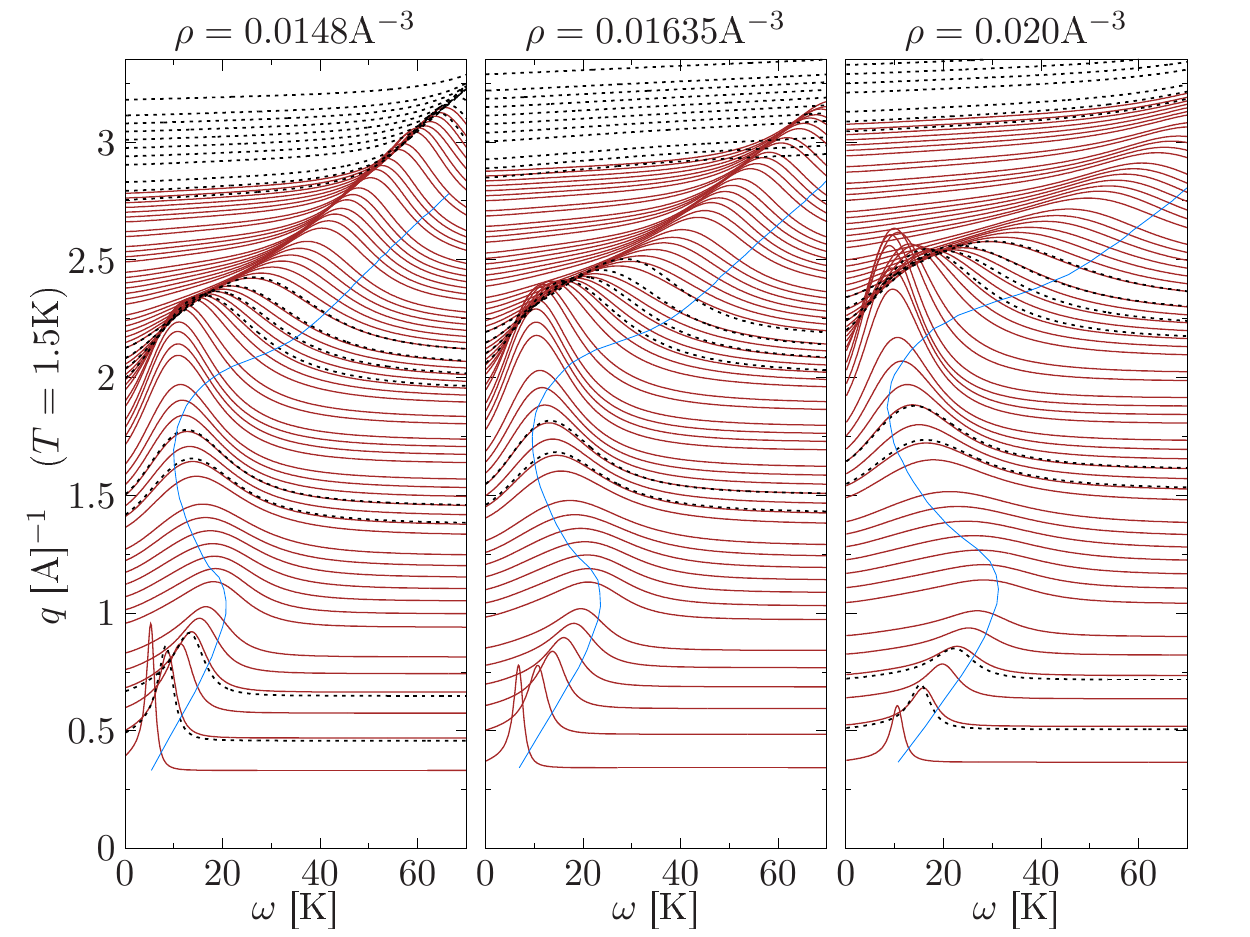}
\vspace{-0.2cm}
\caption{The dynamic structure factor $S(q,\omega)$ of the $^{3}$He fluid at three densities, $\rho=0.0148; 0.01635;0.020$ [$\AA^{-3}$] and $T=1.5$K. 
The solid and the dashed lines are the reconstruction results for different system sizes: $N=100\; (q\leq 2.75 \AA^{-1})$ 
and  $N=38\; (q\geq 2.75 \AA^{-1})$, respectively. A few wavenumber values such that $q\leq 2.2 \AA^{-1}$,  when both curves overlap, are selected to demonstrate that the reconstruction results are robust with respect to the finite-size effects (Appendix A3). The solid blue lines are the poles of the density response function. They represent the collective branch $\Omega_1(q)$ of quasiparticle excitations.}
\label{fig:DSF_T15}
\end{figure}

\begin{figure}[t]
\hspace{-0.90cm}
\includegraphics[width=0.51\textwidth]{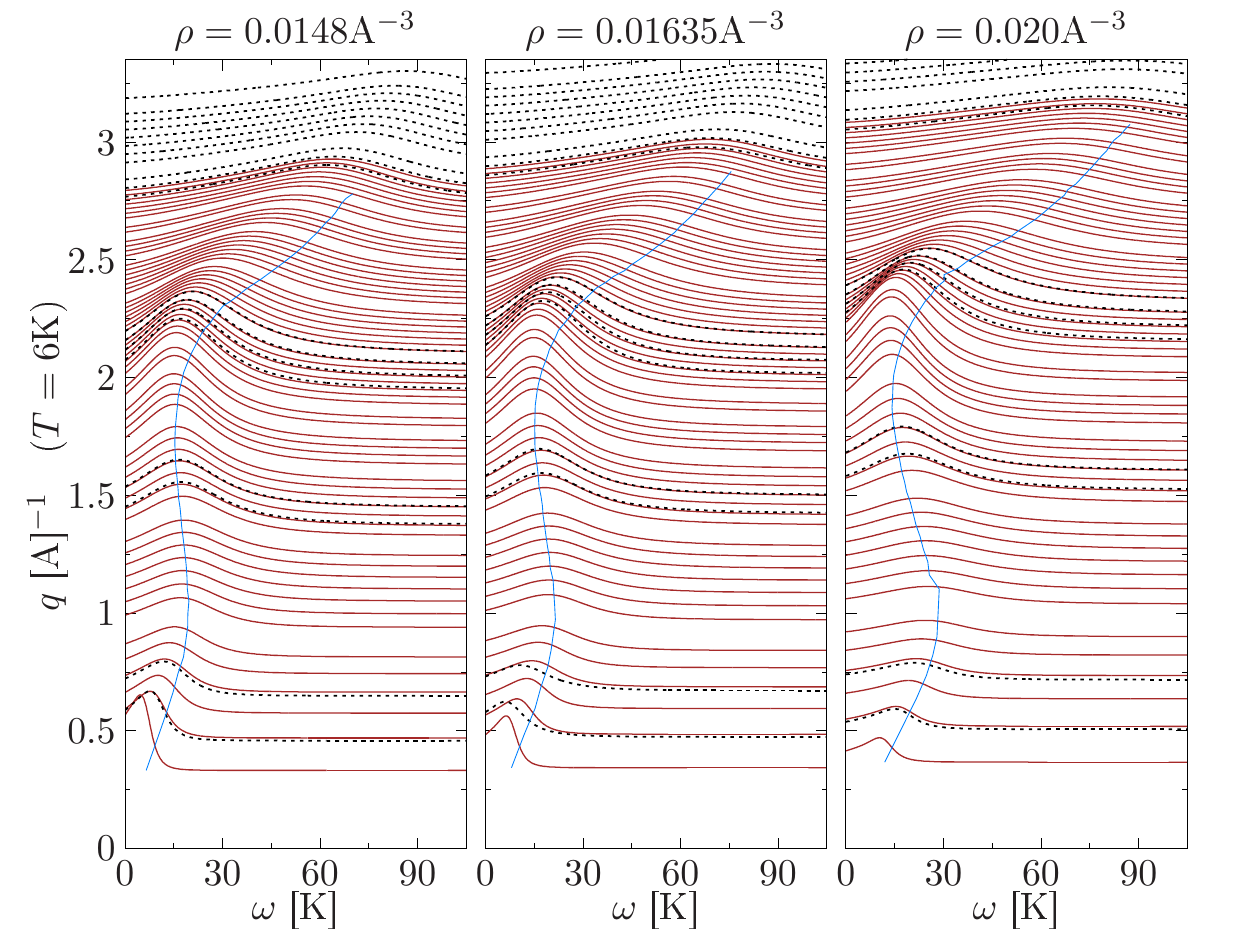}
\vspace{-0.2cm}
\caption{As in Fig.~\ref{fig:DSF_T15} but at a higher temperature $T=6$K. 
The solid  and the dashed lines are the results for $N=100\; (q\lesssim 2.75 \AA^{-1})$ and  $N=38\; (q\gtrsim 2.75 \AA^{-1})$, respectively.}
\label{fig:DSF_T6}
\end{figure}

Here we present our results on the dynamic structure factor (DSF) stemming from the present nine-moment approximation (9MA) in comparison to the lower-energy solution of the explicit fifth-order dispersion equation, 
$\chi^{-1} \left( q,z\right)=0$. 
Since, as we have seen in Fig.~\ref{fig:modesT15} and Fig.~\ref{fig:modesT6}, the decrement of the lower-energy mode $\Delta \Omega_1(q)$ is smaller than the mode frequency $\Omega_1(q)$, it is not surprising that the positions of the maxima of the dynamic structure factor displayed in Figs.~\ref{fig:DSF_T6} and Figs.~\ref{fig:DSF_T15} are quite close to the lower-frequency collective mode comprising the phonon, maxon, and roton branches.         

\section{Conclusions}

Thoroughly verified theoretical and computational results on the static and dynamic properties of the liquid $^3$He are obtained within a novel non-perturbative approach which is a combination of the extended self-consistent method of moments with the Shannon information entropy maximization procedure and the \textit{ab initio} fermionic PIMC simulations (Appendix A), which provide an indispensable information on the frequency power moments of the spectral density and the intermediate scattering function, the Laplace transform of the dynamic structure factor.

Taking into account a finite resolution of the available experimental data, we conclude that a very reasonable agreement is achieved between the latter and the present theoretical predictions on the dynamic structure factor, the dispersion relation of the lower-energy eigenmode and its decrement in the phonon and the roton wavenumber segments.


Our theoretical results directly imply the existence in 
the low temperature $^3$He fluid of the propagating {\it roton} mode.
This mode is also encountered in other strongly correlated non-fermionic liquids like superfluid $^4$He~\cite{vitali.2010prb}, Yukawa/Coulomb classical plasmas and ultra-cold Bose gases~\cite{filinov.2012pra,PhysRevA.87.033624,filinov.2016pra}. It is understood to be of a fundamental physical origin~\cite{Kalman_2010, Kalman_2012, Dornheim2022} related with the incipient localization of the particles due to the interactions~\cite{Noz2004}.  
However, in contrast to the bosonic $^4$He, the present {\it roton-like} excitation 
decays faster due to its 
localization in the wavenumber range spanned by the particle-hole continuum where the interactions with the particle-hole excitations open an additional decay channel. Hence, the present \textit{ab initio} results for the non-monotonic wavenumber dependence of the roton decrement, cf. Figs.~\ref{fig:modesT15} and~\ref{fig:modesT6}, constitute an important theoretical issue to be further verified both experimentally and theoretically. In this connection, we mention the recent experimental validation of the re-emergence of the roton-like mode~\cite{Godfrin2012} in a monolayer of liquid $^3$He once the dispersion curve leaves the particle-hole band.

Investigation of a physical nature of the higher-energy asymptotic behaviour of $S(q,\omega)$ which stems from the explicit dispersion equation (the poles of the density-density 
response function with a large imaginary part), constitutes the second important issue and will be carried out elsewhere.

The presented version of the method of moments in combination with \textit{ab initio} path integral Monte Carlo
simulations due to its mathematical and non-perturbative character, can be applied to the solution 
of other physical problems 
dealing with the dynamic properties of 
quantum fluids under the warm-dense matter conditions and beyond. 

\section{Acknowledgments}

The authors acknowledge the support by the Deutsche Forschungsgemeinschaft via Project No. BO1366-15 and the grant \#AP09260349 of the Ministry of Education and Science, Kazakhstan. 

\section*{Appendices}

\subsection{Details of PIMC simulations}

Within the path-integral representation ab initio simulations of quantum degenerate systems are based on the high-temperature factorization of the $N$-body density matrix (DM) introduced originally by R.~Feynmann~\cite{Feynmanbook}. Since then, many efficient numerical implementation schemes have been developed starting with the pioneer works by D.~Ceperley devoted to the bosonic simulations of $^4$He~\cite{RevModPhys.67.279}. In contrast, similar detailed analysis of fermionic $^3$He at finite temperatures~\cite{Boronat2000} are missing due to the {\it fermion sign problem}~\cite{CeperleyFermi,Troyer2005}, when the statistical error of the estimated thermodynamic observables, $\bar{A}=\Avr{A}\pm \delta A$, with $\delta A \sim 1/\Avr{S(N,\beta)}$, is enhanced due to an exponential decay of the average sign $\Avr{S(N,\beta)}$ with the particle number $N$ and the inverse temperature $\beta$, see Fig.~\ref{fig:avrS} and Appendix A2 for more details. This leads to extreme computational costs in Monte-Carlo ensemble sampling scaled as $\sim 1/\Avr{S}^2$ and  prohibits the PIMC simulations for the $\{N,\beta\}$ combinations with $\Avr{S} \lesssim 10^{-3}$.

Trying to overcome this principle difficulty, here we employ the recently developed fermionic propagator approach (FP-PIMC)~\cite{filinov-etal.2021ctpp}, where in the representation of the many-body density matrix (DM) the standard bosonic propagators~\cite{RevModPhys.67.279} are substituted with the antisymmetric ones in the form of Slater determinants. As a result in the modified partition function, the summation over different permutation classes~\cite{RevModPhys.67.279} is performed analytically. In this representation, the $N$-body DM contains the Slater determinants, $\mathbb{M}^{s}(p-1,p)$, between each successive imaginary times $\tau_{p}-\tau_{p-1}=\epsilon$, with $p=1,\ldots P$ and $\epsilon=\beta/P$. The average sign now reduces to the composite sign of the determinants 
\begin{eqnarray}
\Avr{S}=\Avr{\prod\limits_{p=1}^P\sgn \, \mathbb{M}^{\uparrow}(p-1,p) \cdot \sgn  \mathbb{M}^{\downarrow}(p-1,p)},\label{def_Sign}    
\end{eqnarray}
while their absolute value enters explicitly in the probability density~\cite{filinov-etal.2021ctpp}. Here we explicitly use the fact that $^3\text{He}$ is the spin-$1/2$ fermion and consider the spin-unpolarized system, $N^{\uparrow}=N^{\downarrow}=N/2$. 

The efficiency of the fermionic propagator approach has been demonstrated by several authors: Takahashi and Imada~\cite{Takahashi1984}, V.Filinov \textit{et al.}~\cite{filinov_ppcf_01} and Lyubartsev~\cite{Lyubartsev_2005}. Chin~\cite{Chin2015} used the determinant propagators  to simulate relatively large ensembles of electrons in 3D quantum dots. A similar approach termed ``permutation-blocking PIMC" (PBPIMC) was recently developed by Dornheim \textit{et al.} \cite{dornheim_njp15} and then applied to the uniform electron gas at warm dense matter conditions~\cite{dornheim_physrep_18}.

The efficiency of the FP-PIMC depends on the imaginary time step $\epsilon=\beta/P$ crucially. 
The larger time step (smaller $P$-value)
in the high-temperature factorization increases the average sign $\Avr{S}$ and extends the applicability range of fermionic simulations to a higher degeneracy. To reduce a number of $P$ factors in the DM, in the present work we take advantage of the fourth-order factorization scheme proposed by 
Chin \textit{et al.}~\cite{Chin2002} and Sakkos \textit{et al.}~\cite{Sakkos2009}:
\begin{eqnarray}
  &&e^{-\beta\hat H}
     =\prod\limits_{p=1}^P e^{-\epsilon(\hat K+\hat V) } \label{4thorder} \\ 
      &&\approx \prod\limits_{p=1}^P e^{-\epsilon \hat W_{1}} e^{-t_1 \epsilon \hat K}  e^{- \epsilon \hat W_{2}} e^{-t_1 \epsilon \hat K} e^{\epsilon \hat W_{1}} e^{-t_0 \epsilon \hat K} + O(\epsilon^{4})\,,
     \nonumber
\end{eqnarray}
with the choice $\epsilon=\beta/P,\,  (2t_1+t_0=1),\, t_0=1/6$, and $\hat K (\hat V)$ being the kinetic (potential) energy operator. The detailed analysis is provided below.

Finally, we note that the standard periodic boundary conditions (PBC) were employed due to the short-range nature of the interatomic $^3\text{He}$ potential (Fig.~\ref{fig:DiffP}d). In the pair interaction only atoms within the original simulation cell are included using the minimum-image convention. This does affect the estimated potential energy per atom for system sizes $N< 100$, but has a negligible effect on the static properties (see Appendex A3) used as the input in the self-consistent method of moment introduced in Sec.~\ref{SSMMSec}.


\subsubsection{Convergence analysis}\label{ConvergSec}

\begin{figure}[]
\hspace{-0.2cm}
\includegraphics[width=0.515\textwidth]{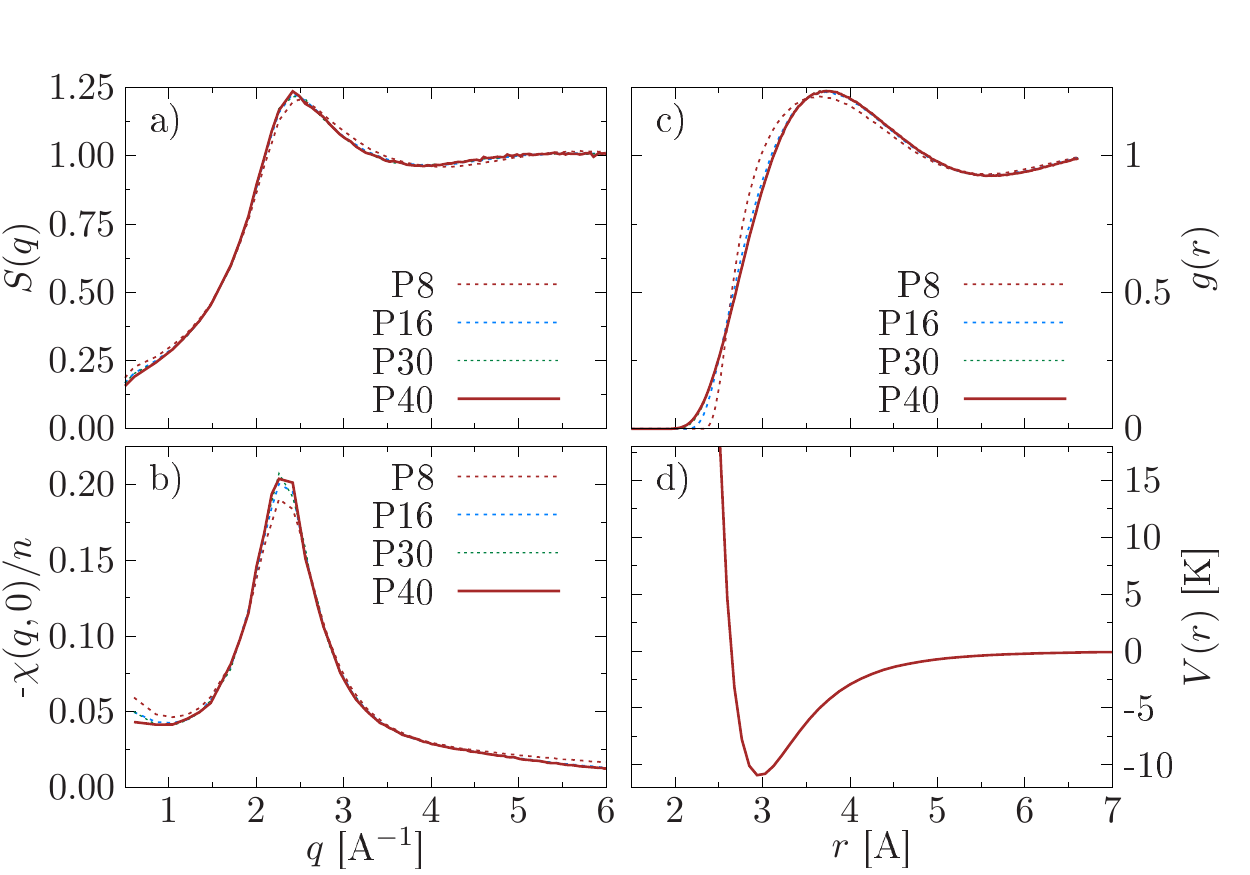}
\vspace{-0.5cm}
\caption{a) The static structure factor, $S(q)$; b) the static density response function, $-\chi(q)/n$, 
c) the radial distribution function $g(r)$ at the saturated vapor pressure density $\rho=0.01635 \AA^{-3}$ (SVP) 
and temperature $T=2$~$K$ for different numbers of factorization factors $8\leq P\leq 40$, and d) the $^3$He 
interatomic interaction potential $V(r)$ used in the PIMC simulations (the explicit analytical 
form is taken from Ref.~\cite{JChemPhys79} and is specified in Appendix B).}
\label{fig:DiffP}
\end{figure}

\begin{figure}[]
\hspace{-0.2cm}
\includegraphics[width=0.515\textwidth]{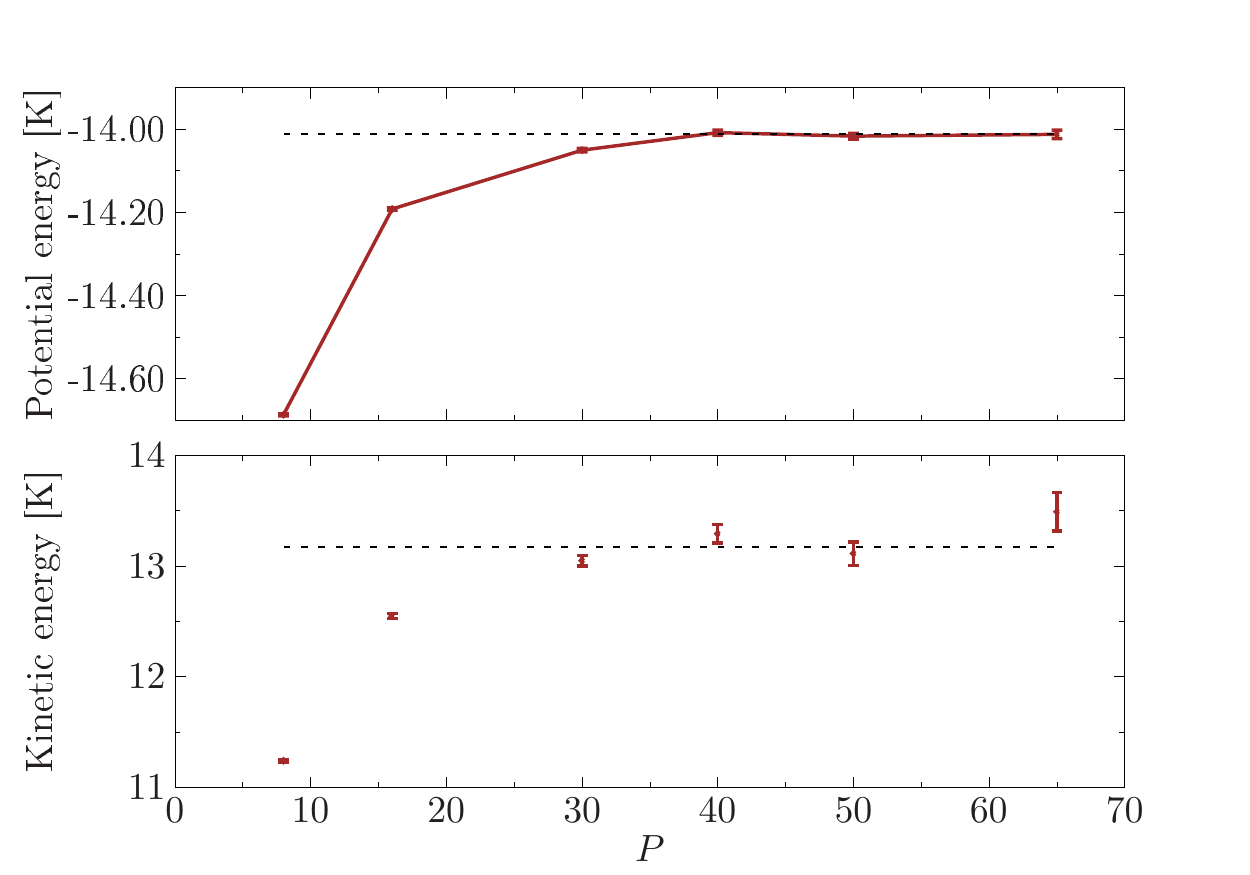}
\vspace{-.8cm}
\caption{The potential energy (estimated with the minimum-image convention) and the kinetic energy vs. 
the number of factorization factors, $P$. 
The simulation parameters ($\rho=0.01635 \AA^{-3}$, $T=2$~K, the number of atoms $N=38$) are those of Fig.~\ref{fig:DiffP}.}
\label{fig:EnergyDiffP}
\end{figure}

An important practical issue in PIMC simulations is the $P$-convergence of the estimated thermodynamic properties. This means that the corresponding systematic errors due to the neglected high-order commutators in the $N$-body DM, the terms of the order $O(\epsilon^4)$ in Eq.~(\ref{4thorder}) between the non-commuting operators, $[\hat K, \hat V]\neq 0$,
 remain much smaller~\cite{Chin2002,Sakkos2009} than the Monte Carlo statistical error, and the results obtained with $P\geq P_0$ agree well within their error bars.     

The results of our $P$-convergence test for $^3\text{He}$ at $T=2$K are presented
in Fig.~\ref{fig:DiffP}. The static structure factor $S(q)$ demonstrates the fastest convergence. 
The DM factorization error in the static density response function $\chi(q,0)$ is still noticeable for $P=8,16$, 
but vanishes fast for $P\geq 30$. The strongest $P$-dependence is observed in the short-range behaviour of the radial 
distribution function $g(r)$ at $r\lesssim 2.5 \AA$, see Fig.~\ref{fig:DiffP}c, where $^3$He atoms experience the 
influence of a strong repulsive core of the interaction potential $V(r)$, see Fig.~\ref{fig:DiffP}d. 
At the temperature $T=2$K this range of interatomic distances is classically forbidden and only 
becomes accessible due to the quantum-mechanical effects taken into account in the PIMC representation of the density matrix by the high-temperature factorization. This requires the employment of a higher number of propagators $P$.    

More quantitatively, the analysed $P$-dependence can be verified by the estimation of the thermodynamic properties. 
The convergence of the kinetic and the potential energy versus $P$ is demonstrated in Fig.~\ref{fig:EnergyDiffP}. 
The revealed factorization errors imply the usage of at least $P_0=40$ propagators for the SVP $^3$He at $T=2$K. 
With the increase in $P$ the kinetic energy estimator demonstrates 
larger statistical error bars compared to those in the potential energy, but this is the expected theoretical effect~\cite{RevModPhys.67.279}. The dotted line in Fig.~\ref{fig:DiffP} shows the extrapolation to the limiting case when $P \gg 1$. Both data sets for $P\geq P_0$ do agree with the extrapolated values within the statistical deviations. In the rest of the simulations we used $P_0=40$ for $T_0 \geq 2$ $K$. At lower temperatures, the number of propagators was increased employing the scaling, $P=P_0 \cdot (T_0/T$).  

Finally, we note that the estimated average kinetic energy per atom (for $N=38$ in the PBC), $\epsilon_{k,T=2K}=13.2(\pm 0.2)$K, is found to be in a good agreement with the experimental value $\epsilon^{\text{ex}}_{k}=12.05(\pm 1.33)$K (Ref.~\cite{Andreani_2006}) and the zero-temperature QMC data, $12.7$K (Refs.~\cite{Boronat2004,Boronat2000}). The low temperature results are presented in Fig.~\ref{fig:avrS} and demonstrate even better agreement, $\epsilon_{k,T=1.5K}=12.62(\pm 0.26)$ $(N=14)$. 

\subsubsection{Fermion sign problem}\label{avrS}

\begin{figure}[]
\hspace{-0.6cm}
\includegraphics[width=0.51\textwidth]{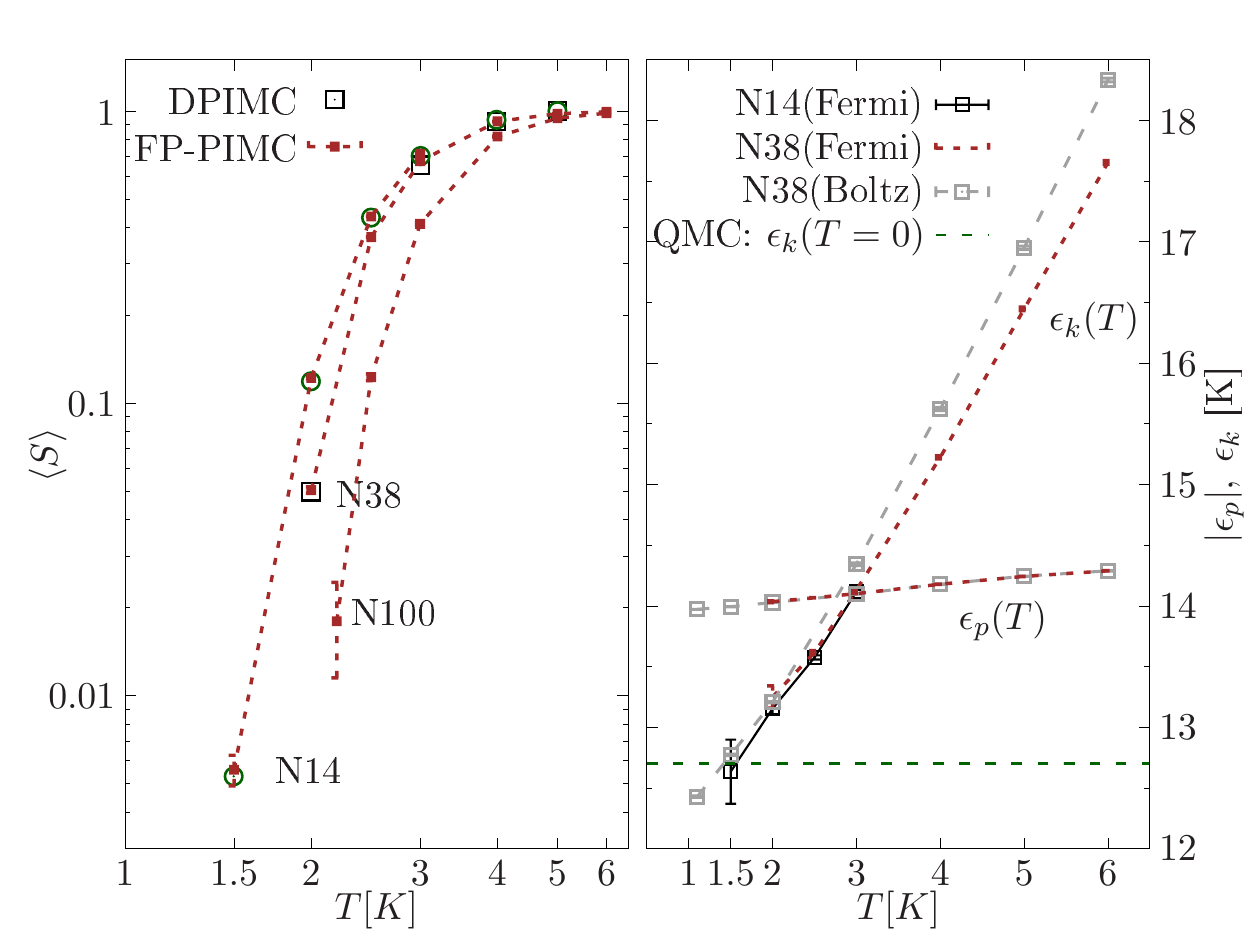}
\vspace{-0.6cm}
\caption{{\it Left}: The average sign $\Avr{S}$ in the FP-PIMC (dashed brown curves with symbols) and the DPIMC~\cite{dornheim-etal.2022sr} (black squares) simulations for $N=14,38$ and $N=100$ of $^3$He system at SVP as a function of temperature $T$. {\it Right}: The temperature dependence of the kinetic ($\epsilon_k$) and the potential ($|\epsilon_p|$) energies for $N=38$ ($T\geq 2$K) and $N=14$ (Fermi) ($1.5 \text{K} \leq T\leq 3$K) estimated using the Fermi and the Boltzmann statistics. The green horizontal line, $\epsilon_k(T=0)$, is the ground state fermionic QMC result~\cite{Boronat2004,Boronat2000}. A noticeable difference in the kinetic energy obtained with the Boltzmann and the Fermi statistics is observed only for $T\gtrsim 2.5$K. This discrepancy would then lead to noticeable exchange-correlation effects in the kinetic contribution $K(q)$ to the fourth moment $\mu_4(q)$, see Eqs.~(\ref{C4Ka})-(\ref{C4K}).
}
\label{fig:avrS}
\end{figure}

In this section we discuss the severity of the fermion sign problem~\cite{CeperleyFermi,Troyer2005} in our $^3$He simulations.
The left-hand panel in Fig.~\ref{fig:avrS} demonstrates the decay of the average sign $\Avr{S(N,\beta)}$, Eq.~(\ref{def_Sign}), using the fourth-order anti-symmetric propagators with $P=40$, versus the temperature $T$ and the particle number $N$ (brown dashed lines). For the comparison, the similar average sign is shown from the recent PIMC simulations~\cite{dornheim-etal.2022sr} with the $P\sim 1000$ bosonic-propagators (black squares) and where the fermionic antisymmetry is taken into account by the correct permutation parity~\cite{RevModPhys.67.279,CeperleyFermi}.  
Both results for $\Avr{S}$ nearly coincide and demonstrate a similar decay with the increase in the degeneracy parameter, $\theta=T/T_F\leq 0.43$ ($T\leq 2$K). This trend "prohibits" the simulations at lower temperatures due to an exponential decay of the average sign $\Avr{S}$.

Surprisingly, for cold helium we observe no noticeable improvement in the {\it fermion sign problem} using more sophisticated antisymmetric propagators, in contrast to the uniform electron gas case~\cite{dornheim_physrep_18,filinov-etal.2021ctpp}. The main reason is the need for a large number of high-temperature factors ($P=40$) to reproduce the short-range correlation
between helium atoms in the density matrix (Appendix A1). As a result, the Slater determinants, being evaluated at the effectively much high temperature $\tilde T=T\cdot 3P$ (the fourth-order propagators, Eq.~(\ref{4thorder}), contain three additional imaginary time steps), cannot reproduce the Pauli exclusion principle and  degeneracy effects present at the original "physical" temperature $T$. They are recovered only by the cancellation of the positive and negative contributions to the DM~\cite{CeperleyFermi}. 

\subsubsection{Finite-size effects}\label{finiteSF}

\begin{figure}[]
\hspace{-0.5cm}
\includegraphics[width=0.51\textwidth]{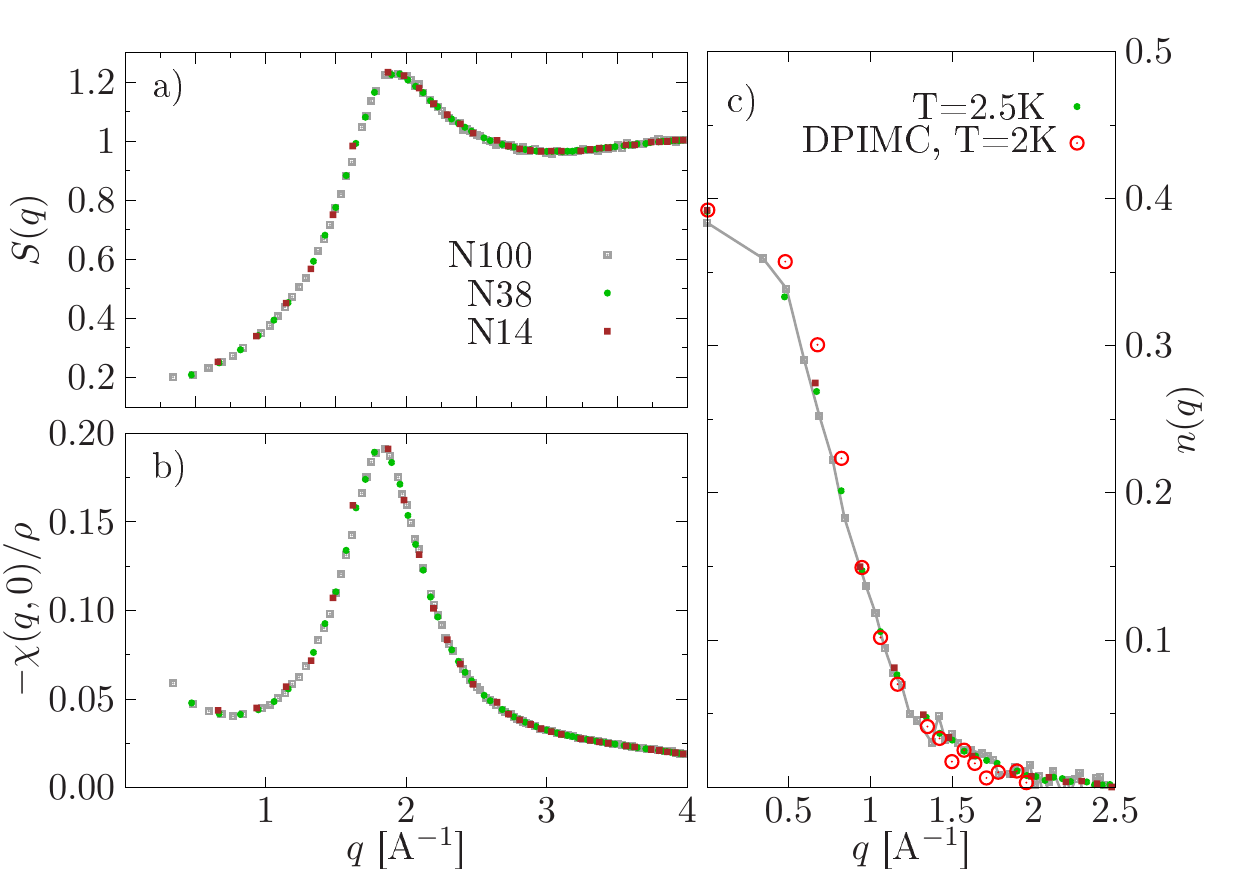}
\vspace{-0.6cm}
\caption{a), b) The finite-size effects ($N=14,38,100$) in the FP-PIMC 
simulations at $T=2.5$K (SVP) for 
a) the static structure factor $S(q)$; b) the 
static density response $\chi(q)$; and c) the momentum 
distribution $n(q)$ for $N=14,38,100$ compared to  the 
DPIMC~\cite{dornheim-etal.2022sr} 
(red circles, $N=38$ and $T=2$K).
}
\label{fig:FSplot}
\end{figure}

The results of numerical simulations discussed in the main text reveal a reasonable agreement with the experimental data (see Fig.~\ref{fig:Sk} and Fig.~\ref{fig:DSFExp}), even though the theoretical expectation values have been obtained for a finite system size $N\leq 100$, with additional limitations on $N$ imposed by the fermion sign problem (Appendix A2). To understand the importance of the finite-size effects for our results, the static structure factor (and thus the radial distribution function as well) and the static density response function are estimated in Fig.~\ref{fig:FSplot}a,b for $N=14,38,100$. Both quantities are critical for the present analysis as they contribute to the power moments $\mu_0(q)$ and $\mu_4(q)$ (see Sec.~\ref{FreqMom}), and, therefore, directly influence the reconstructed dynamic properties. We notice no differences between the $N=38$ and $N=100$ cases in a broad range of the $q$-vector. Even the smallest system ($N=14$) reproduces quite accurately (within few percent) the behaviour at small and large $q$ values, including
the amplitude and the width of the main peak in $S(q)$ and $\chi(q,0)$. As an additional test of our simulations, in Fig.~\ref{fig:FSplot}c we present the momentum distribution function $n(q)$ evaluated at $T=2.5$K as the Fourier transform of the one-particle off-diagonal density matrix~\cite{filinov.2012pra}. The agreement for $N=14,38,100$ is excellent. Some minor deviations are observed in the occupation of the zero-momentum state $n(0)$ for the smaller system $N=14$ (which is expected due to the shell effects, $q_n=2\pi n/L$). A similar distribution at $T=2$K from Ref.~\cite{dornheim-etal.2022sr} is included for comparison and demonstrates a reasonable agreement. Both results show the temperature effect resolved as a slight depletion at lower momenta induced by the thermal fluctuations (see Fig.~\ref{fig:nkDiffT} for the present $n(q)$ at $T=2$K and $N=38$). To conclude, no noticeable finite-size effects are present in our simulations with $N\geq 38$.

\subsubsection{Effect of Fermi statistics in $^3$He}

\begin{figure}[]
\hspace{-0.5cm}
\includegraphics[width=0.51\textwidth]{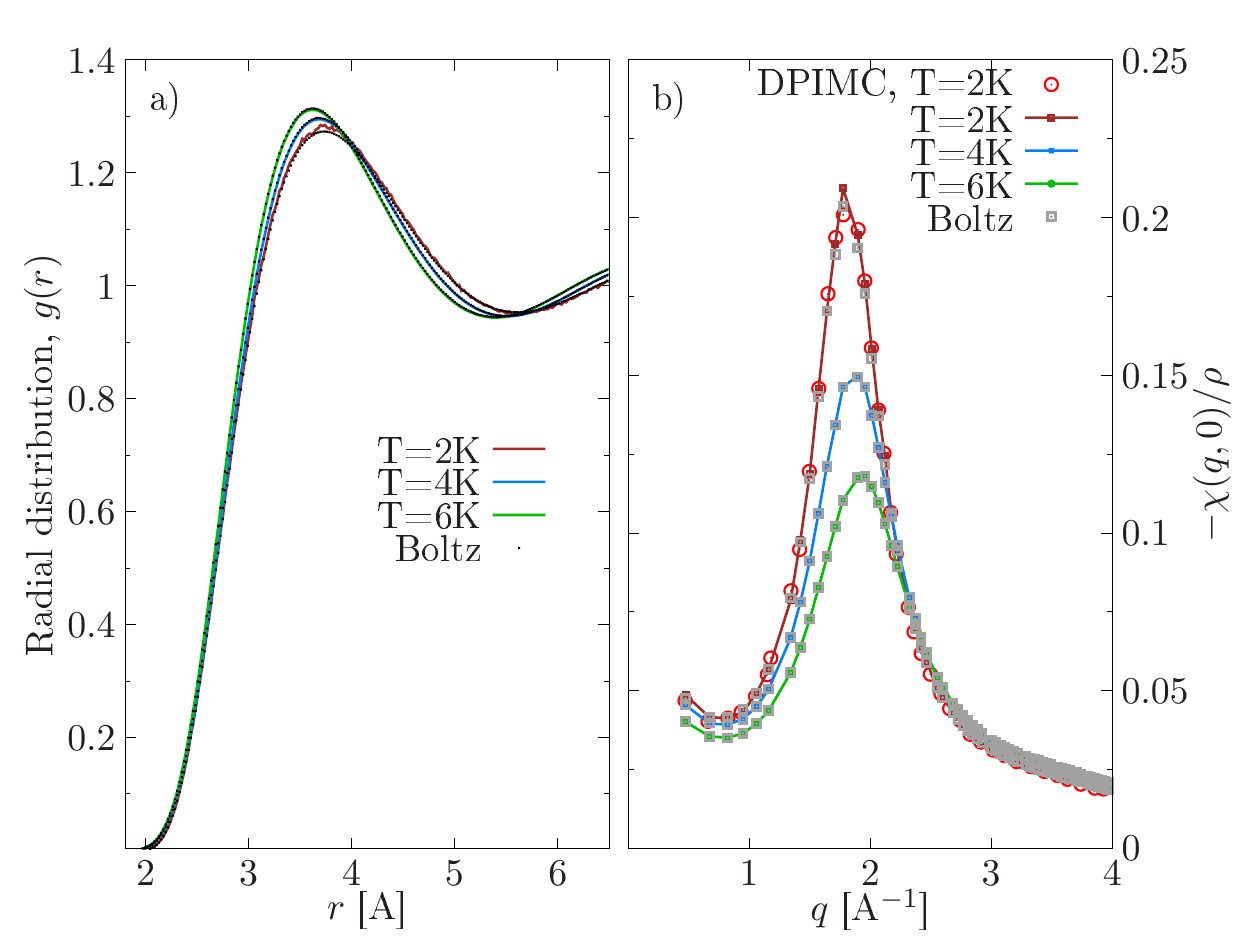}
\vspace{-0.6cm}
\caption{The effect of Fermi statistics vs. the Boltzmann model (distinguishable particles) on the pair distribution function $g(r)$ (panel "a") and the static density response function $\chi(q)$ (panel "b") for three temperatures, $T=2,4$ and $6$K, and $N=38$ of $^3$He at SVP. The $\chi(q)$ results are compared to those of DPIMC~\cite{dornheim-etal.2022sr} (red circles, $N=38$ and $T=2$K). 
The results confirm the negligible effect of quantum statistics on the $^3$He static properties, being in agreement with 
the similar findings for bosonic $^4$He systems~\cite{RevModPhys.67.279}. The same conclusion directly applies to the related frequency moment $\mu_0(q)$ and the interaction contribution $U(q)$ in $\mu_4(q)$, see Eqs.~(\ref{C4Ka}) and~(\ref{Qr}), which for $T < 2$K can be substituted with those estimated in the Boltzmann model (neglecting the exchange effects).
}
\label{fig:gpairFB}
\end{figure}

\begin{figure}[]
\hspace{-0.5cm}
\includegraphics[width=0.51\textwidth]{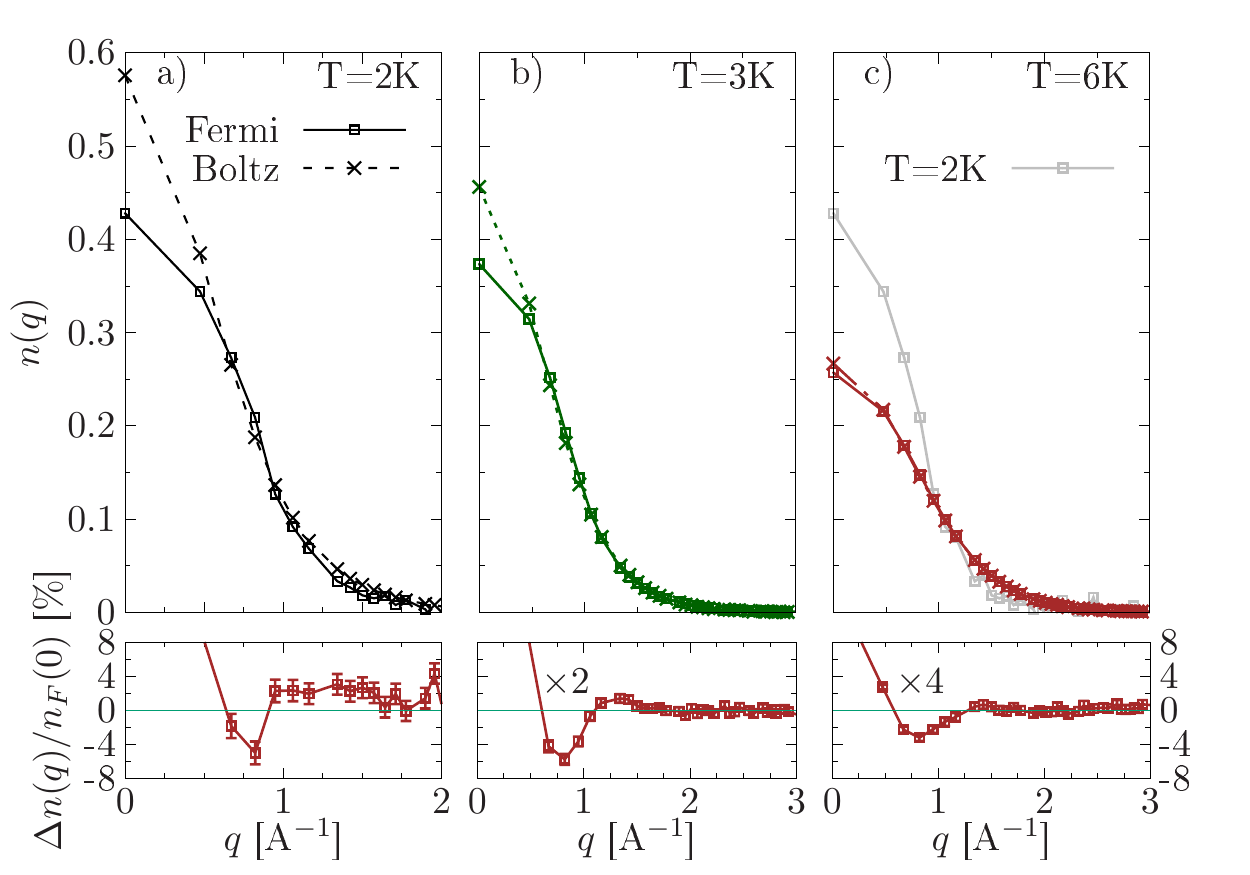}
\vspace{-0.6cm}
\caption{As in Fig.~\ref{fig:gpairFB} but for the momentum distribution function $n_{F,B}(q,T)$ shown at three temperature values and $N=38$. Here, in contrast to the previous case, we do observe a noticeable effect of quantum statistics. The additional exchange interaction (the Pauli blocking effect) leads to a stronger depletion of the low-momentum states compared to the Boltzmann case. For $T=2$K (first panel) our data are in a full agreement with those of Ref.~\cite{dornheim-etal.2022sr}.
The lower panels show the corresponding difference (in percent),  $\Delta n(q)=n_B(q)-n_F(q)$, normalized to the occupation of the zero momentum state $n_F(0)$. The deviations in the second (third) panel are scaled by the factor $\times 2(\times 4)$. The largest deviations are observed at $q=0$ (not shown): $\Delta n(0)/n_F(0)\approx 34\%(2\text{K}),22\%(3\text{K})$ and $3.8\%(6\text{K})$.   
}
\label{fig:nkDiffT}
\end{figure}

As we point out in the main text and in Appendix B, the present reconstruction procedure of the dynamic properties involves the frequency moments $\mu_0(q), \mu_4(q)$. In turn, they are determined via the pair distribution function $g(r)$ and the static response function $\chi(q)$.
The detailed comparison of both quantities evaluated with Fermi statistics and within the Boltzmann model (distinguishable particles) is demonstrated in Fig.~\ref{fig:gpairFB}, and confirm that the influence of the exchange-correlation effects on these static characteristics is negligible in the analysed temperature range $2\text{K}\leq T\leq 6$K. Similar observations have been reported for bosonic $^4$He~\cite{RevModPhys.67.279} and fermionic $^3$He systems~\cite{dornheim-etal.2022sr}.
This fact permits to employ the Boltzmann model at the lowest temperature $T=1.5$K, where the fermionic simulations face with the fermion sign problem for $N > 14$, see Fig.~\ref{fig:avrS}.  On the other hand, the importance of quantum statistics is clearly seen in the momentum distribution function $n_{F}(q,T)$ compared to $n_{B}(q,T)$ in Fig.~\ref{fig:nkDiffT}. 


\subsection{Fourth frequency moment $\mu _{4}\left( q\right)$}

In order to evaluate the fourth moment \eqref{moms} we write it in the form obtained using the commutation relations~\cite{Puff,filinov.2016pra,filinov.2012pra} and split the resulting expression into the kinetic and interaction contributions:

\begin{eqnarray}
  &&\mu_4(q)=\frac{\rho \epsilon_q}{\hbar^4}\left[K(q)+U(q)\right],\label{C4Ka}\\
  &&K(q)=\epsilon_q^2/4+2 \epsilon_q E_q,\label{C4K}
\end{eqnarray}
where $\epsilon_q=\hbar^2 q^2/m$, $E_q$ is the average kinetic energy per atom in the interacting system; the interaction contribution factor $U(q)$ is estimated in the coordinate space and is uniquely determined by the radial distribution function $g(r)$ and the interparticle interaction energy, $V_{ij}(r)$ (Ref.~\cite{filinov.2016pra})
\begin{eqnarray*}
U(q)=\frac{2\hbar^2}{N m} \sum\limits_{i<j}^N \left\langle R_{ij}(q) (\mathbf e_q \cdot  \vec \nabla_i) (\mathbf e_q \cdot \vec \nabla_j) V(\mathbf r_{ij})\right\rangle \label{C4I} ,
\end{eqnarray*}
where $R_{ij}(q)=exp\left[i \mathbf q (\mathbf r_i-\mathbf r_j)-1 \right]$ 
and $\mathbf e_q $ is the unit vector in the direction of the vector $\mathbf q$ . 

For a spatially isotropic and monoatomic system the above expression simplifies into

\begin{eqnarray}
U(q)=\frac{\rho \hbar^2}{m} \int \db \mathbf r \ g(r) \left[e^{i \mathbf q\cdot \mathbf r}-1 \right] 
(\mathbf e_q \cdot \vec \nabla)^2 V(r) . \label{Qr}
\end{eqnarray}

Further, if the pair interaction potential is radially symmetric, the angular averaging in Eq.~(\ref{Qr}) can be performed analytically. By choosing the z axis along the wavevector $\mathbf{q}$, the general expression involving the angular averaged force tensor reduces to the evaluation of a single matrix element
\begin{eqnarray}
    \Avr{f_{zz}}_{\psi}=\Avr{\tilde z^2}_{\psi} \frac{\partial^2 V(r)}{\partial r^2}+\Avr{1-\tilde{z}^2}_{\psi}\frac{1}{r} \, \frac{\partial V}{\partial r}, \label{Vzz}
\end{eqnarray}
with $\tilde z=\cos \psi$ and $\psi=\angle(z,\mathbf r)$ being the polar angle. The explicit evaluation of the prefactors (for a system with a radially symmetric pair potential) leads to the following results:
\begin{eqnarray}
  &&\Avr{\tilde{z}^2}_{\psi}=\int\limits_{-1}^1 \db \tilde z \, \tilde z^2 \, (e^{iq r \tilde z}-1)= \nonumber \\ 
  && \frac{4}{3}\left[j_2(qr)+\frac{1}{2}\left(1-j_0(qr)\right)\right],\\
  &&\Avr{1-\tilde{z}^2}_{\psi} =
  \frac{4}{3}\left[1- j_0(qr)- j_2(qr)\right]  
\end{eqnarray}
with $j_n(x)$ being the spherical Bessel function. 

Below we apply the above result to the $^3$He system with the semiempirical potential~\cite{JChemPhys79} presented in the following general form vs. the reduced variable $x=r/r_m$ and with $V_0=1$K:
\begin{eqnarray}
  &&V(x)/V_0=F(x)\cdot G(x)+ A(x),\label{V1} \\
  &&G(x)=\sum_{n=6,8,10} C_n f^n(x), \; f^n(x)=1/x^n,\\
  && F(x)=e^{-(D f(x)-1)^2}, \; x<D,\\
  && A(x)=A e^{-\alpha x + \beta x^2}=
  A \exp\left[\sum\limits_{n=-2}^{-1} B_n f^n(x)\right], \label{Vint}
\end{eqnarray}
where $r_m$ specifies the position of the minimum in the above Van-der-Waals-type interaction potential.

The force tensor reduces to 
\begin{eqnarray}
    &&\Avr{f^{\text{He}}_{zz}}(x)=\Avr{F_{zz}}(x)\cdot G(x) + 2 \Avr{F_{z} \cdot G_z}(x) \nonumber \\
    &&+ F(x) \cdot \Avr{G_{zz}}(x)+\Avr{A_{zz}}(x). \label{fzz}
\end{eqnarray}
Next, it is convenient to present the angular average of the above expression exclusively in terms of the corresponding averages of the first, $f_z$, and the second derivatives, $f_{zz}$, of the Coulomb interaction, $f(x)=1/x$,
\begin{eqnarray}
\Avr{f_z^2}_{\psi}=\Avr{\tilde z^2}/x^4 , \; \Avr{f_{zz}}_{\psi}=- 4 j_2(x)/x^3     
\end{eqnarray}
Below we list the explicit expressions involved in Eq.~(\ref{fzz}): 
\begin{eqnarray*}
&&A_z=A(x) A_1(x) f_z(x),\\
&&\Avr{A_{zz}}=A(x) \left[A_1^2(x)+A_2(x)\right] \Avr{f_z^2}+ A(x)A_1(x) \Avr{f_{zz}}  
\end{eqnarray*}
with
\begin{eqnarray*}
&&A_1(x)=\sum\limits_{n=-2}^{-1} B_n n f^{n-1}(x),\\
&&A_2(x)=\sum\limits_{n=-2}^{-1} B_n n (n-1) f^{n-1}(x).
\end{eqnarray*}
The next factor
\begin{eqnarray*}
&&F_z=F(x) F_1(x) \cdot f_z(x),\\
&&\Avr{F_{zz}}=F(x)\left[F_1^2(x)-2 D^2\right]\Avr{f_z^2}+F(x)F_1(x)\cdot \Avr{f_{zz}} 
\end{eqnarray*}
with
\begin{eqnarray*}
F_1(x)=-2 \left(D/x-1 \right) D.
\end{eqnarray*}

The long-range tail of the Van der Waals-type potential produces the following contribution
\begin{eqnarray*}
&&G_{z}=G_1(x) \cdot f_z(x),\; G_1(x)=\sum\limits_{n=6,8,10} n C_n f^{n-1}(x)\\
&&\Avr{G_{zz}}=\sum\limits_{n=6,8,10} n C_n f^{n-1}(x)  \left[\Avr{f_{zz}}+(n-1) x \Avr{f_z^2} \right].
\end{eqnarray*}
The crossterm in Eq.~(\ref{fzz}) contains the force-squared contribution
\begin{eqnarray*}
2 \Avr{F_{z} \cdot G_z}=2\, F(x) F_1(x) G_1(x)\cdot  \Avr{f_z^2}. 
\end{eqnarray*}
The above expressions can be used to evaluate the angular-averaged total force tensor $\Avr{f_{zz}^{\text{He}}}$  for the $^3$He system, and, finally, to obtain the correlation contribution to the fourth moment as a one-dimensional integral 
\begin{eqnarray*}
U(q)=2 \pi \rho \frac{\hbar^2}{m} \int\limits_0^{\infty} \db r  \, (r/r_m)^2 g(r) \, V_0\Avr{f_{zz}^{\text{He}}(q)}(r/r_m), 
\end{eqnarray*}
with $g(x)=g(r)$ being the radial distribution function evaluated via the PIMC at 
given thermodynamic conditions $\{\rho[\AA^{-3}],T[K]\}$. 

By introduction of the length units $l=1 \AA$ ($r \rightarrow \tilde r= r/l$), 
the above expression reduces to 
\begin{eqnarray*}
U(q)=2 \pi \tilde \rho E_0 V_0 \int\limits_0^{\infty} \db \tilde r \, (\tilde r/\tilde r_m)^2 g(\tilde r) \Avr{f_{zz}^{\text{He}}(q)}(\tilde r/\tilde r_m)
\end{eqnarray*}
with $\tilde \rho=\rho l^3$, $V_0[K]=1$,  $E_0=m_e/m_{He}(a_B/l)^2 \cdot 1 Ha=16.08$K, 
and its dimension is that of the energy squared.

\begin{figure}[]
\hspace{-0.1cm}
\includegraphics[width=0.52\textwidth]{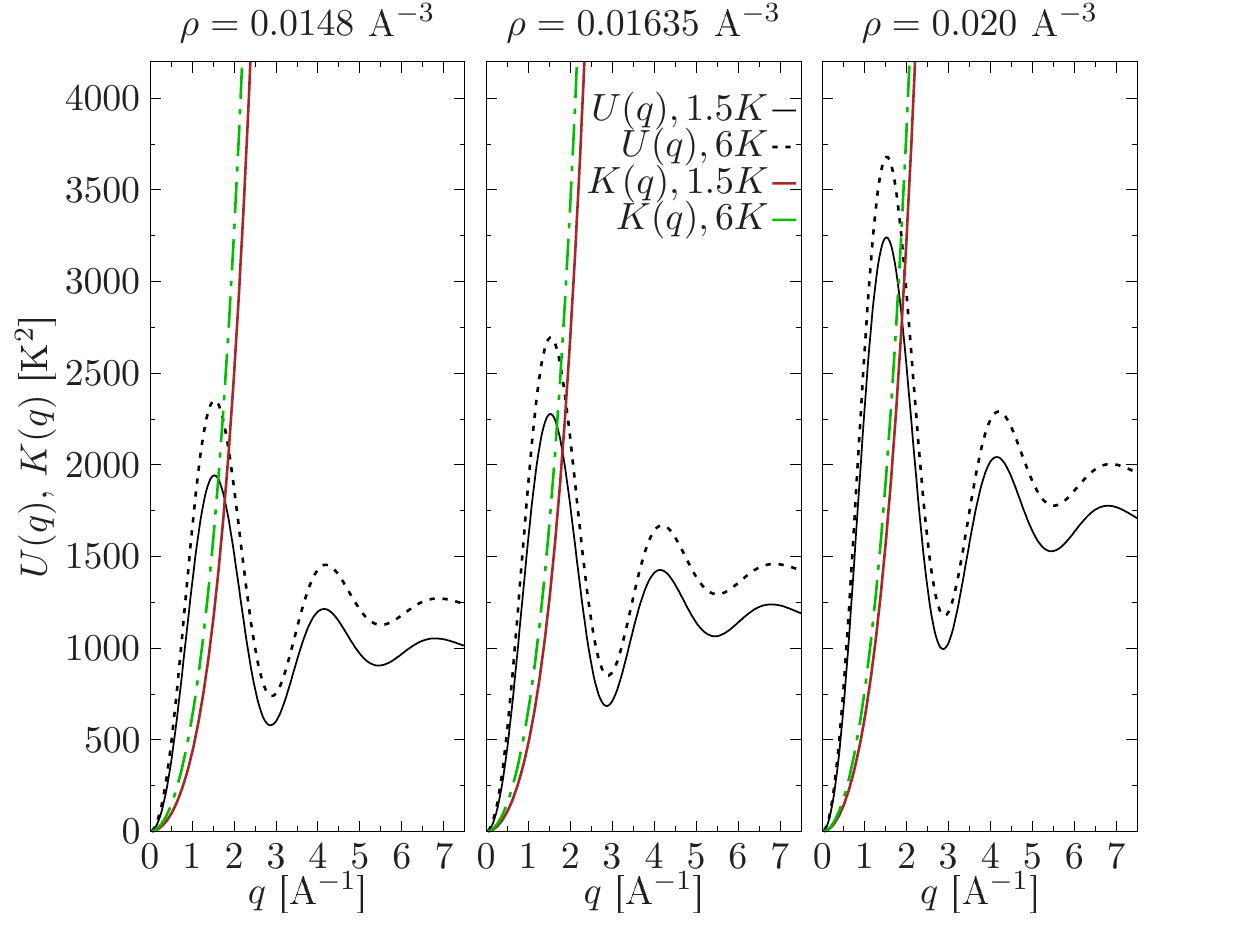}
\vspace{-0.8cm}
\caption{Estimation of the kinetic and interaction contributions to the fourth frequency moment $\mu_4(q)$, see Eqs.~(\ref{C4K}) and~(\ref{C4I}).
}
\label{fig:C4mom}
\end{figure}

The results for the kinetic and interaction contributions to the fourth-moment at three densities and temperatures $T=1.5$K and $6$K are presented in Fig.~\ref{fig:C4mom}.
Both the kinetic and the interaction contributions to the moment $\mu_{4}$ demonstrate only a weak temperature dependence.

\subsection{The dynamic Nevanlinna parameter function}

In order to construct the dynamical Nevanlinna parameter function $Q_{2}\left( q,z\right) $ introduced above, we consider the cases of five and nine 
moments and construct the auxiliary functions $N_{2}$ and  $N_{4}$ . By virtue of the Nevanlinna theorem both of them determine the spectral function, see equation \eqref{Mnu}. Hence, we
can equalize them: 
\begin{equation}
\frac{E_{3}+Q_{2}E_{2}}{D_{3}+Q_{2}D_{2}}=\frac{E_{5}+Q_{4}E_{4}}{%
D_{5}+Q_{4}D_{4}}\ ,  \label{25}
\end{equation}%
and arrive to the expression for the dynamical five-moment Nevanlinna function
in terms of the nine-moment one:
\begin{equation}
Q_{2}=-\frac{D_{3}E_{5}-E_{3}D_{5}+\left( D_{3}E_{4}-D_{4}E_{3}\right) Q_{4}%
}{D_{2}E_{5}-E_{2}D_{5}+\left( D_{2}E_{4}-E_{2}D_{4}\right) Q_{4}}\ .  \label{400}
\end{equation}%
Then, we apply for the function $Q_{4}$ the static approximation and once more take into account the effective absence in the system of the diffusion zero-frequency mode:
\begin{equation}
Q_{4}\left( q,0\right) =ih_{4}\left( q,\tilde{\omega}\right) =\frac{i\omega
_{3}^{2}\left( \omega _{2}^{2}-\omega _{1}^{2}\right) \left( \omega
_{4}^{2}-\omega _{3}^{2}\right) }{\omega _{1}\sqrt{2\left( \omega
_{3}^{2}-\omega _{2}^{2}\right) ^{3}\left( \omega _{3}^{2}-\omega
_{1}^{2}\right) }}\ .
\end{equation}%
Thus we arrive to the closed expression for the dynamical Nevanlinna parameter function $Q_{2}$ provided in the main text. 
Notice that the nine-moment expression (\ref{400}) simplifies into the static five-moment solution (\ref{sa}) as soon as we consider 
two successive limiting transitions: $\omega_{4}\left( q\right) \rightarrow
\infty $ and $\omega_{3}\left( q\right) \rightarrow \infty$.

In the above procedure we employ only the following polynomials~\cite{dornheim-etal.nl.2020prl}:
\begin{equation}
\begin{array}{c}
D_{2}\left( z;q\right) =\left( z^{2}-\omega _{1}^{2}\right) \ ,\quad
D_{3}\left( z;q\right) =z\left( z^{2}-\omega _{2}^{2}\right) \ ,\quad  \\ 
D_{2}\left( z;q\right) =\left( z^{2}-\omega _{1}^{2}\right) \ ,\quad
D_{3}\left( z;q\right) =z\left( z^{2}-\omega _{2}^{2}\right) \ ,\quad  \\ 
E_{2}\left( z;q\right) =\mu_{0}z\ ,\quad E_{3}\left( z;q\right)
=\mu_{0}\left( z^{2}-\left( \omega _{2}^{2}-\omega _{1}^{2}\right) \right) \
,\quad  \\ 
E_{4}\left( z;q\right) =\mu_{0}\left( z^{3}+b_{1}z\right) \ ,
\quad E_{5}\left(
z;q\right) =\mu_{0}\left( z^{4}+d_{2}z^{2}+d_{0}\right) \ .%
\end{array}
\label{DE}
\end{equation}

The following notations are used here: 
\begin{equation*}
\begin{array}{c}
b_{1}=\frac{\omega _{1}^{4}-2\omega _{1}^{2}\omega _{2}^{2}+\omega
_{2}^{2}\omega _{3}^{2}}{\omega _{1}^{2}-\omega _{2}^{2}}\ ,\quad d_{2}=%
\frac{\omega _{1}^{2}\left( \omega _{2}^{2}-\omega _{3}^{2}\right) +\omega
_{3}^{2}\left( \omega _{4}^{2}-\omega _{2}^{2}\right) }{\omega
_{2}^{2}-\omega _{3}^{2}}\ , \\ 
d_{0}=\omega _{1}^{2}\omega _{2}^{2}+\omega _{3}^{2}\frac{\omega
_{1}^{2}\left( \omega _{4}^{2}-\omega _{2}^{2}\right) +\omega _{2}^{2}\left(
\omega _{3}^{2}-\omega _{4}^{2}\right) }{\omega _{2}^{2}-\omega _{3}^{2}}\ .%
\end{array}%
\end{equation*}%

\bibliography{ref}

\begin{thebibliography}{83}%
\makeatletter
\providecommand \@ifxundefined [1]{%
 \@ifx{#1\undefined}
}%
\providecommand \@ifnum [1]{%
 \ifnum #1\expandafter \@firstoftwo
 \else \expandafter \@secondoftwo
 \fi
}%
\providecommand \@ifx [1]{%
 \ifx #1\expandafter \@firstoftwo
 \else \expandafter \@secondoftwo
 \fi
}%
\providecommand \natexlab [1]{#1}%
\providecommand \enquote  [1]{``#1''}%
\providecommand \bibnamefont  [1]{#1}%
\providecommand \bibfnamefont [1]{#1}%
\providecommand \citenamefont [1]{#1}%
\providecommand \href@noop [0]{\@secondoftwo}%
\providecommand \href [0]{\begingroup \@sanitize@url \@href}%
\providecommand \@href[1]{\@@startlink{#1}\@@href}%
\providecommand \@@href[1]{\endgroup#1\@@endlink}%
\providecommand \@sanitize@url [0]{\catcode `\\12\catcode `\$12\catcode
  `\&12\catcode `\#12\catcode `\^12\catcode `\_12\catcode `\%12\relax}%
\providecommand \@@startlink[1]{}%
\providecommand \@@endlink[0]{}%
\providecommand \url  [0]{\begingroup\@sanitize@url \@url }%
\providecommand \@url [1]{\endgroup\@href {#1}{\urlprefix }}%
\providecommand \urlprefix  [0]{URL }%
\providecommand \Eprint [0]{\href }%
\providecommand \doibase [0]{http://dx.doi.org/}%
\providecommand \selectlanguage [0]{\@gobble}%
\providecommand \bibinfo  [0]{\@secondoftwo}%
\providecommand \bibfield  [0]{\@secondoftwo}%
\providecommand \translation [1]{[#1]}%
\providecommand \BibitemOpen [0]{}%
\providecommand \bibitemStop [0]{}%
\providecommand \bibitemNoStop [0]{.\EOS\space}%
\providecommand \EOS [0]{\spacefactor3000\relax}%
\providecommand \BibitemShut  [1]{\csname bibitem#1\endcsname}%
\let\auto@bib@innerbib\@empty
\bibitem [{\citenamefont {Albergamo}\ \emph {et~al.}(2007)\citenamefont
  {Albergamo}, \citenamefont {Verbeni}, \citenamefont {Huotari}, \citenamefont
  {Vank\'o},\ and\ \citenamefont {Monaco}}]{Albergamo2007}%
  \BibitemOpen
  \bibfield  {author} {\bibinfo {author} {\bibfnamefont {F.}~\bibnamefont
  {Albergamo}}, \bibinfo {author} {\bibfnamefont {R.}~\bibnamefont {Verbeni}},
  \bibinfo {author} {\bibfnamefont {S.}~\bibnamefont {Huotari}}, \bibinfo
  {author} {\bibfnamefont {G.}~\bibnamefont {Vank\'o}}, \ and\ \bibinfo
  {author} {\bibfnamefont {G.}~\bibnamefont {Monaco}},\ }\bibfield  {title}
  {\enquote {\bibinfo {title} {Zero sound mode in normal liquid
  $^{3}\mathrm{He}$},}\ }\href {\doibase 10.1103/PhysRevLett.99.205301}
  {\bibfield  {journal} {\bibinfo  {journal} {Phys. Rev. Lett.}\ }\textbf
  {\bibinfo {volume} {99}},\ \bibinfo {pages} {205301} (\bibinfo {year}
  {2007})}\BibitemShut {NoStop}%
\bibitem [{\citenamefont {Fåk}\ \emph {et~al.}(1994)\citenamefont {Fåk},
  \citenamefont {Guckelsberger}, \citenamefont {Scherm},\ and\ \citenamefont
  {Stunault}}]{fak1994}%
  \BibitemOpen
  \bibfield  {author} {\bibinfo {author} {\bibfnamefont {B}~\bibnamefont
  {Fåk}}, \bibinfo {author} {\bibfnamefont {K.}~\bibnamefont {Guckelsberger}},
  \bibinfo {author} {\bibfnamefont {R.}~\bibnamefont {Scherm}}, \ and\ \bibinfo
  {author} {\bibfnamefont {A.}~\bibnamefont {Stunault}},\ }\bibfield  {title}
  {\enquote {\bibinfo {title} {Spin fluctuations and zero-sound in normal
  liquid $^{3}\mathrm{He}$ studied by neutron scattering},}\ }\href {\doibase
  https://doi.org/10.1007/BF00754303} {\bibfield  {journal} {\bibinfo
  {journal} {Journal of Low Temperature Physics}\ }\textbf {\bibinfo {volume}
  {97}},\ \bibinfo {pages} {445--487} (\bibinfo {year} {1994})}\BibitemShut
  {NoStop}%
\bibitem [{\citenamefont {Nozières}(2004)}]{Noz2004}%
  \BibitemOpen
  \bibfield  {author} {\bibinfo {author} {\bibfnamefont {P.}~\bibnamefont
  {Nozières}},\ }\bibfield  {title} {\enquote {\bibinfo {title} {Is the roton
  in superfluid 4he the ghost of a bragg spot?}}\ }\href {\doibase
  10.1023/B:JOLT.0000044234.82957.2f} {\bibfield  {journal} {\bibinfo
  {journal} {Journal of Low Temperature Physics}\ }\textbf {\bibinfo {volume}
  {137}},\ \bibinfo {pages} {45--67} (\bibinfo {year} {2004})}\BibitemShut
  {NoStop}%
\bibitem [{\citenamefont {Levin}\ \emph {et~al.}(2012)\citenamefont {Levin},
  \citenamefont {Fetter},\ and\ \citenamefont {Stamper-Kurn}}]{book_gases}%
  \BibitemOpen
  \bibfield  {author} {\bibinfo {author} {\bibfnamefont {K.}~\bibnamefont
  {Levin}}, \bibinfo {author} {\bibfnamefont {A.}~\bibnamefont {Fetter}}, \
  and\ \bibinfo {author} {\bibfnamefont {Dan}\ \bibnamefont {Stamper-Kurn}},\
  }\href@noop {} {\emph {\bibinfo {title} {{Ultracold Bosonic and Fermionic
  Gases}}}}\ (\bibinfo  {publisher} {Elsevier},\ \bibinfo {year}
  {2012})\BibitemShut {NoStop}%
\bibitem [{\citenamefont {Kremp}\ \emph {et~al.}(2005)\citenamefont {Kremp},
  \citenamefont {Schlanges},\ and\ \citenamefont {Kraeft}}]{book_plasma}%
  \BibitemOpen
  \bibfield  {author} {\bibinfo {author} {\bibfnamefont {D.}~\bibnamefont
  {Kremp}}, \bibinfo {author} {\bibfnamefont {M.}~\bibnamefont {Schlanges}}, \
  and\ \bibinfo {author} {\bibfnamefont {W.D.}\ \bibnamefont {Kraeft}},\
  }\href@noop {} {\emph {\bibinfo {title} {{Quantum Statistics of Nonideal
  Plasmas}}}}\ (\bibinfo  {publisher} {Springer-Verlag Berlin Heidelberg},\
  \bibinfo {year} {2005})\BibitemShut {NoStop}%
\bibitem [{\citenamefont {Filinov}\ and\ \citenamefont
  {Bonitz}(2012)}]{filinov.2012pra}%
  \BibitemOpen
  \bibfield  {author} {\bibinfo {author} {\bibfnamefont {A.}~\bibnamefont
  {Filinov}}\ and\ \bibinfo {author} {\bibfnamefont {M.}~\bibnamefont
  {Bonitz}},\ }\bibfield  {title} {\enquote {\bibinfo {title} {Collective and
  single-particle excitations in two-dimensional dipolar bose gases},}\ }\href
  {\doibase 10.1103/PhysRevA.86.043628} {\bibfield  {journal} {\bibinfo
  {journal} {Phys. Rev. A}\ }\textbf {\bibinfo {volume} {86}},\ \bibinfo
  {pages} {043628} (\bibinfo {year} {2012})}\BibitemShut {NoStop}%
\bibitem [{\citenamefont {Hufnagl}\ and\ \citenamefont
  {Zillich}(2013)}]{PhysRevA.87.033624}%
  \BibitemOpen
  \bibfield  {author} {\bibinfo {author} {\bibfnamefont {D.}~\bibnamefont
  {Hufnagl}}\ and\ \bibinfo {author} {\bibfnamefont {R.~E.}\ \bibnamefont
  {Zillich}},\ }\bibfield  {title} {\enquote {\bibinfo {title} {Stability and
  excitations of a bilayer of strongly correlated dipolar bosons},}\ }\href
  {\doibase 10.1103/PhysRevA.87.033624} {\bibfield  {journal} {\bibinfo
  {journal} {Phys. Rev. A}\ }\textbf {\bibinfo {volume} {87}},\ \bibinfo
  {pages} {033624} (\bibinfo {year} {2013})}\BibitemShut {NoStop}%
\bibitem [{\citenamefont {Filinov}(2016)}]{filinov.2016pra}%
  \BibitemOpen
  \bibfield  {author} {\bibinfo {author} {\bibfnamefont {A.}~\bibnamefont
  {Filinov}},\ }\bibfield  {title} {\enquote {\bibinfo {title} {Correlation
  effects and collective excitations in bosonic bilayers: Role of quantum
  statistics, superfluidity, and the dimerization transition},}\ }\href
  {\doibase 10.1103/PhysRevA.94.013603} {\bibfield  {journal} {\bibinfo
  {journal} {Phys. Rev. A}\ }\textbf {\bibinfo {volume} {94}},\ \bibinfo
  {pages} {013603} (\bibinfo {year} {2016})}\BibitemShut {NoStop}%
\bibitem [{\citenamefont {Dornheim}\ \emph {et~al.}(2021)\citenamefont
  {Dornheim}, \citenamefont {Moldabekov},\ and\ \citenamefont
  {Vorberger}}]{dornheim.2021cpp}%
  \BibitemOpen
  \bibfield  {author} {\bibinfo {author} {\bibfnamefont {Tobias}\ \bibnamefont
  {Dornheim}}, \bibinfo {author} {\bibfnamefont {Zhandos~A.}\ \bibnamefont
  {Moldabekov}}, \ and\ \bibinfo {author} {\bibfnamefont {Jan}\ \bibnamefont
  {Vorberger}},\ }\bibfield  {title} {\enquote {\bibinfo {title} {Nonlinear
  electronic density response of the ferromagnetic uniform electron gas at warm
  dense matter conditions},}\ }\href {\doibase
  https://doi.org/10.1002/ctpp.202100098} {\bibfield  {journal} {\bibinfo
  {journal} {Contributions to Plasma Physics}\ }\textbf {\bibinfo {volume}
  {61}},\ \bibinfo {pages} {e202100098} (\bibinfo {year} {2021})}\BibitemShut
  {NoStop}%
\bibitem [{\citenamefont {Godfrin}\ \emph {et~al.}(2012)\citenamefont
  {Godfrin}, \citenamefont {Meschke}, \citenamefont {Lauter}, \citenamefont
  {Sultan}, \citenamefont {Böhm}, \citenamefont {Krotscheck},\ and\
  \citenamefont {Panholzer}}]{Godfrin2012}%
  \BibitemOpen
  \bibfield  {author} {\bibinfo {author} {\bibfnamefont {Henri}\ \bibnamefont
  {Godfrin}}, \bibinfo {author} {\bibfnamefont {Matthias}\ \bibnamefont
  {Meschke}}, \bibinfo {author} {\bibfnamefont {Hans-Jochen}\ \bibnamefont
  {Lauter}}, \bibinfo {author} {\bibfnamefont {Ahmad}\ \bibnamefont {Sultan}},
  \bibinfo {author} {\bibfnamefont {Helga~M.}\ \bibnamefont {Böhm}}, \bibinfo
  {author} {\bibfnamefont {Eckhard}\ \bibnamefont {Krotscheck}}, \ and\
  \bibinfo {author} {\bibfnamefont {Martin}\ \bibnamefont {Panholzer}},\
  }\bibfield  {title} {\enquote {\bibinfo {title} {Observation of a roton
  collective mode in a two-dimensional fermi liquid},}\ }\href {\doibase
  10.1038/nature10919} {\bibfield  {journal} {\bibinfo  {journal} {Nature}\
  }\textbf {\bibinfo {volume} {483}},\ \bibinfo {pages} {576--579} (\bibinfo
  {year} {2012})}\BibitemShut {NoStop}%
\bibitem [{\citenamefont {Feenberg}(1969)}]{CBF-book}%
  \BibitemOpen
  \bibfield  {author} {\bibinfo {author} {\bibfnamefont {E.}~\bibnamefont
  {Feenberg}},\ }\href@noop {} {\emph {\bibinfo {title} {{Theory of Quantum
  Fluids}}}}\ (\bibinfo  {publisher} {Academic, New York},\ \bibinfo {year}
  {1969})\BibitemShut {NoStop}%
\bibitem [{\citenamefont {Krotscheck}(1982)}]{PhysRevA.26.3536}%
  \BibitemOpen
  \bibfield  {author} {\bibinfo {author} {\bibfnamefont {E.}~\bibnamefont
  {Krotscheck}},\ }\bibfield  {title} {\enquote {\bibinfo {title} {Effective
  interactions, linear response, and correlated rings: A study of chain
  diagrams in correlated basis functions},}\ }\href {\doibase
  10.1103/PhysRevA.26.3536} {\bibfield  {journal} {\bibinfo  {journal} {Phys.
  Rev. A}\ }\textbf {\bibinfo {volume} {26}},\ \bibinfo {pages} {3536--3556}
  (\bibinfo {year} {1982})}\BibitemShut {NoStop}%
\bibitem [{\citenamefont {Krotscheck}\ and\ \citenamefont
  {Panholzer}(2011)}]{Krotscheck2011}%
  \BibitemOpen
  \bibfield  {author} {\bibinfo {author} {\bibfnamefont {E.}~\bibnamefont
  {Krotscheck}}\ and\ \bibinfo {author} {\bibfnamefont {M.}~\bibnamefont
  {Panholzer}},\ }\bibfield  {title} {\enquote {\bibinfo {title} {Theoretical
  analysis of neutron and x-ray scattering data on 3he},}\ }\href {\doibase
  10.1007/s10909-010-0308-y} {\bibfield  {journal} {\bibinfo  {journal}
  {Journal of Low Temperature Physics}\ }\textbf {\bibinfo {volume} {163}},\
  \bibinfo {pages} {1--12} (\bibinfo {year} {2011})}\BibitemShut {NoStop}%
\bibitem [{\citenamefont {Schoerkhuber}(2002)}]{Schoerkhuber-book}%
  \BibitemOpen
  \bibfield  {author} {\bibinfo {author} {\bibfnamefont {K.W.}\ \bibnamefont
  {Schoerkhuber}},\ }\href
  {http://inis.iaea.org/search/search.aspx?orig_q=RN:35072499} {\emph {\bibinfo
  {title} {{Multi-particle-hole excitations in many-body Fermi systems}}}},\
  Trans. of Math. Monographs 50, Amer. Math. Soc.\ (\bibinfo  {publisher}
  {Available from Universitaet Linz Bibliothek, 4040 Linz-Auhof (AT)},\
  \bibinfo {year} {2002})\BibitemShut {NoStop}%
\bibitem [{\citenamefont {Ceperley}(1995)}]{RevModPhys.67.279}%
  \BibitemOpen
  \bibfield  {author} {\bibinfo {author} {\bibfnamefont {D.~M.}\ \bibnamefont
  {Ceperley}},\ }\bibfield  {title} {\enquote {\bibinfo {title} {Path integrals
  in the theory of condensed helium},}\ }\href {\doibase
  10.1103/RevModPhys.67.279} {\bibfield  {journal} {\bibinfo  {journal} {Rev.
  Mod. Phys.}\ }\textbf {\bibinfo {volume} {67}},\ \bibinfo {pages} {279--355}
  (\bibinfo {year} {1995})}\BibitemShut {NoStop}%
\bibitem [{\citenamefont {Dornheim}\ \emph
  {et~al.}(2018{\natexlab{a}})\citenamefont {Dornheim}, \citenamefont {Groth},\
  and\ \citenamefont {Bonitz}}]{dornheim_physrep_18}%
  \BibitemOpen
  \bibfield  {author} {\bibinfo {author} {\bibfnamefont {Tobias}\ \bibnamefont
  {Dornheim}}, \bibinfo {author} {\bibfnamefont {Simon}\ \bibnamefont {Groth}},
  \ and\ \bibinfo {author} {\bibfnamefont {Michael}\ \bibnamefont {Bonitz}},\
  }\bibfield  {title} {\enquote {\bibinfo {title} {The uniform electron gas at
  warm dense matter conditions},}\ }\href {\doibase
  10.1016/j.physrep.2018.04.001} {\bibfield  {journal} {\bibinfo  {journal}
  {Phys. Rep.}\ }\textbf {\bibinfo {volume} {744}},\ \bibinfo {pages} {1 -- 86}
  (\bibinfo {year} {2018}{\natexlab{a}})}\BibitemShut {NoStop}%
\bibitem [{\citenamefont {Filinov}\ \emph {et~al.}(2021)\citenamefont
  {Filinov}, \citenamefont {Levashov},\ and\ \citenamefont
  {Bonitz}}]{filinov-etal.2021ctpp}%
  \BibitemOpen
  \bibfield  {author} {\bibinfo {author} {\bibfnamefont {Alexey}\ \bibnamefont
  {Filinov}}, \bibinfo {author} {\bibfnamefont {Pavel~R.}\ \bibnamefont
  {Levashov}}, \ and\ \bibinfo {author} {\bibfnamefont {Michael}\ \bibnamefont
  {Bonitz}},\ }\bibfield  {title} {\enquote {\bibinfo {title} {Thermodynamics
  of the uniform electron gas: Fermionic path integral monte carlo simulations
  in the restricted grand canonical ensemble},}\ }\href {\doibase
  https://doi.org/10.1002/ctpp.202100112} {\bibfield  {journal} {\bibinfo
  {journal} {Contributions to Plasma Physics}\ }\textbf {\bibinfo {volume}
  {61}},\ \bibinfo {pages} {e202100112} (\bibinfo {year} {2021})}\BibitemShut
  {NoStop}%
\bibitem [{\citenamefont {Mishchenko}\ \emph {et~al.}(2000)\citenamefont
  {Mishchenko}, \citenamefont {Prokof'ev}, \citenamefont {Sakamoto},\ and\
  \citenamefont {Svistunov}}]{mishchenko.prb2000}%
  \BibitemOpen
  \bibfield  {author} {\bibinfo {author} {\bibfnamefont {A.~S.}\ \bibnamefont
  {Mishchenko}}, \bibinfo {author} {\bibfnamefont {N.~V.}\ \bibnamefont
  {Prokof'ev}}, \bibinfo {author} {\bibfnamefont {A.}~\bibnamefont {Sakamoto}},
  \ and\ \bibinfo {author} {\bibfnamefont {B.~V.}\ \bibnamefont {Svistunov}},\
  }\bibfield  {title} {\enquote {\bibinfo {title} {Diagrammatic quantum monte
  carlo study of the fr\"ohlich polaron},}\ }\href {\doibase
  10.1103/PhysRevB.62.6317} {\bibfield  {journal} {\bibinfo  {journal} {Phys.
  Rev. B}\ }\textbf {\bibinfo {volume} {62}},\ \bibinfo {pages} {6317--6336}
  (\bibinfo {year} {2000})}\BibitemShut {NoStop}%
\bibitem [{\citenamefont {Prokof’ev}\ and\ \citenamefont
  {Svistunov}(2013)}]{SCC13}%
  \BibitemOpen
  \bibfield  {author} {\bibinfo {author} {\bibfnamefont {N.~V.}\ \bibnamefont
  {Prokof’ev}}\ and\ \bibinfo {author} {\bibfnamefont {B.~V.}\ \bibnamefont
  {Svistunov}},\ }\bibfield  {title} {\enquote {\bibinfo {title} {Spectral
  analysis by the method of consistent constraints},}\ }\href {\doibase
  10.1134/S002136401311009X} {\bibfield  {journal} {\bibinfo  {journal} {JETP
  Letters}\ }\textbf {\bibinfo {volume} {97}},\ \bibinfo {pages} {649--653}
  (\bibinfo {year} {2013})}\BibitemShut {NoStop}%
\bibitem [{\citenamefont {Vitali}\ \emph {et~al.}(2010)\citenamefont {Vitali},
  \citenamefont {Rossi}, \citenamefont {Reatto},\ and\ \citenamefont
  {Galli}}]{vitali.2010prb}%
  \BibitemOpen
  \bibfield  {author} {\bibinfo {author} {\bibfnamefont {E.}~\bibnamefont
  {Vitali}}, \bibinfo {author} {\bibfnamefont {M.}~\bibnamefont {Rossi}},
  \bibinfo {author} {\bibfnamefont {L.}~\bibnamefont {Reatto}}, \ and\ \bibinfo
  {author} {\bibfnamefont {D.~E.}\ \bibnamefont {Galli}},\ }\bibfield  {title}
  {\enquote {\bibinfo {title} {Ab initio low-energy dynamics of superfluid and
  solid $^{4}\text{H}\text{e}$},}\ }\href {\doibase 10.1103/PhysRevB.82.174510}
  {\bibfield  {journal} {\bibinfo  {journal} {Phys. Rev. B}\ }\textbf {\bibinfo
  {volume} {82}},\ \bibinfo {pages} {174510} (\bibinfo {year}
  {2010})}\BibitemShut {NoStop}%
\bibitem [{\citenamefont {Dornheim}\ \emph
  {et~al.}(2018{\natexlab{b}})\citenamefont {Dornheim}, \citenamefont {Groth},
  \citenamefont {Vorberger},\ and\ \citenamefont {Bonitz}}]{dornheim.2018prl}%
  \BibitemOpen
  \bibfield  {author} {\bibinfo {author} {\bibfnamefont {T.}~\bibnamefont
  {Dornheim}}, \bibinfo {author} {\bibfnamefont {S.}~\bibnamefont {Groth}},
  \bibinfo {author} {\bibfnamefont {J.}~\bibnamefont {Vorberger}}, \ and\
  \bibinfo {author} {\bibfnamefont {M.}~\bibnamefont {Bonitz}},\ }\bibfield
  {title} {\enquote {\bibinfo {title} {Ab initio path integral monte carlo
  results for the dynamic structure factor of correlated electrons: From the
  electron liquid to warm dense matter},}\ }\href {\doibase
  10.1103/PhysRevLett.121.255001} {\bibfield  {journal} {\bibinfo  {journal}
  {Phys. Rev. Lett.}\ }\textbf {\bibinfo {volume} {121}},\ \bibinfo {pages}
  {255001} (\bibinfo {year} {2018}{\natexlab{b}})}\BibitemShut {NoStop}%
\bibitem [{\citenamefont {Silver}\ \emph {et~al.}(1990)\citenamefont {Silver},
  \citenamefont {Sivia},\ and\ \citenamefont {Gubernatis}}]{silver.1990prb}%
  \BibitemOpen
  \bibfield  {author} {\bibinfo {author} {\bibfnamefont {R.~N.}\ \bibnamefont
  {Silver}}, \bibinfo {author} {\bibfnamefont {D.~S.}\ \bibnamefont {Sivia}}, \
  and\ \bibinfo {author} {\bibfnamefont {J.~E.}\ \bibnamefont {Gubernatis}},\
  }\bibfield  {title} {\enquote {\bibinfo {title} {Maximum-entropy method for
  analytic continuation of quantum monte carlo data},}\ }\href {\doibase
  10.1103/PhysRevB.41.2380} {\bibfield  {journal} {\bibinfo  {journal} {Phys.
  Rev. B}\ }\textbf {\bibinfo {volume} {41}},\ \bibinfo {pages} {2380--2389}
  (\bibinfo {year} {1990})}\BibitemShut {NoStop}%
\bibitem [{\citenamefont {Arkhipov}\ \emph {et~al.}(2017)\citenamefont
  {Arkhipov}, \citenamefont {Askaruly}, \citenamefont {Davletov}, \citenamefont
  {Dubovtsev}, \citenamefont {Donk\'o}, \citenamefont {Hartmann}, \citenamefont
  {Korolov}, \citenamefont {Conde},\ and\ \citenamefont
  {Tkachenko}}]{arkhipov-etal.2017prl}%
  \BibitemOpen
  \bibfield  {author} {\bibinfo {author} {\bibfnamefont {Yu.~V.}\ \bibnamefont
  {Arkhipov}}, \bibinfo {author} {\bibfnamefont {A.}~\bibnamefont {Askaruly}},
  \bibinfo {author} {\bibfnamefont {A.~E.}\ \bibnamefont {Davletov}}, \bibinfo
  {author} {\bibfnamefont {D.~Yu.}\ \bibnamefont {Dubovtsev}}, \bibinfo
  {author} {\bibfnamefont {Z.}~\bibnamefont {Donk\'o}}, \bibinfo {author}
  {\bibfnamefont {P.}~\bibnamefont {Hartmann}}, \bibinfo {author}
  {\bibfnamefont {I.}~\bibnamefont {Korolov}}, \bibinfo {author} {\bibfnamefont
  {L.}~\bibnamefont {Conde}}, \ and\ \bibinfo {author} {\bibfnamefont {I.~M.}\
  \bibnamefont {Tkachenko}},\ }\bibfield  {title} {\enquote {\bibinfo {title}
  {Direct determination of dynamic properties of coulomb and yukawa classical
  one-component plasmas},}\ }\href {\doibase 10.1103/PhysRevLett.119.045001}
  {\bibfield  {journal} {\bibinfo  {journal} {Phys. Rev. Lett.}\ }\textbf
  {\bibinfo {volume} {119}},\ \bibinfo {pages} {045001} (\bibinfo {year}
  {2017})}\BibitemShut {NoStop}%
\bibitem [{\citenamefont {Arkhipov}\ \emph {et~al.}(2020)\citenamefont
  {Arkhipov}, \citenamefont {Ashikbayeva}, \citenamefont {Askaruly},
  \citenamefont {Davletov}, \citenamefont {Dubovtsev}, \citenamefont
  {Santybayev}, \citenamefont {Syzganbayeva}, \citenamefont {Conde},\ and\
  \citenamefont {Tkachenko}}]{arkhipov-etal.2020pre}%
  \BibitemOpen
  \bibfield  {author} {\bibinfo {author} {\bibfnamefont {Yu.~V.}\ \bibnamefont
  {Arkhipov}}, \bibinfo {author} {\bibfnamefont {A.}~\bibnamefont
  {Ashikbayeva}}, \bibinfo {author} {\bibfnamefont {A.}~\bibnamefont
  {Askaruly}}, \bibinfo {author} {\bibfnamefont {A.~E.}\ \bibnamefont
  {Davletov}}, \bibinfo {author} {\bibfnamefont {D.~Yu.}\ \bibnamefont
  {Dubovtsev}}, \bibinfo {author} {\bibfnamefont {Kh.~S.}\ \bibnamefont
  {Santybayev}}, \bibinfo {author} {\bibfnamefont {S.~A.}\ \bibnamefont
  {Syzganbayeva}}, \bibinfo {author} {\bibfnamefont {L.}~\bibnamefont {Conde}},
  \ and\ \bibinfo {author} {\bibfnamefont {I.~M.}\ \bibnamefont {Tkachenko}},\
  }\bibfield  {title} {\enquote {\bibinfo {title} {Dynamic characteristics of
  three-dimensional strongly coupled plasmas},}\ }\href {\doibase
  10.1103/PhysRevE.102.053215} {\bibfield  {journal} {\bibinfo  {journal}
  {Phys. Rev. E}\ }\textbf {\bibinfo {volume} {102}},\ \bibinfo {pages}
  {053215} (\bibinfo {year} {2020})}\BibitemShut {NoStop}%
\bibitem [{\citenamefont {Giuliani}\ and\ \citenamefont
  {Vignale}(2008)}]{GV-book}%
  \BibitemOpen
  \bibfield  {author} {\bibinfo {author} {\bibfnamefont {G.}~\bibnamefont
  {Giuliani}}\ and\ \bibinfo {author} {\bibfnamefont {G.}~\bibnamefont
  {Vignale}},\ }\href@noop {} {\emph {\bibinfo {title} {{Quantum Theory of the
  Electron Liquid}}}}\ (\bibinfo  {publisher} {Cambridge University Press, New
  York},\ \bibinfo {year} {2008})\BibitemShut {NoStop}%
\bibitem [{\citenamefont {Glyde}(1994)}]{Glyde-book}%
  \BibitemOpen
  \bibfield  {author} {\bibinfo {author} {\bibfnamefont {H.~R.}\ \bibnamefont
  {Glyde}},\ }\href@noop {} {\emph {\bibinfo {title} {{Excitations in Liquid
  and Solid Helium}}}}\ (\bibinfo  {publisher} {Clarendon, New York},\ \bibinfo
  {year} {1994})\BibitemShut {NoStop}%
\bibitem [{\citenamefont {Dobbs}(2000)}]{Dobbs-book}%
  \BibitemOpen
  \bibfield  {author} {\bibinfo {author} {\bibfnamefont {E.~R.}\ \bibnamefont
  {Dobbs}},\ }\href@noop {} {\emph {\bibinfo {title} {{Helium Three}}}}\
  (\bibinfo  {publisher} {(Oxford University, Oxford},\ \bibinfo {year}
  {2000})\BibitemShut {NoStop}%
\bibitem [{\citenamefont {Ara}\ \emph {et~al.}(2022)\citenamefont {Ara},
  \citenamefont {Filinov},\ and\ \citenamefont {Tkachenko}}]{Ara_2022}%
  \BibitemOpen
  \bibfield  {author} {\bibinfo {author} {\bibfnamefont {J.}~\bibnamefont
  {Ara}}, \bibinfo {author} {\bibfnamefont {A.V.}\ \bibnamefont {Filinov}}, \
  and\ \bibinfo {author} {\bibfnamefont {I.M.}\ \bibnamefont {Tkachenko}},\
  }\bibfield  {title} {\enquote {\bibinfo {title} {Classical and quantum warm
  dense electron gas dynamic characteristics: analytic predictions},}\ }\href
  {\doibase 10.1088/1742-6596/2270/1/012041} {\bibfield  {journal} {\bibinfo
  {journal} {Journal of Physics: Conference Series}\ }\textbf {\bibinfo
  {volume} {2270}},\ \bibinfo {pages} {012041} (\bibinfo {year}
  {2022})}\BibitemShut {NoStop}%
\bibitem [{\citenamefont {Landau}(1957{\natexlab{a}})}]{Landau1}%
  \BibitemOpen
  \bibfield  {author} {\bibinfo {author} {\bibfnamefont {L.D.}\ \bibnamefont
  {Landau}},\ }\bibfield  {title} {\enquote {\bibinfo {title} {The theory of a
  fermi liquid},}\ }\href@noop {} {\bibfield  {journal} {\bibinfo  {journal}
  {Sov. Phys. JETP,}\ }\textbf {\bibinfo {volume} {3}},\ \bibinfo {pages}
  {921--925} (\bibinfo {year} {1957}{\natexlab{a}})}\BibitemShut {NoStop}%
\bibitem [{\citenamefont {Landau}(1957{\natexlab{b}})}]{Landau2}%
  \BibitemOpen
  \bibfield  {author} {\bibinfo {author} {\bibfnamefont {L.D.}\ \bibnamefont
  {Landau}},\ }\bibfield  {title} {\enquote {\bibinfo {title} {Oscillations in
  a fermi liquid},}\ }\href@noop {} {\bibfield  {journal} {\bibinfo  {journal}
  {Sov. Phys. JETP,}\ }\textbf {\bibinfo {volume} {5}},\ \bibinfo {pages}
  {101--108} (\bibinfo {year} {1957}{\natexlab{b}})}\BibitemShut {NoStop}%
\bibitem [{\citenamefont {Abrikosov}\ and\ \citenamefont
  {Khalatnikov}(1958)}]{Abrikosov:1958}%
  \BibitemOpen
  \bibfield  {author} {\bibinfo {author} {\bibfnamefont {A.~A.}\ \bibnamefont
  {Abrikosov}}\ and\ \bibinfo {author} {\bibfnamefont {I.~M.}\ \bibnamefont
  {Khalatnikov}},\ }\bibfield  {title} {\enquote {\bibinfo {title} {Theory of
  the fermi fluid (the properties of liquid he3 at low temperatures)},}\ }\href
  {\doibase 10.1070/PU1958v001n01ABEH003086} {\bibfield  {journal} {\bibinfo
  {journal} {Phys. Usp.}\ }\textbf {\bibinfo {volume} {1}},\ \bibinfo {pages}
  {68--90} (\bibinfo {year} {1958})}\BibitemShut {NoStop}%
\bibitem [{\citenamefont {Pines}\ and\ \citenamefont
  {P.~Nozieres}(2018)}]{PN-book}%
  \BibitemOpen
  \bibfield  {author} {\bibinfo {author} {\bibfnamefont {D.}~\bibnamefont
  {Pines}}\ and\ \bibinfo {author} {\bibfnamefont {P.}~\bibnamefont
  {P.~Nozieres}},\ }\href@noop {} {\emph {\bibinfo {title} {{Theory of Quantum
  Liquids}}}}\ (\bibinfo  {publisher} {CRC Press, Boca Raton, Florida, USA},\
  \bibinfo {year} {2018})\BibitemShut {NoStop}%
\bibitem [{\citenamefont {Dornheim}\ \emph {et~al.}(2017)\citenamefont
  {Dornheim}, \citenamefont {Moldabekov}, \citenamefont {Vorberger},\ and\
  \citenamefont {Militzer}}]{dornheim-etal.2022sr}%
  \BibitemOpen
  \bibfield  {author} {\bibinfo {author} {\bibfnamefont {Tobias}\ \bibnamefont
  {Dornheim}}, \bibinfo {author} {\bibfnamefont {Zhandos~A.}\ \bibnamefont
  {Moldabekov}}, \bibinfo {author} {\bibfnamefont {Jan}\ \bibnamefont
  {Vorberger}}, \ and\ \bibinfo {author} {\bibfnamefont {Burkhard}\
  \bibnamefont {Militzer}},\ }\bibfield  {title} {\enquote {\bibinfo {title}
  {Path integral monte carlo approach to the structural properties and
  collective excitations of liquid $^{3}\mathrm{He}$ without fixed nodes},}\
  }\href {\doibase 10.1002/ctpp.201700096} {\bibfield  {journal} {\bibinfo
  {journal} {Scientific Reports}\ }\textbf {\bibinfo {volume} {12}},\ \bibinfo
  {pages} {708} (\bibinfo {year} {2017})}\BibitemShut {NoStop}%
\bibitem [{\citenamefont {Filinov}\ \emph {et~al.}(2022)\citenamefont
  {Filinov}, \citenamefont {Syrovatka},\ and\ \citenamefont
  {Levashov}}]{VFilinovHe3}%
  \BibitemOpen
  \bibfield  {author} {\bibinfo {author} {\bibfnamefont {V.~S.}\ \bibnamefont
  {Filinov}}, \bibinfo {author} {\bibfnamefont {R.~S.}\ \bibnamefont
  {Syrovatka}}, \ and\ \bibinfo {author} {\bibfnamefont {P.~R.}\ \bibnamefont
  {Levashov}},\ }\bibfield  {title} {\enquote {\bibinfo {title} {Solution of
  the ‘sign problem’ in the path integral monte carlo simulations of
  strongly correlated fermi systems: thermodynamic properties of helium-3},}\
  }\href {\doibase 10.1080/00268976.2022.2102549} {\bibfield  {journal}
  {\bibinfo  {journal} {Molecular Physics}\ }\textbf {\bibinfo {volume}
  {120}},\ \bibinfo {pages} {e2102549} (\bibinfo {year} {2022})}\BibitemShut
  {NoStop}%
\bibitem [{\citenamefont {Arkhipov}\ \emph
  {et~al.}(2010{\natexlab{a}})\citenamefont {Arkhipov}, \citenamefont
  {Askaruly}, \citenamefont {Ballester}, \citenamefont {Davletov},
  \citenamefont {Tkachenko},\ and\ \citenamefont
  {Zwicknagel}}]{arkhipov-etal.2010pre}%
  \BibitemOpen
  \bibfield  {author} {\bibinfo {author} {\bibfnamefont {Yu.~V.}\ \bibnamefont
  {Arkhipov}}, \bibinfo {author} {\bibfnamefont {A.}~\bibnamefont {Askaruly}},
  \bibinfo {author} {\bibfnamefont {D.}~\bibnamefont {Ballester}}, \bibinfo
  {author} {\bibfnamefont {A.~E.}\ \bibnamefont {Davletov}}, \bibinfo {author}
  {\bibfnamefont {I.~M.}\ \bibnamefont {Tkachenko}}, \ and\ \bibinfo {author}
  {\bibfnamefont {G.}~\bibnamefont {Zwicknagel}},\ }\bibfield  {title}
  {\enquote {\bibinfo {title} {Dynamic properties of one-component strongly
  coupled plasmas: The sum-rule approach},}\ }\href {\doibase
  10.1103/PhysRevE.81.026402} {\bibfield  {journal} {\bibinfo  {journal} {Phys.
  Rev. E}\ }\textbf {\bibinfo {volume} {81}},\ \bibinfo {pages} {026402}
  (\bibinfo {year} {2010}{\natexlab{a}})}\BibitemShut {NoStop}%
\bibitem [{\citenamefont {Ara}\ \emph {et~al.}(2021)\citenamefont {Ara},
  \citenamefont {Coloma},\ and\ \citenamefont {Tkachenko}}]{ara-etal.2021pop}%
  \BibitemOpen
  \bibfield  {author} {\bibinfo {author} {\bibfnamefont {J.}~\bibnamefont
  {Ara}}, \bibinfo {author} {\bibfnamefont {Ll.}\ \bibnamefont {Coloma}}, \
  and\ \bibinfo {author} {\bibfnamefont {I.~M.}\ \bibnamefont {Tkachenko}},\
  }\bibfield  {title} {\enquote {\bibinfo {title} {Static properties of a warm
  dense uniform electron gas},}\ }\href {\doibase 10.1063/5.0062259} {\bibfield
   {journal} {\bibinfo  {journal} {Physics of Plasmas}\ }\textbf {\bibinfo
  {volume} {28}},\ \bibinfo {pages} {112704} (\bibinfo {year}
  {2021})}\BibitemShut {NoStop}%
\bibitem [{\citenamefont {Aziz}\ \emph {et~al.}(1979)\citenamefont {Aziz},
  \citenamefont {Nain}, \citenamefont {Carley}, \citenamefont {Taylor},\ and\
  \citenamefont {McConville}}]{JChemPhys79}%
  \BibitemOpen
  \bibfield  {author} {\bibinfo {author} {\bibfnamefont {R.~A.}\ \bibnamefont
  {Aziz}}, \bibinfo {author} {\bibfnamefont {V.~P.~S.}\ \bibnamefont {Nain}},
  \bibinfo {author} {\bibfnamefont {J.~S.}\ \bibnamefont {Carley}}, \bibinfo
  {author} {\bibfnamefont {W.~L.}\ \bibnamefont {Taylor}}, \ and\ \bibinfo
  {author} {\bibfnamefont {G.~T.}\ \bibnamefont {McConville}},\ }\bibfield
  {title} {\enquote {\bibinfo {title} {An accurate intermolecular potential for
  helium},}\ }\href {\doibase 10.1063/1.438007} {\bibfield  {journal} {\bibinfo
   {journal} {J. Chem. Phys.}\ }\textbf {\bibinfo {volume} {70}},\ \bibinfo
  {pages} {4330—4341} (\bibinfo {year} {1979})},\ \Eprint
  {http://arxiv.org/abs/https://doi.org/10.1063/1.438007}
  {https://doi.org/10.1063/1.438007} \BibitemShut {NoStop}%
\bibitem [{\citenamefont {Shohat}\ and\ \citenamefont
  {Tamarkin}(1943)}]{shohat-book}%
  \BibitemOpen
  \bibfield  {author} {\bibinfo {author} {\bibfnamefont {J.A.}\ \bibnamefont
  {Shohat}}\ and\ \bibinfo {author} {\bibfnamefont {J.D.}\ \bibnamefont
  {Tamarkin}},\ }\href@noop {} {\emph {\bibinfo {title} {{The Problem of
  Moments}}}},\ Amer. Math. Soc.\ (\bibinfo  {publisher} {Providence, R.I.},\
  \bibinfo {year} {1943})\BibitemShut {NoStop}%
\bibitem [{\citenamefont {Akhiezer}(1965)}]{akhiezer-book}%
  \BibitemOpen
  \bibfield  {author} {\bibinfo {author} {\bibfnamefont {N.I.}\ \bibnamefont
  {Akhiezer}},\ }\href@noop {} {\emph {\bibinfo {title} {{The Classical Moment
  Problem}}}}\ (\bibinfo  {publisher} {Hafner Publishing Company, New York},\
  \bibinfo {year} {1965})\BibitemShut {NoStop}%
\bibitem [{\citenamefont {Krein}\ and\ \citenamefont
  {Nudel'man}(1977)}]{krein-book}%
  \BibitemOpen
  \bibfield  {author} {\bibinfo {author} {\bibfnamefont {M.G.}\ \bibnamefont
  {Krein}}\ and\ \bibinfo {author} {\bibfnamefont {A.A.}\ \bibnamefont
  {Nudel'man}},\ }\href@noop {} {\emph {\bibinfo {title} {{The Markov moment
  problem and extremal problems}}}},\ Trans. of Math. Monographs 50, Amer.
  Math. Soc.\ (\bibinfo  {publisher} {Providence, R.I.},\ \bibinfo {year}
  {1977})\BibitemShut {NoStop}%
\bibitem [{\citenamefont {Varentsov}\ \emph {et~al.}(2005)\citenamefont
  {Varentsov}, \citenamefont {Tkachenko},\ and\ \citenamefont
  {Hoffmann}}]{PhysRevE.71.066501}%
  \BibitemOpen
  \bibfield  {author} {\bibinfo {author} {\bibfnamefont {Dmitry}\ \bibnamefont
  {Varentsov}}, \bibinfo {author} {\bibfnamefont {Igor~M.}\ \bibnamefont
  {Tkachenko}}, \ and\ \bibinfo {author} {\bibfnamefont {Dieter H.~H.}\
  \bibnamefont {Hoffmann}},\ }\bibfield  {title} {\enquote {\bibinfo {title}
  {Statistical approach to beam shaping},}\ }\href {\doibase
  10.1103/PhysRevE.71.066501} {\bibfield  {journal} {\bibinfo  {journal} {Phys.
  Rev. E}\ }\textbf {\bibinfo {volume} {71}},\ \bibinfo {pages} {066501}
  (\bibinfo {year} {2005})}\BibitemShut {NoStop}%
\bibitem [{\citenamefont {Tkachenko}\ \emph {et~al.}(2012)\citenamefont
  {Tkachenko}, \citenamefont {Arkhipov},\ and\ \citenamefont
  {Askaruly}}]{tkachenko-book}%
  \BibitemOpen
  \bibfield  {author} {\bibinfo {author} {\bibfnamefont {I.M.}\ \bibnamefont
  {Tkachenko}}, \bibinfo {author} {\bibfnamefont {Y.V.}\ \bibnamefont
  {Arkhipov}}, \ and\ \bibinfo {author} {\bibfnamefont {A.}~\bibnamefont
  {Askaruly}},\ }\href@noop {} {\emph {\bibinfo {title} {{The Method of Moments
  and its Applications in Plasma Physics}}}}\ (\bibinfo  {publisher} {Lambert,
  Saarbrücken},\ \bibinfo {year} {2012})\BibitemShut {NoStop}%
\bibitem [{\citenamefont {Adamyan}\ and\ \citenamefont
  {Tkachenko}(2014)}]{pamm.2014}%
  \BibitemOpen
  \bibfield  {author} {\bibinfo {author} {\bibfnamefont {Vadym}\ \bibnamefont
  {Adamyan}}\ and\ \bibinfo {author} {\bibfnamefont {Igor}\ \bibnamefont
  {Tkachenko}},\ }\bibfield  {title} {\enquote {\bibinfo {title} {Local moment
  problem},}\ }\href {\doibase https://doi.org/10.1002/pamm.201410471}
  {\bibfield  {journal} {\bibinfo  {journal} {PAMM}\ }\textbf {\bibinfo
  {volume} {14}},\ \bibinfo {pages} {981--982} (\bibinfo {year}
  {2014})}\BibitemShut {NoStop}%
\bibitem [{\citenamefont {Alcober}\ \emph {et~al.}(2008)\citenamefont
  {Alcober}, \citenamefont {Tkachenko},\ and\ \citenamefont
  {Urrea}}]{chapter2}%
  \BibitemOpen
  \bibfield  {author} {\bibinfo {author} {\bibfnamefont {J.}~\bibnamefont
  {Alcober}}, \bibinfo {author} {\bibfnamefont {I.M.}\ \bibnamefont
  {Tkachenko}}, \ and\ \bibinfo {author} {\bibfnamefont {M.}~\bibnamefont
  {Urrea}},\ }\href {http://www.imse08.unican.es/daily_program.htm} {\emph
  {\bibinfo {title} {{Construction of solutions of the Hamburger-Löwner mixed
  interpolation problem for the Nevanlinna-class functions}}}}\ (\bibinfo
  {publisher} {The 10th International Conference on Integral Methods in Science
  and Engineering (IMSE 2008), Santander, Spain, July, 2008, Book of Abstracts,
  p. 199},\ \bibinfo {year} {2008})\BibitemShut {NoStop}%
\bibitem [{\citenamefont {Nevanlinna}(1922)}]{nevanlinna-book}%
  \BibitemOpen
  \bibfield  {author} {\bibinfo {author} {\bibfnamefont {R.}~\bibnamefont
  {Nevanlinna}},\ }\href@noop {} {\emph {\bibinfo {title} {{Asymptotische
  Entwicklungen beschr\"{a}nkter Funktionen und das Stieltjessche
  Momentenproblem}}}},\ STK\ (\bibinfo  {publisher} {Helsinki},\ \bibinfo
  {year} {1922})\BibitemShut {NoStop}%
\bibitem [{\citenamefont {Urrea}\ \emph {et~al.}(2001)\citenamefont {Urrea},
  \citenamefont {Tkachenko},\ and\ \citenamefont {Córdoba}}]{UTFC}%
  \BibitemOpen
  \bibfield  {author} {\bibinfo {author} {\bibfnamefont {M.}~\bibnamefont
  {Urrea}}, \bibinfo {author} {\bibfnamefont {I.~M.}\ \bibnamefont
  {Tkachenko}}, \ and\ \bibinfo {author} {\bibfnamefont {P.~Fernández~De}\
  \bibnamefont {Córdoba}},\ }\bibfield  {title} {\enquote {\bibinfo {title}
  {The nevanlinna theorem of the classical theory of moments revisited},}\
  }\href {\doibase doi:10.1515/JAA.2001.209} {\bibfield  {journal} {\bibinfo
  {journal} {Journal of Applied Analysis}\ }\textbf {\bibinfo {volume} {7}},\
  \bibinfo {pages} {209--224} (\bibinfo {year} {2001})}\BibitemShut {NoStop}%
\bibitem [{\citenamefont {Krein}(1947{\natexlab{a}})}]{sb1}%
  \BibitemOpen
  \bibfield  {author} {\bibinfo {author} {\bibfnamefont {M.}~\bibnamefont
  {Krein}},\ }\bibfield  {title} {\enquote {\bibinfo {title} {The theory of
  self-adjoint extensions of somi-bcunded hermitian transformations and its
  applications. i},}\ }\href@noop {} {\bibfield  {journal} {\bibinfo  {journal}
  {Rec. Math. [Mat. Sbornik] N.S.}\ }\textbf {\bibinfo {volume} {20(62)}},\
  \bibinfo {pages} {431--495} (\bibinfo {year}
  {1947}{\natexlab{a}})}\BibitemShut {NoStop}%
\bibitem [{\citenamefont {Krein}(1947{\natexlab{b}})}]{sb2}%
  \BibitemOpen
  \bibfield  {author} {\bibinfo {author} {\bibfnamefont {M.}~\bibnamefont
  {Krein}},\ }\bibfield  {title} {\enquote {\bibinfo {title} {The theory of
  self-adjoint extensions of somi-bcunded hermitian transformations and its
  applications. ii},}\ }\href@noop {} {\bibfield  {journal} {\bibinfo
  {journal} {Rec. Math. [Mat. Sbornik] N.S.}\ }\textbf {\bibinfo {volume}
  {21(63)}},\ \bibinfo {pages} {365--404} (\bibinfo {year}
  {1947}{\natexlab{b}})}\BibitemShut {NoStop}%
\bibitem [{\citenamefont {Tkachenko}(2018)}]{PST18}%
  \BibitemOpen
  \bibfield  {author} {\bibinfo {author} {\bibfnamefont {I.M.}\ \bibnamefont
  {Tkachenko}},\ }\bibfield  {title} {\enquote {\bibinfo {title} {New
  developments in the application of the method of moments in plasma
  physics},}\ }\href@noop {} {\bibfield  {journal} {\bibinfo  {journal}
  {Physical Sciences and Technology}\ }\textbf {\bibinfo {volume} {5}},\
  \bibinfo {pages} {16--35} (\bibinfo {year} {2018})}\BibitemShut {NoStop}%
\bibitem [{\citenamefont {Ortner}\ and\ \citenamefont
  {Tkachenko}(1992)}]{PhysRevA.46.7882}%
  \BibitemOpen
  \bibfield  {author} {\bibinfo {author} {\bibfnamefont {J.}~\bibnamefont
  {Ortner}}\ and\ \bibinfo {author} {\bibfnamefont {I.~M.}\ \bibnamefont
  {Tkachenko}},\ }\bibfield  {title} {\enquote {\bibinfo {title} {Dielectric
  permeability of quasi-two-dimensional one-component plasmas},}\ }\href
  {\doibase 10.1103/PhysRevA.46.7882} {\bibfield  {journal} {\bibinfo
  {journal} {Phys. Rev. A}\ }\textbf {\bibinfo {volume} {46}},\ \bibinfo
  {pages} {7882--7888} (\bibinfo {year} {1992})}\BibitemShut {NoStop}%
\bibitem [{\citenamefont {Adamjan}\ \emph {et~al.}(1993)\citenamefont
  {Adamjan}, \citenamefont {Tkachenko}, \citenamefont {Mu\~noz
  Cobo~Gonz\'alez},\ and\ \citenamefont
  {Verd\'u~Mart\'{\i}n}}]{PhysRevE.48.2067}%
  \BibitemOpen
  \bibfield  {author} {\bibinfo {author} {\bibfnamefont {S.~V.}\ \bibnamefont
  {Adamjan}}, \bibinfo {author} {\bibfnamefont {I.~M.}\ \bibnamefont
  {Tkachenko}}, \bibinfo {author} {\bibfnamefont {J.~L.}\ \bibnamefont {Mu\~noz
  Cobo~Gonz\'alez}}, \ and\ \bibinfo {author} {\bibfnamefont {G.}~\bibnamefont
  {Verd\'u~Mart\'{\i}n}},\ }\bibfield  {title} {\enquote {\bibinfo {title}
  {Dynamic and static correlations in model coulomb systems},}\ }\href
  {\doibase 10.1103/PhysRevE.48.2067} {\bibfield  {journal} {\bibinfo
  {journal} {Phys. Rev. E}\ }\textbf {\bibinfo {volume} {48}},\ \bibinfo
  {pages} {2067--2072} (\bibinfo {year} {1993})}\BibitemShut {NoStop}%
\bibitem [{\citenamefont {Arkhipov}\ \emph {et~al.}(2007)\citenamefont
  {Arkhipov}, \citenamefont {Askaruly}, \citenamefont {Ballester},
  \citenamefont {Davletov}, \citenamefont {Meirkanova},\ and\ \citenamefont
  {Tkachenko}}]{PhysRevE.76.026403}%
  \BibitemOpen
  \bibfield  {author} {\bibinfo {author} {\bibfnamefont {Yu.~V.}\ \bibnamefont
  {Arkhipov}}, \bibinfo {author} {\bibfnamefont {A.}~\bibnamefont {Askaruly}},
  \bibinfo {author} {\bibfnamefont {D.}~\bibnamefont {Ballester}}, \bibinfo
  {author} {\bibfnamefont {A.~E.}\ \bibnamefont {Davletov}}, \bibinfo {author}
  {\bibfnamefont {G.~M.}\ \bibnamefont {Meirkanova}}, \ and\ \bibinfo {author}
  {\bibfnamefont {I.~M.}\ \bibnamefont {Tkachenko}},\ }\bibfield  {title}
  {\enquote {\bibinfo {title} {Collective and static properties of model
  two-component plasmas},}\ }\href {\doibase 10.1103/PhysRevE.76.026403}
  {\bibfield  {journal} {\bibinfo  {journal} {Phys. Rev. E}\ }\textbf {\bibinfo
  {volume} {76}},\ \bibinfo {pages} {026403} (\bibinfo {year}
  {2007})}\BibitemShut {NoStop}%
\bibitem [{\citenamefont {Arkhipov}\ \emph
  {et~al.}(2010{\natexlab{b}})\citenamefont {Arkhipov}, \citenamefont
  {Askaruly}, \citenamefont {Ballester}, \citenamefont {Davletov},
  \citenamefont {Tkachenko},\ and\ \citenamefont
  {Zwicknagel}}]{PhysRevE.81.026402}%
  \BibitemOpen
  \bibfield  {author} {\bibinfo {author} {\bibfnamefont {Yu.~V.}\ \bibnamefont
  {Arkhipov}}, \bibinfo {author} {\bibfnamefont {A.}~\bibnamefont {Askaruly}},
  \bibinfo {author} {\bibfnamefont {D.}~\bibnamefont {Ballester}}, \bibinfo
  {author} {\bibfnamefont {A.~E.}\ \bibnamefont {Davletov}}, \bibinfo {author}
  {\bibfnamefont {I.~M.}\ \bibnamefont {Tkachenko}}, \ and\ \bibinfo {author}
  {\bibfnamefont {G.}~\bibnamefont {Zwicknagel}},\ }\bibfield  {title}
  {\enquote {\bibinfo {title} {Dynamic properties of one-component strongly
  coupled plasmas: The sum-rule approach},}\ }\href {\doibase
  10.1103/PhysRevE.81.026402} {\bibfield  {journal} {\bibinfo  {journal} {Phys.
  Rev. E}\ }\textbf {\bibinfo {volume} {81}},\ \bibinfo {pages} {026402}
  (\bibinfo {year} {2010}{\natexlab{b}})}\BibitemShut {NoStop}%
\bibitem [{\citenamefont {Arkhipov}\ \emph {et~al.}(2014)\citenamefont
  {Arkhipov}, \citenamefont {Ashikbayeva}, \citenamefont {Askaruly},
  \citenamefont {Davletov},\ and\ \citenamefont
  {Tkachenko}}]{PhysRevE.90.053102}%
  \BibitemOpen
  \bibfield  {author} {\bibinfo {author} {\bibfnamefont {Yu.~V.}\ \bibnamefont
  {Arkhipov}}, \bibinfo {author} {\bibfnamefont {A.~B.}\ \bibnamefont
  {Ashikbayeva}}, \bibinfo {author} {\bibfnamefont {A.}~\bibnamefont
  {Askaruly}}, \bibinfo {author} {\bibfnamefont {A.~E.}\ \bibnamefont
  {Davletov}}, \ and\ \bibinfo {author} {\bibfnamefont {I.~M.}\ \bibnamefont
  {Tkachenko}},\ }\bibfield  {title} {\enquote {\bibinfo {title} {Dielectric
  function of dense plasmas, their stopping power, and sum rules},}\ }\href
  {\doibase 10.1103/PhysRevE.90.053102} {\bibfield  {journal} {\bibinfo
  {journal} {Phys. Rev. E}\ }\textbf {\bibinfo {volume} {90}},\ \bibinfo
  {pages} {053102} (\bibinfo {year} {2014})}\BibitemShut {NoStop}%
\bibitem [{\citenamefont {Filinov}\ \emph {et~al.}()\citenamefont {Filinov},
  \citenamefont {Ara},\ and\ \citenamefont {Tkachenko}}]{UEGarXiv}%
  \BibitemOpen
  \bibfield  {author} {\bibinfo {author} {\bibfnamefont {A.}~\bibnamefont
  {Filinov}}, \bibinfo {author} {\bibfnamefont {J.}~\bibnamefont {Ara}}, \ and\
  \bibinfo {author} {\bibfnamefont {I.M.}\ \bibnamefont {Tkachenko}},\
  }\bibfield  {title} {\enquote {\bibinfo {title} {Analysis of dynamical
  effects in the uniform electron liquids with the self-consistent method of
  moments complemented by the shannon information entropy and the path-integral
  monte-carlo simulations},}\ }\href {http://arxiv.org/abs/2302.07082}
  {\bibinfo  {journal} {http://arxiv.org/abs/2302.07082}\ }\BibitemShut
  {NoStop}%
\bibitem [{\citenamefont {Shannon}(1948)}]{shannon.1948}%
  \BibitemOpen
\bibfield  {journal} {  }\bibfield  {author} {\bibinfo {author} {\bibfnamefont
  {C.~E.}\ \bibnamefont {Shannon}},\ }\bibfield  {title} {\enquote {\bibinfo
  {title} {A mathematical theory of communication},}\ }\href {\doibase
  https://doi.org/10.1002/j.1538-7305.1948.tb01338.x} {\bibfield  {journal}
  {\bibinfo  {journal} {Bell System Technical Journal}\ }\textbf {\bibinfo
  {volume} {27}},\ \bibinfo {pages} {379--423} (\bibinfo {year}
  {1948})}\BibitemShut {NoStop}%
\bibitem [{\citenamefont {Khinchin}(1953)}]{khinchin.53umn}%
  \BibitemOpen
  \bibfield  {author} {\bibinfo {author} {\bibfnamefont {A.~Ya.}\ \bibnamefont
  {Khinchin}},\ }\bibfield  {title} {\enquote {\bibinfo {title} {The entropy
  concept in probability theory},}\ }\href@noop {} {\bibfield  {journal}
  {\bibinfo  {journal} {Uspekhi Mat. Nauk [in Russian]}\ }\textbf {\bibinfo
  {volume} {{\bf 8}}},\ \bibinfo {pages} {3} (\bibinfo {year}
  {1953})}\BibitemShut {NoStop}%
\bibitem [{\citenamefont {Zubarev}(1974)}]{zubarev-book}%
  \BibitemOpen
  \bibfield  {author} {\bibinfo {author} {\bibfnamefont {D.N.}\ \bibnamefont
  {Zubarev}},\ }\href@noop {} {\emph {\bibinfo {title} {{Nonequilibrium
  Statistical Mechanics}}}}\ (\bibinfo  {publisher} {Consultants Bureau,
  London},\ \bibinfo {year} {1974})\BibitemShut {NoStop}%
\bibitem [{\citenamefont {Jaynes}(1957)}]{jaynes.1957pr}%
  \BibitemOpen
  \bibfield  {author} {\bibinfo {author} {\bibfnamefont {E.~T.}\ \bibnamefont
  {Jaynes}},\ }\bibfield  {title} {\enquote {\bibinfo {title} {Information
  theory and statistical mechanics},}\ }\href {\doibase
  10.1103/PhysRev.106.620} {\bibfield  {journal} {\bibinfo  {journal} {Phys.
  Rev.}\ }\textbf {\bibinfo {volume} {106}},\ \bibinfo {pages} {620--630}
  (\bibinfo {year} {1957})}\BibitemShut {NoStop}%
\bibitem [{\citenamefont {Achter}\ and\ \citenamefont
  {Meyer}(1969)}]{Achter1969}%
  \BibitemOpen
  \bibfield  {author} {\bibinfo {author} {\bibfnamefont {Eugene~K.}\
  \bibnamefont {Achter}}\ and\ \bibinfo {author} {\bibfnamefont {Lothar}\
  \bibnamefont {Meyer}},\ }\bibfield  {title} {\enquote {\bibinfo {title}
  {X-ray scattering from liquid helium},}\ }\href {\doibase
  10.1103/PhysRev.188.291} {\bibfield  {journal} {\bibinfo  {journal} {Phys.
  Rev.}\ }\textbf {\bibinfo {volume} {188}},\ \bibinfo {pages} {291--300}
  (\bibinfo {year} {1969})}\BibitemShut {NoStop}%
\bibitem [{\citenamefont {Hallock}(1972)}]{Hallock1972}%
  \BibitemOpen
  \bibfield  {author} {\bibinfo {author} {\bibfnamefont {Robert~B.}\
  \bibnamefont {Hallock}},\ }\bibfield  {title} {\enquote {\bibinfo {title}
  {Liquid structure factor measurements on $^{3}\mathrm{He}$},}\ }\href
  {\doibase https://doi.org/10.1007/BF00655490} {\bibfield  {journal} {\bibinfo
   {journal} {Journal of Low Temperature Physics}\ }\textbf {\bibinfo {volume}
  {9}},\ \bibinfo {pages} {109--121} (\bibinfo {year} {1972})}\BibitemShut
  {NoStop}%
\bibitem [{\citenamefont {Pitaevskii}(1967)}]{Pitaevskii:1967}%
  \BibitemOpen
  \bibfield  {author} {\bibinfo {author} {\bibfnamefont {L.~P.}\ \bibnamefont
  {Pitaevskii}},\ }\bibfield  {title} {\enquote {\bibinfo {title} {Zero sound
  in liquid $^3\text{He}$},}\ }\href {\doibase 10.1070/PU1967v010n01ABEH003201}
  {\bibfield  {journal} {\bibinfo  {journal} {Phys. Usp.}\ }\textbf {\bibinfo
  {volume} {10}},\ \bibinfo {pages} {100--101} (\bibinfo {year}
  {1967})}\BibitemShut {NoStop}%
\bibitem [{\citenamefont {Sköld}\ \emph {et~al.}(1980)\citenamefont {Sköld},
  \citenamefont {Pelizzari}, \citenamefont {Mason}, \citenamefont {Mitchell},\
  and\ \citenamefont {White}}]{Skold1980}%
  \BibitemOpen
  \bibfield  {author} {\bibinfo {author} {\bibfnamefont {K.}~\bibnamefont
  {Sköld}}, \bibinfo {author} {\bibfnamefont {C.A.}\ \bibnamefont
  {Pelizzari}}, \bibinfo {author} {\bibfnamefont {R.}~\bibnamefont {Mason}},
  \bibinfo {author} {\bibfnamefont {E.W.J.}\ \bibnamefont {Mitchell}}, \ and\
  \bibinfo {author} {\bibfnamefont {J.W.}\ \bibnamefont {White}},\ }\bibfield
  {title} {\enquote {\bibinfo {title} {Elementary excitations in liquid
  $^{3}\mathrm{He}$},}\ }\href@noop {} {\bibfield  {journal} {\bibinfo
  {journal} {Philos. Trans. R. Soc. Lond.}\ }\textbf {\bibinfo {volume}
  {290}},\ \bibinfo {pages} {605–616} (\bibinfo {year} {1980})}\BibitemShut
  {NoStop}%
\bibitem [{\citenamefont {Kalman}\ \emph {et~al.}(2010)\citenamefont {Kalman},
  \citenamefont {Hartmann}, \citenamefont {Golden}, \citenamefont {Filinov},\
  and\ \citenamefont {Donkó}}]{Kalman_2010}%
  \BibitemOpen
  \bibfield  {author} {\bibinfo {author} {\bibfnamefont {G.~J.}\ \bibnamefont
  {Kalman}}, \bibinfo {author} {\bibfnamefont {P.}~\bibnamefont {Hartmann}},
  \bibinfo {author} {\bibfnamefont {K.~I.}\ \bibnamefont {Golden}}, \bibinfo
  {author} {\bibfnamefont {A.}~\bibnamefont {Filinov}}, \ and\ \bibinfo
  {author} {\bibfnamefont {Z.}~\bibnamefont {Donkó}},\ }\bibfield  {title}
  {\enquote {\bibinfo {title} {Correlational origin of the roton minimum},}\
  }\href {\doibase 10.1209/0295-5075/90/55002} {\bibfield  {journal} {\bibinfo
  {journal} {Europhysics Letters}\ }\textbf {\bibinfo {volume} {90}},\ \bibinfo
  {pages} {55002} (\bibinfo {year} {2010})}\BibitemShut {NoStop}%
\bibitem [{\citenamefont {Kalman}\ \emph {et~al.}(2012)\citenamefont {Kalman},
  \citenamefont {Kyrkos}, \citenamefont {Golden}, \citenamefont {Hartmann},\
  and\ \citenamefont {Donko}}]{Kalman_2012}%
  \BibitemOpen
  \bibfield  {author} {\bibinfo {author} {\bibfnamefont {G.J.}\ \bibnamefont
  {Kalman}}, \bibinfo {author} {\bibfnamefont {S.}~\bibnamefont {Kyrkos}},
  \bibinfo {author} {\bibfnamefont {K.I.}\ \bibnamefont {Golden}}, \bibinfo
  {author} {\bibfnamefont {P.}~\bibnamefont {Hartmann}}, \ and\ \bibinfo
  {author} {\bibfnamefont {Z.}~\bibnamefont {Donko}},\ }\bibfield  {title}
  {\enquote {\bibinfo {title} {The roton minimum: Is it a general feature of
  strongly correlated liquids?}}\ }\href {\doibase
  https://doi.org/10.1002/ctpp.201100095} {\bibfield  {journal} {\bibinfo
  {journal} {Contributions to Plasma Physics}\ }\textbf {\bibinfo {volume}
  {52}},\ \bibinfo {pages} {219--223} (\bibinfo {year} {2012})}\BibitemShut
  {NoStop}%
\bibitem [{\citenamefont {Glyde}\ \emph {et~al.}(2000)\citenamefont {Glyde},
  \citenamefont {F\aa{}k}, \citenamefont {Dijk}, \citenamefont {Godfrin},
  \citenamefont {Guckelsberger},\ and\ \citenamefont
  {Scherm}}]{PhysRevB.61.1421}%
  \BibitemOpen
  \bibfield  {author} {\bibinfo {author} {\bibfnamefont {H.~R.}\ \bibnamefont
  {Glyde}}, \bibinfo {author} {\bibfnamefont {B.}~\bibnamefont {F\aa{}k}},
  \bibinfo {author} {\bibfnamefont {N.~H.~van}\ \bibnamefont {Dijk}}, \bibinfo
  {author} {\bibfnamefont {H.}~\bibnamefont {Godfrin}}, \bibinfo {author}
  {\bibfnamefont {K.}~\bibnamefont {Guckelsberger}}, \ and\ \bibinfo {author}
  {\bibfnamefont {R.}~\bibnamefont {Scherm}},\ }\bibfield  {title} {\enquote
  {\bibinfo {title} {Effective mass, spin fluctuations, and zero sound in
  liquid ${}^{3}\mathrm{He}$},}\ }\href {\doibase 10.1103/PhysRevB.61.1421}
  {\bibfield  {journal} {\bibinfo  {journal} {Phys. Rev. B}\ }\textbf {\bibinfo
  {volume} {61}},\ \bibinfo {pages} {1421--1432} (\bibinfo {year}
  {2000})}\BibitemShut {NoStop}%
\bibitem [{\citenamefont {Boninsegni}\ \emph {et~al.}(2006)\citenamefont
  {Boninsegni}, \citenamefont {Prokof'ev},\ and\ \citenamefont
  {Svistunov}}]{worm2006}%
  \BibitemOpen
  \bibfield  {author} {\bibinfo {author} {\bibfnamefont {Massimo}\ \bibnamefont
  {Boninsegni}}, \bibinfo {author} {\bibfnamefont {Nikolay}\ \bibnamefont
  {Prokof'ev}}, \ and\ \bibinfo {author} {\bibfnamefont {Boris}\ \bibnamefont
  {Svistunov}},\ }\bibfield  {title} {\enquote {\bibinfo {title} {Worm
  algorithm for continuous-space path integral monte carlo simulations},}\
  }\href {\doibase 10.1103/PhysRevLett.96.070601} {\bibfield  {journal}
  {\bibinfo  {journal} {Phys. Rev. Lett.}\ }\textbf {\bibinfo {volume} {96}},\
  \bibinfo {pages} {070601} (\bibinfo {year} {2006})}\BibitemShut {NoStop}%
\bibitem [{\citenamefont {Dornheim}\ \emph {et~al.}(2022)\citenamefont
  {Dornheim}, \citenamefont {Moldabekov}, \citenamefont {Vorberger},
  \citenamefont {Kählert},\ and\ \citenamefont {Bonitz}}]{Dornheim2022}%
  \BibitemOpen
  \bibfield  {author} {\bibinfo {author} {\bibfnamefont {Tobias}\ \bibnamefont
  {Dornheim}}, \bibinfo {author} {\bibfnamefont {Zhandos}\ \bibnamefont
  {Moldabekov}}, \bibinfo {author} {\bibfnamefont {Jan}\ \bibnamefont
  {Vorberger}}, \bibinfo {author} {\bibfnamefont {Hanno}\ \bibnamefont
  {Kählert}}, \ and\ \bibinfo {author} {\bibfnamefont {Michael}\ \bibnamefont
  {Bonitz}},\ }\bibfield  {title} {\enquote {\bibinfo {title} {Electronic pair
  alignment and roton feature in the warm dense electron gas},}\ }\href
  {\doibase 10.1038/s42005-022-01078-9} {\bibfield  {journal} {\bibinfo
  {journal} {Communications Physics}\ }\textbf {\bibinfo {volume} {5}},\
  \bibinfo {pages} {304} (\bibinfo {year} {2022})}\BibitemShut {NoStop}%
\bibitem [{\citenamefont {Feynman}\ and\ \citenamefont
  {Hibbs}(2010)}]{Feynmanbook}%
  \BibitemOpen
  \bibfield  {author} {\bibinfo {author} {\bibfnamefont {R.P.}\ \bibnamefont
  {Feynman}}\ and\ \bibinfo {author} {\bibfnamefont {A.R.}\ \bibnamefont
  {Hibbs}},\ }\href@noop {} {\emph {\bibinfo {title} {{Quantum Mechanics and
  Path Integrals}}}}\ (\bibinfo  {publisher} {Dover Publications Inc.},\
  \bibinfo {year} {2010})\BibitemShut {NoStop}%
\bibitem [{\citenamefont {Casulleras}\ and\ \citenamefont
  {Boronat}(2000)}]{Boronat2000}%
  \BibitemOpen
  \bibfield  {author} {\bibinfo {author} {\bibfnamefont {J.}~\bibnamefont
  {Casulleras}}\ and\ \bibinfo {author} {\bibfnamefont {J.}~\bibnamefont
  {Boronat}},\ }\bibfield  {title} {\enquote {\bibinfo {title} {Progress in
  monte carlo calculations of fermi systems: Normal liquid
  $^{3}\mathrm{He}$},}\ }\href {\doibase 10.1103/PhysRevLett.84.3121}
  {\bibfield  {journal} {\bibinfo  {journal} {Phys. Rev. Lett.}\ }\textbf
  {\bibinfo {volume} {84}},\ \bibinfo {pages} {3121--3124} (\bibinfo {year}
  {2000})}\BibitemShut {NoStop}%
\bibitem [{\citenamefont {Ceperley}(1996)}]{CeperleyFermi}%
  \BibitemOpen
  \bibfield  {author} {\bibinfo {author} {\bibfnamefont {D.M.}\ \bibnamefont
  {Ceperley}},\ }\bibfield  {title} {\enquote {\bibinfo {title} {Path integral
  monte carlo methods for fermions},}\ }in\ \href@noop {} {\emph {\bibinfo
  {booktitle} {Monte Carlo and Molecular Dynamics of Condensed Matter
  Systems}}},\ \bibinfo {editor} {edited by\ \bibinfo {editor} {\bibfnamefont
  {K.}~\bibnamefont {Binder}}\ and\ \bibinfo {editor} {\bibfnamefont
  {G.}~\bibnamefont {Ciccotti}}}\ (\bibinfo  {publisher} {Italian Physical
  Society},\ \bibinfo {address} {Bologna},\ \bibinfo {year} {1996})\BibitemShut
  {NoStop}%
\bibitem [{\citenamefont {Troyer}\ and\ \citenamefont
  {Wiese}(2005)}]{Troyer2005}%
  \BibitemOpen
  \bibfield  {author} {\bibinfo {author} {\bibfnamefont {Matthias}\
  \bibnamefont {Troyer}}\ and\ \bibinfo {author} {\bibfnamefont {Uwe-Jens}\
  \bibnamefont {Wiese}},\ }\bibfield  {title} {\enquote {\bibinfo {title}
  {Computational complexity and fundamental limitations to fermionic quantum
  monte carlo simulations},}\ }\href {\doibase 10.1103/PhysRevLett.94.170201}
  {\bibfield  {journal} {\bibinfo  {journal} {Phys. Rev. Lett.}\ }\textbf
  {\bibinfo {volume} {94}},\ \bibinfo {pages} {170201} (\bibinfo {year}
  {2005})}\BibitemShut {NoStop}%
\bibitem [{\citenamefont {Takahashi}\ and\ \citenamefont
  {Imada}(1984)}]{Takahashi1984}%
  \BibitemOpen
  \bibfield  {author} {\bibinfo {author} {\bibfnamefont {Minoru}\ \bibnamefont
  {Takahashi}}\ and\ \bibinfo {author} {\bibfnamefont {Masatoshi}\ \bibnamefont
  {Imada}},\ }\bibfield  {title} {\enquote {\bibinfo {title} {Monte carlo
  calculation of quantum systems},}\ }\href {\doibase 10.1143/JPSJ.53.963}
  {\bibfield  {journal} {\bibinfo  {journal} {Journal of the Physical Society
  of Japan}\ }\textbf {\bibinfo {volume} {53}},\ \bibinfo {pages} {963--974}
  (\bibinfo {year} {1984})},\ \Eprint
  {http://arxiv.org/abs/https://doi.org/10.1143/JPSJ.53.963}
  {https://doi.org/10.1143/JPSJ.53.963} \BibitemShut {NoStop}%
\bibitem [{\citenamefont {Filinov}\ \emph {et~al.}(2001)\citenamefont
  {Filinov}, \citenamefont {Bonitz}, \citenamefont {Ebeling},\ and\
  \citenamefont {Fortov}}]{filinov_ppcf_01}%
  \BibitemOpen
  \bibfield  {author} {\bibinfo {author} {\bibfnamefont {V~S}\ \bibnamefont
  {Filinov}}, \bibinfo {author} {\bibfnamefont {M}~\bibnamefont {Bonitz}},
  \bibinfo {author} {\bibfnamefont {W}~\bibnamefont {Ebeling}}, \ and\ \bibinfo
  {author} {\bibfnamefont {V~E}\ \bibnamefont {Fortov}},\ }\bibfield  {title}
  {\enquote {\bibinfo {title} {Thermodynamics of hot dense {H}-plasmas: path
  integral {M}onte {C}arlo simulations and analytical approximations},}\ }\href
  {http://stacks.iop.org/0741-3335/43/i=6/a=301} {\bibfield  {journal}
  {\bibinfo  {journal} {Plasma Phys. Control. Fusion}\ }\textbf {\bibinfo
  {volume} {{\bf 43}}},\ \bibinfo {pages} {743} (\bibinfo {year}
  {2001})}\BibitemShut {NoStop}%
\bibitem [{\citenamefont {Lyubartsev}(2005)}]{Lyubartsev_2005}%
  \BibitemOpen
  \bibfield  {author} {\bibinfo {author} {\bibfnamefont {Alexander~P}\
  \bibnamefont {Lyubartsev}},\ }\bibfield  {title} {\enquote {\bibinfo {title}
  {Simulation of excited states and the sign problem in the path integral monte
  carlo method},}\ }\href {\doibase 10.1088/0305-4470/38/30/003} {\bibfield
  {journal} {\bibinfo  {journal} {Journal of Physics A: Mathematical and
  General}\ }\textbf {\bibinfo {volume} {38}},\ \bibinfo {pages} {6659--6674}
  (\bibinfo {year} {2005})}\BibitemShut {NoStop}%
\bibitem [{\citenamefont {Chin}(2015)}]{Chin2015}%
  \BibitemOpen
  \bibfield  {author} {\bibinfo {author} {\bibfnamefont {Siu~A.}\ \bibnamefont
  {Chin}},\ }\bibfield  {title} {\enquote {\bibinfo {title} {High-order
  path-integral monte carlo methods for solving quantum dot problems},}\ }\href
  {\doibase 10.1103/PhysRevE.91.031301} {\bibfield  {journal} {\bibinfo
  {journal} {Phys. Rev. E}\ }\textbf {\bibinfo {volume} {91}},\ \bibinfo
  {pages} {031301} (\bibinfo {year} {2015})}\BibitemShut {NoStop}%
\bibitem [{\citenamefont {Dornheim}\ \emph {et~al.}(2015)\citenamefont
  {Dornheim}, \citenamefont {Groth}, \citenamefont {Filinov},\ and\
  \citenamefont {Bonitz}}]{dornheim_njp15}%
  \BibitemOpen
  \bibfield  {author} {\bibinfo {author} {\bibfnamefont {Tobias}\ \bibnamefont
  {Dornheim}}, \bibinfo {author} {\bibfnamefont {Simon}\ \bibnamefont {Groth}},
  \bibinfo {author} {\bibfnamefont {Alexey}\ \bibnamefont {Filinov}}, \ and\
  \bibinfo {author} {\bibfnamefont {Michael}\ \bibnamefont {Bonitz}},\
  }\bibfield  {title} {\enquote {\bibinfo {title} {{Permutation blocking path
  integral Monte Carlo: a highly efficient approach to the simulation of
  strongly degenerate non-ideal fermions}},}\ }\href
  {http://stacks.iop.org/1367-2630/17/i=7/a=073017} {\bibfield  {journal}
  {\bibinfo  {journal} {New J. Phys.}\ }\textbf {\bibinfo {volume} {17}},\
  \bibinfo {pages} {073017} (\bibinfo {year} {2015})}\BibitemShut {NoStop}%
\bibitem [{\citenamefont {Chin}\ and\ \citenamefont {Chen}(2002)}]{Chin2002}%
  \BibitemOpen
  \bibfield  {author} {\bibinfo {author} {\bibfnamefont {Siu~A.}\ \bibnamefont
  {Chin}}\ and\ \bibinfo {author} {\bibfnamefont {C.~R.}\ \bibnamefont
  {Chen}},\ }\bibfield  {title} {\enquote {\bibinfo {title} {Gradient
  symplectic algorithms for solving the schrödinger equation with
  time-dependent potentials},}\ }\href {\doibase 10.1063/1.1485725} {\bibfield
  {journal} {\bibinfo  {journal} {The Journal of Chemical Physics}\ }\textbf
  {\bibinfo {volume} {117}},\ \bibinfo {pages} {1409--1415} (\bibinfo {year}
  {2002})},\ \Eprint {http://arxiv.org/abs/https://doi.org/10.1063/1.1485725}
  {https://doi.org/10.1063/1.1485725} \BibitemShut {NoStop}%
\bibitem [{\citenamefont {Sakkos}\ \emph {et~al.}(2009)\citenamefont {Sakkos},
  \citenamefont {Casulleras},\ and\ \citenamefont {Boronat}}]{Sakkos2009}%
  \BibitemOpen
  \bibfield  {author} {\bibinfo {author} {\bibfnamefont {K.}~\bibnamefont
  {Sakkos}}, \bibinfo {author} {\bibfnamefont {J.}~\bibnamefont {Casulleras}},
  \ and\ \bibinfo {author} {\bibfnamefont {J.}~\bibnamefont {Boronat}},\
  }\bibfield  {title} {\enquote {\bibinfo {title} {High order chin actions in
  path integral monte carlo},}\ }\href {\doibase 10.1063/1.3143522} {\bibfield
  {journal} {\bibinfo  {journal} {The Journal of Chemical Physics}\ }\textbf
  {\bibinfo {volume} {130}},\ \bibinfo {pages} {204109} (\bibinfo {year}
  {2009})},\ \Eprint {http://arxiv.org/abs/https://doi.org/10.1063/1.3143522}
  {https://doi.org/10.1063/1.3143522} \BibitemShut {NoStop}%
\bibitem [{\citenamefont {Andreani}\ \emph {et~al.}(2006)\citenamefont
  {Andreani}, \citenamefont {Pantalei},\ and\ \citenamefont
  {Senesi}}]{Andreani_2006}%
  \BibitemOpen
  \bibfield  {author} {\bibinfo {author} {\bibfnamefont {C}~\bibnamefont
  {Andreani}}, \bibinfo {author} {\bibfnamefont {C}~\bibnamefont {Pantalei}}, \
  and\ \bibinfo {author} {\bibfnamefont {R}~\bibnamefont {Senesi}},\ }\bibfield
   {title} {\enquote {\bibinfo {title} {Mean kinetic energy of helium atoms in
  fluid $^3\text{He}$ and $^3\text{He}$-$^4\text{He}$ mixtures},}\ }\href
  {\doibase 10.1088/0953-8984/18/24/001} {\bibfield  {journal} {\bibinfo
  {journal} {Journal of Physics: Condensed Matter}\ }\textbf {\bibinfo {volume}
  {18}},\ \bibinfo {pages} {5587} (\bibinfo {year} {2006})}\BibitemShut
  {NoStop}%
\bibitem [{\citenamefont {Mazzanti}\ \emph {et~al.}(2004)\citenamefont
  {Mazzanti}, \citenamefont {Polls}, \citenamefont {Boronat},\ and\
  \citenamefont {Casulleras}}]{Boronat2004}%
  \BibitemOpen
  \bibfield  {author} {\bibinfo {author} {\bibfnamefont {F.}~\bibnamefont
  {Mazzanti}}, \bibinfo {author} {\bibfnamefont {A.}~\bibnamefont {Polls}},
  \bibinfo {author} {\bibfnamefont {J.}~\bibnamefont {Boronat}}, \ and\
  \bibinfo {author} {\bibfnamefont {J.}~\bibnamefont {Casulleras}},\ }\bibfield
   {title} {\enquote {\bibinfo {title} {High-momentum response of liquid
  $^{3}\mathrm{H}\mathrm{e}$},}\ }\href {\doibase
  10.1103/PhysRevLett.92.085301} {\bibfield  {journal} {\bibinfo  {journal}
  {Phys. Rev. Lett.}\ }\textbf {\bibinfo {volume} {92}},\ \bibinfo {pages}
  {085301} (\bibinfo {year} {2004})}\BibitemShut {NoStop}%
\bibitem [{\citenamefont {Puff}(1965)}]{Puff}%
  \BibitemOpen
  \bibfield  {author} {\bibinfo {author} {\bibfnamefont {R.~D.}\ \bibnamefont
  {Puff}},\ }\bibfield  {title} {\enquote {\bibinfo {title} {Application of sum
  rules to the low-temperature interacting boson system},}\ }\href {\doibase
  10.1103/PhysRev.137.A406} {\bibfield  {journal} {\bibinfo  {journal} {Phys.
  Rev.}\ }\textbf {\bibinfo {volume} {137}},\ \bibinfo {pages} {A406--A416}
  (\bibinfo {year} {1965})}\BibitemShut {NoStop}%
\bibitem [{\citenamefont {Dornheim}\ \emph {et~al.}(2020)\citenamefont
  {Dornheim}, \citenamefont {Vorberger},\ and\ \citenamefont
  {Bonitz}}]{dornheim-etal.nl.2020prl}%
  \BibitemOpen
  \bibfield  {author} {\bibinfo {author} {\bibfnamefont {Tobias}\ \bibnamefont
  {Dornheim}}, \bibinfo {author} {\bibfnamefont {Jan}\ \bibnamefont
  {Vorberger}}, \ and\ \bibinfo {author} {\bibfnamefont {Michael}\ \bibnamefont
  {Bonitz}},\ }\bibfield  {title} {\enquote {\bibinfo {title} {Nonlinear
  electronic density response in warm dense matter},}\ }\href {\doibase
  10.1103/PhysRevLett.125.085001} {\bibfield  {journal} {\bibinfo  {journal}
  {Phys. Rev. Lett.}\ }\textbf {\bibinfo {volume} {125}},\ \bibinfo {pages}
  {085001} (\bibinfo {year} {2020})}\BibitemShut {NoStop}%
\end{thebibliography}%

\end{document}